\documentclass[aps,prd,amssymb,eqsecnum,showpacs]{revtex4}
\usepackage{graphicx}
\usepackage{psfrag}

\newcommand{\gz}{\mbox{\em \r{g}\hspace{0.1mm}}}  
\newcommand{\Rz}{\mbox{\em \r{R}}}

\newcommand{\nablaz}{\nabla\hspace{-0.27cm}{}^{\mbox{\r{~}}}{}\hspace{-0.22cm}}

\newcommand{\Gammaz}{\Gamma\hspace{-0.20cm}{}^{\mbox{\r{~}}}{}\hspace{-0.18cm}}

\begin{document}

\title{Boundary Conditions for the Gravitational Field}

\author{Jeffrey Winicour${}^{1,2}$
       }
\affiliation{
${}^{1}$ Department of Physics and Astronomy \\
         University of Pittsburgh, Pittsburgh, PA 15260, USA \\
${}^{2}$ Max-Planck-Institut f\" ur
         Gravitationsphysik, Albert-Einstein-Institut, \\
	 14476 Golm, Germany
	 }

\begin{abstract}

A review of the treatment of boundaries in general relativity is presented
with the emphasis on application to the formulations of Einstein's equations
used in numerical relativity. At present,  it is known how to treat
boundaries in the harmonic formulation of Einstein's equations and a tetrad
formulation of the Einstein-Bianchi system. However, a universal approach
valid for other formulations is not in hand. In particular, there is no
satisfactory boundary theory  for the 3+1 formulations which have been
highly successful in binary black hole simulation. I discuss the underlying
problems that make the initial-boundary value problem much more complicated than
the Cauchy problem. I review the progress that has been made and the
important open questions that remain.

\end{abstract}


\maketitle

Science is a differential equation. Religion is a boundary condition. (Alan
Turing, quoted in J.D. Barrow,``Theories of Everything'').

\section{Introduction}

There are no natural boundaries for the gravitational field analogous to the
conducting boundaries that play a major role in electromagnetism. In
principle, there is no need to introduce any. The behavior of the universe
as a whole can be posed as an initial value (Cauchy) problem. In an initial
value problem, data is given on a spacelike hypersurface ${\cal S}_0$. The
problem is to determine a solution in the future domain of dependence ${\cal
D}^{+}({\cal S}_0)$, which consists of those points whose past directed
characteristics all intersect ${\cal S}_0$. The problem is well-posed if
there exists a unique solution which depends continuously on the initial
data. The pioneering work of Y. Bruhat~\cite{bruhat} showed that the initial
value problem for the (classical) vacuum gravitational field is well-posed.
Assuming that matter fields do not spoil things, this suggests that the
global cosmological problem of treating the universe as a whole can be
solved in a physically meaningful way, i.e. in a way such that the solution
does not undergo uncontrolled variation under a perturbation of the initial
data.  This is indeed the case for the presently accepted cosmological model
of an accelerating universe (positive cosmological constant) where the
conformal boundary at future null infinity ${\cal I}^+$ is spacelike. In a
conformally compactified picture, ${\cal I}^+$ acts as a spacelike cap on
the future evolution domain and no boundary condition is necessary or indeed
allowed.

In practice, of course, treating an isolated system as part of a global
cosmological spacetime is too complicated  a problem without oversimplifying
assumptions such as isotropy or homogeneity. One global approach applicable
to isolated systems is to base the Cauchy problem on the analogue of a
foliation of Minkowski spacetime by the hyperboloidal hypersurfaces
\begin{equation}
    t^2 - x^2 -y^2 - z^2 = T^2 , \quad t \ge T .
\end{equation}  In a Penrose conformally compactified picture~\cite{penrose},
this foliation asymptotes to the light cones and extends to a foliation of
future null infinity ${\cal I}^+$. The analogue in curved spacetime is a
foliation by positive constant mean curvature hypersurfaces. Since no light
rays can enter an asymptotically flat spacetime through ${\cal I}^+$, no
boundary data are needed to evolve the interior spacetime. In addition, the
waveform and polarization of the outgoing radiation can be unambiguously
calculated at ${\cal I}^+$ in terms of the Bondi news function~\cite{bondi}.
This approach was first extensively developed by Friedrich~\cite{helconf}
who formulated a hyperbolic version of the Einstein-Bianchi system of
equations, which is manifestly regular at ${\cal I}^+$, in terms of the
conformally rescaled metric, connection and Weyl curvature. This is
potentially the basis for a very attractive numerical approach to simulate
global problems such as gravitational wave production. For reviews of
progress on the numerical implementation
see~\cite{fraurev,saschrev1,saschrev2}. There has been some success in
simulating model axisymmetric problems~\cite{frauhein}. More recently, there
have been other attempts at the hyperboloidal approach based upon the
Einstein equations for the conformal metric. Zengino{\u g}lu~\cite{zen} has
implemented a code based upon a generalized harmonic formulation in which
the gauge source terms produce a hyperbolic foliation. A mixed
hyperbolic-elliptic system proposed by Moncrief and Rinne~\cite{moncrinn}
has been implemented as an axisymmetric code~\cite{rinn} which produces long
term stable evolutions. Another hyperbolic-elliptic system  based upon a
tetrad approach has been developed by Bardeen, Sarbach and
Buchman~\cite{bardsarbuch}. In spite of the attractiveness of the
hyperboloidal approach and its success with model problems, considerable
work remains to make it applicable to systems of astrophysical interest. 

A different global approach is to match the Cauchy evolution inside a a
finite worldtube to an exterior characteristic evolution extending to ${\cal
I}^+$. In this approach, called Cauchy-characteristic matching, the
characteristic evolution is constructed by using the Cauchy evolution to
supply characteristic data on an inner worldtube, while the characteristic
evolution supplies the outer boundary data for the Cauchy evolution. The
success of Cauchy-characteristic matching depends upon the proper
mathematical and computational treatment of the initial-boundary value
problem (IBVP) for the Cauchy evolution. This approach has been
successfully implemented in the linearized
regime~\cite{harl} but also needs considerable additional work to apply to
astrophysical systems. See~\cite{winrev} for a review.

Instead of a global treatment, the standard approach in numerical
relativity, as in computational studies of other hyperbolic systems, is to
introduce an artificial outer boundary. Ideally,  the outer boundary
treatment is designed to represent a passive external universe by allowing
radiation to cross only in the outgoing direction. This is the primary
application of the IBVP. Other possible
applications, which I will not consider, are the timelike conformal boundary
to a universe with negative cosmological constant and the membranes which
play a role in higher dimensional theories. While there are no natural
boundaries in classical gravitational theory,  boundaries do play a central
role in the  ideas of holographic duality introduced in higher dimensional
attempts at quantum gravity. Such applications are also beyond the scope of
this review as well as beyond my own expertise. Here, I confine my attention
to 4-dimensional spacetime, although the techniques governing a well-posed
IBVP readily extend to hyperbolic systems in any dimension.

\begin{figure}[htb]
\begin{center}
\includegraphics[scale=.4]{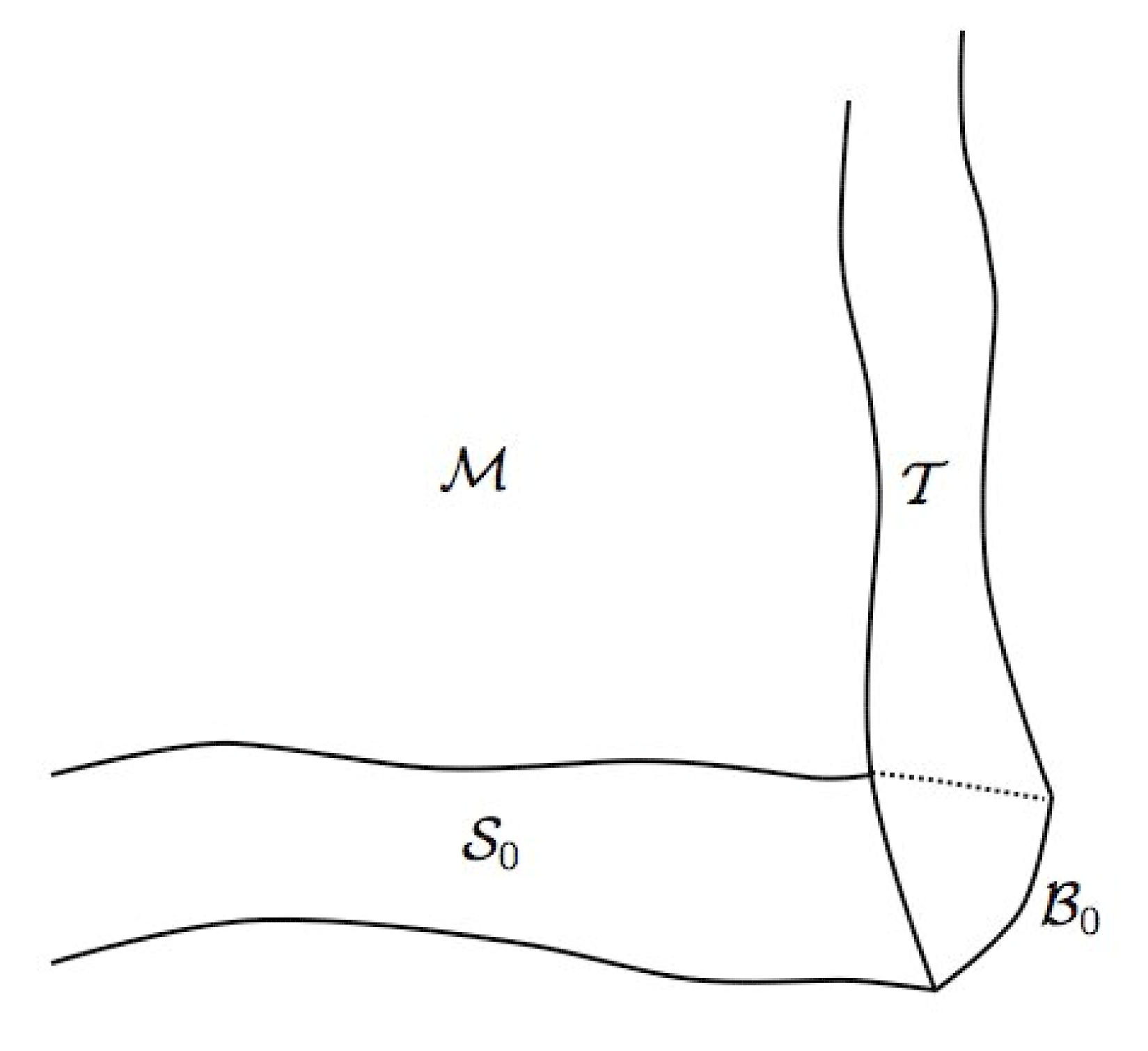}  
\caption{Data on the 3-manifolds ${\cal S}_0$ and ${\cal T}$, which
 intersect in the 2-surface ${\cal B}_0$, locally determine a solution
 in the spacetime manifold ${\cal M}$}.  
\end{center}
\label{fig:bound}
\end{figure}

In the IBVP, data on a timelike boundary ${\cal T}$, which meets ${\cal
S}_0$ in a surface  ${\cal B}_0$, are used to further extend the solution of
the Cauchy problem to the domain of dependence ${\cal D}^{+}({\cal S}_0 \cup
{\cal T})$. In the simulation of an isolated astrophysical system containing
neutron stars and black holes, the outer boundary ${\cal T}$ is coincident
with the boundary of the computational grid and ${\cal B}_0$ is
topologically a sphere surrounding the system. However, for purposes of
treating the underlying mathematical and computational problems, it suffices
to concentrate on the local problem in the neighborhood of some point on the
intersection ${\cal B}_0$ between the Cauchy hypersurface  ${\cal S}_0$  and
the boundary ${\cal T}$. For hyperbolic systems, the global solution in the
spacetime manifold ${\cal M}$ can be obtained by patching together local
solutions. This is because the finite speed of propagation allows {\it
localization} of the problem. The setting for this local problem is depicted
in Fig.~\ref{fig:bound}.

The IBVP for Einstein's equations only received widespread attention after
its importance to the artificial outer boundaries used in numerical
relativity was pointed out~\cite{stewart}. The first {\it strongly}
well-posed IBVP was achieved for a tetrad version of the Einstein-Bianchi
system, expressed in first differential order form, which included the
tetrad,  connection and curvature tensor as evolution fields~\cite{fn}.
Strong well-posedness guarantees the existence of a unique solution which
depends continuously on both the Cauchy data and the boundary data. A
strongly well-posed IBVP was later established for the harmonic formulation
of Einstein's equations as a system of second order quasilinear wave
equations for the metric~\cite{wpgs}.  The results were further generalized
in~\cite{wpe,isol} to apply to a general quasilinear class of symmetric
hyperbolic systems whose boundary conditions have a certain hierarchical
form.

A review of the IBVP in general relativity must per force be of a different
nature than a review of the Cauchy problem. The local properties of the
Cauchy problem are now well understood. Several excellent reviews
exist~\cite{crev1,crev2,crev3,crev4}. For the IBVP, the results are not
comprehensive and  are closely tied to the choice of hyperbolic reduction of
Einstein's equations. There are only a few universal features and, in
particular, there is no satisfactory treatment of the 3+1 formulation which
is extensively used in numerical relativity. For that reason, I will adopt a
presentation which differs somewhat from the standard approach with the
motivation of setting up a bare bones  framework whose flexibility might be
helpful in further investigations. 	

My presentation is also biased by the important role of the IBVP in
numerical relativity, which treats Einstein's equations as a set of partial
differential equations (PDEs) governing the metric in some preferred
coordinate system. On the other hand, from a geometrical perspective, one of
the most fundamental and beautiful results of general relativity is that the
properties of the local Cauchy problem can be summed up in geometric terms
independent of any coordinates or explicit PDEs. This geometric formulation
only came about after the Cauchy problem was well understood from the PDE
point-of-view. The importance of the geometric approach to the numerical
relativist is that it supplies a common starting point for discussing and
comparing different formulations of Einstein's equations. Presently, the PDE
aspects of metric formulations of the IBVP are only understood in the
harmonic formulation. In order to transfer this insight into other
formulations a geometric framework can serve as an important guide. For that
reason, I will shift often between the PDE and geometric approach. When the
emphasis is on the geometric side I will use abstract indices, e.g. $v^a$ to
denote a vector field, and on the PDE side I will use coordinate indices, e.g.
$v^\mu= (v^t,v^i)$, to denote components with respect to spacetime
coordinates $x^\mu= (t,x^i)$.

The standard mathematical approach to the IBVP is to first establish the
well-posedness of the underlying Cauchy problem, and next the local
half-space problem. If these individual problems are well-posed  (in a sense
to be qualified later) then the problems with more general boundaries will
also be well-posed. Thus I start my review in Sec.~\ref{sec:cauchy} by first
providing some brief background material for the Cauchy problem. 

Next, in Sec.~\ref{sec:compl}, I point out the complications in going from
the Cauchy problem to the IBVP. The IBVP for Einstein's equations is not
well understood due to problems arising from the constraint equations. The
motivation for this work stems from the need for an improved understanding
and implementation of boundary conditions in the computational codes being
used to simulate binary black holes. The ability to compute the details of
the gravitational radiation produced by compact astrophysical sources, such
as coalescing black holes, is of major importance to the success of
gravitational wave astronomy. If the simulation of such systems is based upon
a well-posed Cauchy problem but not  a well-posed IBVP then the
results cannot be trusted in the domain of dependence of the outer boundary.
In Sec's~\ref{sec:bare},~\ref{sec:initial} and~\ref{sec:hyperb}, I present the
underlying mathematical theory.

Early computational work in general relativity focused on the Cauchy problem
and the IBVP only received considerable attention after its importance to
stable and accurate simulations was recognized.
I discuss some history of the work on the IBVP in Sec.~\ref{sec:history},
for the purpose of pointing out some of the partial
successes and ideas which may be of future use.

In addition to the mathematical issue of an appropriate boundary condition,
the description of a binary black hole as an isolated system raises the
physical issue of the appropriate outer boundary data. In the absence of an
exterior solution, which could provide this data by matching, the standard
practice is to set this data to zero.  This raises the question, discussed in
Sec.~\ref{sec:absorbc}, of how to formulate a non-reflecting outer boundary
condition in order to avoid spurious incoming radiation. 

Sec's~\ref{sec:fn} and~\ref{sec:harm} describe the two strongly
well-posed formulations of the
IBVP, which are known at the present time.  Neither is based upon a $3+1$
formulation and they both handle the constraints in different ways.  This
prompts the discussion, in Sec.~\ref{sec:constr}, of constraint enforcement
in the $3+1$ formulations. The resolution of issues regarding
geometric uniqueness, discussed in Sec.~\ref{sec:geom}, would shed light on
the universal features of the IBVP that would perhaps guide the way to
a successful $3+1$ treatment.

There is also the computational problem of turning a well-posed IBVP into a
stable and accurate evolution code.  I will not go into the details of the
large range of techniques which are necessary for the successful
implementation of a numerical relativity code. Since initiating this
review,  I have learned of a separate review in progress~\cite{sarbtig} which covers
such numerical techniques in great detail.  Building a numerical relativity
code is a complex undertaking. As observed by Post and Votta~\cite{postvot}
in a study of large scale computational projects, ``the peer review process
in computational science generally doesn't provide as effective a filter as
it does for experiment or theory. Many things that a referee cannot detect
could be wrong with a computational science paper\ldots The few existing
studies of error levels in scientific computer codes indicate that the
defect rate is about seven faults per 1000 lines of Fortran''. They
emphasize that ``New methods of verifying and validating complex codes are
mandatory if computational science is to fulfill its promise for science and
society''. These observations are especially pertinent for numerical
relativity where validation by agreement with experiment is not yet possible.
In that spirit, I discuss the code tests that have been proposed and carried
out for the gravitational IBVP in Sec.~\ref{sec:num}. 

My aim has been to present the background material which might  open new
avenues for a better understanding of the IBVP and lead to progress on some
of  the important open questions posed in Sec.~\ref{sec:quest}. 

\section {The Cauchy problem}
\label{sec:cauchy}

Here I summarize those aspects of the Cauchy problem which are
fundamental to the IBVP. For more detail, see~\cite{crev1,crev2,crev3,crev4}. 

In contrast to Newtonian theory, which describes gravity in terms of an
elliptic Poisson equation that propagates the gravitational field
instantaneously, the retarded interactions implicit in general relativity
give rise to new features such as gravitational waves. Wave propagation
results from the mathematical property that Einstein's equations can be
reduced to a  hyperbolic system of PDEs. However, the coordinate freedom in
Einstein's theory admits gauge waves which propagate with arbitrarily high
speeds, including speeds faster than light.  Einstein's equations are not
a priori a hyperbolic system in which propagation speeds must be bounded and
for which an initial value problem can be posed. 

This crucial step in going from Einstein's equations to a hyperbolic system
has been highlighted by Friedrich as the process of hyperbolic
reduction~\cite{Friedrich}. The first and most famous example of hyperbolic
reduction was through the introduction of harmonic coordinates, which led to
the classic result that the Cauchy problem for the harmonic formulation of
Einstein's equations is well-posed~\cite{bruhat}. Here I summarize the
hyperbolic reduction of Einstein's equations in in terms of generalized
harmonic coordinates $x^\alpha=(t,x^i)=(t,x,y,z)$, which are functionally
independent solutions of the curved space scalar wave equation
\begin{equation}
    \Box x^\mu  = \frac{1}{\sqrt{-g}}\partial_\alpha
        (\sqrt{-g}g^{\alpha\beta}\partial_\beta x^\mu) =-\hat \Gamma^\mu ,
\end{equation}
where $\hat \Gamma^\mu$ are gauge source functions~\cite{Friedrich}. In
terms of the connection $\Gamma^\mu_{\alpha\beta}$, these harmonic
conditions are
\begin{equation}
   {\cal C}^\mu :=\Gamma^\mu -\hat \Gamma^\mu =0,
\label{eq:harmcond}
\end{equation}
where
\begin{equation}
     \Gamma^\mu = g^{\alpha\beta}\Gamma^\mu_{\alpha\beta}= 
    -\frac{1}{\sqrt {-g}}\partial_\alpha ( \sqrt{-g}g^{\alpha\mu} ).
\end{equation}
 
The hyperbolic reduction of the Einstein tensor results from setting
\begin{equation}
     E^{\mu\nu}:= G^{\mu\nu} -\nabla^{(\mu}{\cal C}^{\nu)} 
       +\frac{1}{2}g^{\mu\nu}\nabla_\rho{\cal C}^\rho =0 ,
       \label{eq:creduced}
\end{equation}
where $C^\nu$ is treated formally as a vector field in constructing the
``covariant'' derivatives $\nabla^{\mu}C^{\nu}$. (In generalized harmonic
formulations based upon a background connection, $C^\nu$ is a legitimate
vector field. See Sec.~\ref{sec:harm}.)

When the harmonic gauge source functions have the functional dependence
$\hat \Gamma^\nu(x,g)$, the principal part of
(\ref{eq:creduced}) reduces to the wave operator acting on the densitized
metric, i.e.
\begin{equation}
     E^{\mu\nu}= \frac{1}{2\sqrt{-g}}\partial_\alpha \bigg (g^{\alpha\beta}
   \partial_\beta( \sqrt{-g}g^{\mu\nu})\bigg )  + \text{lower order terms}.
           \label{eq:reduced}
\end{equation}
Thus the harmonic evolution equations (\ref{eq:creduced}) are quasilinear
wave equations for the components of the densitized metric
$\sqrt{-g}g^{\mu\nu}$. The well-posedness of the Cauchy problem for the
system (\ref{eq:creduced}) then follows from known results for systems of
quasilinear wave equations. (It is important to bear in mind that such
results are local in time since there is no general theory for the global
existence of solutions to nonlinear equations.)

In turn, the well-posedness of the Cauchy problem for the harmonic Einstein
equations  also follows provided that the harmonic conditions ${\cal
C}^\mu=0$ are preserved under the evolution. The proof of constraint
preservation results from applying the contracted Bianchi identity $\nabla_\mu
G^{\mu\nu} =0$ to (\ref{eq:creduced}). This leads to a homogeneous wave
equation for ${\cal C}^\mu$,
\begin{equation}
    \nabla^\rho \nabla_\rho \, {\cal C}^\mu +R^\mu_\rho \,{\cal C}^\rho=0.
  \label{eq:bianchi}
\end{equation}
If the initial data enforce
\begin{equation}
    {\cal C}^\mu |_{{\cal S}_0}= 0
    \label{eq:c0}
 \end{equation}
 and 
 \begin{equation}   
    \partial_t {\cal C}^\mu |_{{\cal S}_0}=0
    \label{eq:ct0}
\end{equation}
then the unique solution of (\ref{eq:bianchi}) is ${\cal C}^\rho=0$. It is
easy to satisfy  (\ref{eq:c0})  by algebraically determining the initial
values of $\partial_t g^{\mu t}$ in terms of the  initial values of
$g^{\mu\nu}$ and their spatial derivatives. In order to see how to satisfy
(\ref{eq:ct0}) note that the reduced equations (\ref{eq:reduced}) imply
\begin{equation}
     G^{\mu\nu} n_\nu=n_\nu \nabla^{(\mu}{\cal C}^{\nu)} 
       -\frac{1}{2}n^\mu \nabla_\rho{\cal C}^\rho,
\label{eq:Gn}
\end{equation}
where
$$ n_\nu= -\frac{1}{\sqrt{-g^{tt}}}\partial_\nu t
$$
is the unit timelike normal to the Cauchy hyperurfaces. Thus if
\begin{equation}
     G^{\mu\nu} n_\nu|_{{\cal S}_0} =0,
     \label{eq:hamomc}
\end{equation}
i.e. if the Hamiltonian and momentum constraints are satisfied by the
initial data, and if the reduced equations (\ref{eq:creduced}) are satisfied
then it follows that 
\begin{equation}
     [n_\nu \nabla^{(\mu}{\cal C}^{\nu)} 
       -\frac{1}{2}n^\mu \nabla_\rho{\cal C}^\rho ]|_{{\cal S}_0} =0.
\label{eq:ndc}
\end{equation}
It is easy to check that (\ref{eq:ndc}) implies that  $\partial_t {\cal
C}^\mu |_{{\cal S}_0}=0$ provided ${\cal C}^\mu |_{{\cal S}_0}= 0$. 

As a result, the traditional Hamiltonian and momentum constraints on the
initial data, along with the reduced evolution equations
(\ref{eq:creduced})), imply that the initial conditions (\ref{eq:c0}) and 
(\ref{eq:ct0}) required for preserving the harmonic conditions are satisfied.
Conversely, if the Hamiltonian and momentum constraints are satisfied
initially, then (\ref{eq:Gn}) ensures that they will be preserved under
harmonic evolution. Thus the conditions ${\cal C}^\nu =0$ can be considered
as the constraints of the generalized harmonic formulation.

The formalism also allows constraint adjustments by which 
(\ref{eq:creduced}) is modified by
\begin{equation}
     E^{\mu\nu}:= G^{\mu\nu} -\nabla^{(\mu}{\cal C}^{\nu)} 
       +\frac{1}{2}g^{\mu\nu}\nabla_\rho{\cal C}^\rho 
       +A^{\mu\nu}_\sigma {\cal C}^\sigma=0,
       \label{eq:creduced2}
\end{equation}
where the coefficients
$A^{\mu\nu}_\sigma$ have the dependence $A^{\mu\nu}_\sigma(x,g,\partial
g)$. Such constraint adjustments have proved to be important in
applying constraint damping~\cite{constrdamp} in the simulation of
black holes~\cite{pret1,pret2,pret3} and in suppressing long wavelength instabilities
in a shifted gauge wave test~\cite{babev} (see Sec.~\ref{sec:num}).
However, they do not change the principal part
of the reduced equations and have no effect on well-posedness.

Historically, the first Cauchy
codes were based upon the ``3+1'' or Arnowitt-Deser-Misner (ADM)
formulation of the Einstein equations~\cite{adm}. 
The ADM formulation introduces a Cauchy foliation of
space-time by a time coordinate $t$ and expresses the 4-dimensional
metric as
\begin{equation}
  ds^2 = -\alpha^2 dt^2 + h_{ij} \left(dx^i + \beta^i dt\right)
                                 \left(dx^j + \beta^j dt\right) ,
       \label{eq:admmet}                          
\end{equation}
where $h_{ij}$ is the induced 3-metric of the $t=const$ foliation,
$\alpha$ is the lapse and $\beta^i$ the shift, with the unit  normal
to the foliation given by $n^\mu=(1,-\beta^i)/\alpha$.

The field equations are written in first differential form in terms
of the extrinsic curvature of the Cauchy foliation
$$
  k_{ij} = \frac{1}{2}{\cal L}_n g_{ij}.
$$
This can be accomplished in many ways. In one of the earliest schemes
proposed for numerical relativity by York~\cite{york},
the requirement that the 6 spatial components of the Ricci tensor vanish, i.e
$R_{ij}=0$, yields a set of evolution equations for the 3-metric and extrinsic curvature, 
\begin{eqnarray}
   \partial_t g_{ij} -{\cal L}_\beta g_{ij} &=& -2\alpha k_{ij} \\
   \partial_t k_{ij} -{\cal L}_\beta k_{ij} &=& -D_i D_j\alpha
    + \alpha\left({\cal R}_{ij} + k k_{ij} - 2 k_i^l k_{lj} \right) ,
    \label{eq:adm}
\end{eqnarray}
where $D_i$ is the connection and ${\cal R}_{ij}$
is the Ricci tensor associated with $h_{ij}$.
The Hamiltonian and momentum constraints take the form
\begin{eqnarray}
   2 G^{\mu\nu}n_\mu n_\nu ={\cal R} - k_{ij} k^{ij} + k^2 &=& 0 
    \label{eq:hamc} \\
    G^{\mu i}n_\mu = D_j \left( k^{ij} - h^{ij} k\right) &=&0,
     \label{eq:momc}
\end{eqnarray}
where ${\cal R}=h^{ij}{\cal R}_{ij}$ and $k=h^{ij}k_{ij}$.

Codes presently used for the simulation of binary black holes
apply the constraints to the initial Cauchy data
but do not enforce them during the evolution.
The choice of evolution equations may be modified by mixing in
combinations of the constraint equations. In addition, the evolution equations
must be supplemented by equations governing the lapse
and shift. There is a lot of freedom in how all this
can be done. The choices affect whether the Cauchy problem is
well-posed.

\section{Complications of the IBVP}
\label{sec:compl}

The difficulties
underlying the IBVP have recently been discussed
in~\cite{disem,hjuerg,juerg,reulsarrev}. There are several
chief complications which do not arise in the Cauchy problem.

\begin{enumerate}

\item The first complication stems from a well-known property of the
flat-space scalar wave boundary problem
\begin{equation}
 (\partial_t^2 -\nabla^2)\Phi=0 \, , \quad  x \ge 0, \,  t\ge 0.
 \label{eq:swe1}
\end{equation}
The light rays are the characteristics of the equation.
There are two characteristics associated with each
direction, e.g. the characteristics in the $\pm x$ direction. Both of
these characteristics cross the initial hypersurface $t=0$ but only one crosses
the boundary at $x=0$. As a result,
although the initial Cauchy data consist of the two pieces of information
$\Phi|_{t=0}$ and $\partial_t \Phi|_{t=0}$, only half as much boundary
data can be freely prescribed at $x=0$, e.g the Dirichlet data
$q_D=\partial_t \Phi|_{x=0}$, or the Neumann data $q_N=\partial_x \Phi|_{x=0}$ or
the Sommerfeld data $q_S=(\partial_t - \partial_x) \Phi|_{x=0} $.
Sommerfeld data is based upon
the derivative of $\Phi$  in the characteristic direction determined by
the outward normal to the boundary.
(In the first differential order formalism $(\partial_t - \partial_x) \Phi$
is an {\it ingoing variable} at the boundary.)
The choices $q_D=q_N=q_S$ do not lead to the same solution. In order
to obtain a given
physical solution, this implies that the boundary data cannot be
prescribed before the boundary condition is specified, i.e. the 
boundary data for the solution
depends upon the boundary condition, unlike the situation
for the Cauchy problem.  The analogue in the gravitational case is the
inability to prescribe both the metric and its normal derivative on a
timelike boundary, which implies the inability to freely prescribe both
the intrinsic 3-metric of the boundary and its extrinsic curvature. This
leads to a further complication regarding constraint enforcement at the
boundary, i.e. the Hamiltonian and momentum constraints (\ref{eq:hamc})
and (\ref{eq:momc}) cannot be enforced directly because they couple the
metric and its normal derivative. 

\item For computational purposes, a Sommerfeld boundary condition is preferable
because it allows numerical noise to propagate across the boundary. Thus
discretization error can leave the numerical grid, whereas Dirichlet and Neumann
boundary conditions would reflect the error and trap it in the grid. 
The Sommerfeld condition on a metric component supplies the
value of the derivative $K^\alpha \partial_\alpha g_{\mu\nu}$ in an outgoing null
direction $K^\alpha$. However, the boundary does
not pick out a unique outgoing null direction at a given point but,
instead, essentially a half cone of null directions. This complicates the geometric
formulation of a Sommerfeld boundary condition. In addition, constraint
preservation does not allow free specification of Sommerfeld data for all
components of the metric, as will be seen later in formulating the Sommerfeld
conditions (\ref{eq:hk}) - (\ref{eq:hl}).

\item The correct boundary data for the gravitational field is generally
not known except in special cases, e.g. when simulating an exact solution.
This differs from electromagnetic theory where, say, homogeneous Dirichlet
or Neumann data for the various components of the electromagnetic field correctly
describe the data for reflection from a mirror. The tacit assumption in
the simulation of an isolated system is that homogeneous Sommerfeld
data gives rise to minimal back reflection of gravitational waves from
the outer boundary. But this is an approximation which only becomes
exact in the limit of an infinite sized boundary.

\item Another major complication arises from the gauge freedom. In the
evolution  of the Cauchy data it is necessary to introduce a foliation of
the spacetime by Cauchy hypersurfaces ${\cal S}_t$, with unit timelike
normal $n_a$. The evolution of the spacetime metric
\begin{equation}
     g_{ab}=-n_a n_b +h_{ab}
     \label{eq:gdecom}
\end{equation}     
is carried out along the flow of an evolution vector field $t^a$
which is related
to the normal by the lapse $\alpha$ and shift $\beta^a$ by
\begin{equation}
    t^a= \alpha n^a + \beta^a \, , \quad  \beta^a n_a =0.
    \label{eq:lapshif}
\end{equation}         
The choice of foliation is part of the gauge freedom in the resulting solution
but does not enter into the specification of the initial data.
In the current treatments of the IBVP, the foliation is
coupled with the formulation of the boundary condition.
As a result, some gauge information enters into the
boundary condition and boundary data.

\item The partial derivative $\partial_\alpha g_{\mu\nu} $ entering into
the construction of the boundary condition for the metric has by itself
no intrinsic geometric interpretation, unless, say, a background
connection or a preferred vector field is introduced. 

\item In general, the boundary moves with respect to the initial Cauchy
hypersurface in the sense that the spacelike unit  outer normal $N_a$ to
${\cal T}$ is not orthogonal to the timelike unit normal $n_a$ to ${\cal
S}_0$. The initial velocity of the boundary is characterized by the
hyperbolic angle $\Theta$, where
\begin{equation}
              N^a n_a =\sinh \Theta
              \label{eq:hangle}
\end{equation}
Specification of $\Theta$ on the edge ${\cal B}_0$ must be included in
the data.

The coordinate specification of the location of the boundary is pure
gauge since it does not determine its location geometrically in the
sense that a curve is determined geometrically  by its its acceleration,
given its initial position and velocity. Given  ${\cal B}_0$ and $\Theta$,
the future location of the boundary should be determined in a
geometrically unique way. In the Friedrich-Nagy system, the motion of the
boundary is determined by specifying its mean extrinsic curvature. But
this is tantamount to a piece of Neumann data. Can this be accomplished
via a non-reflecting boundary condition of the Sommerfeld type?

\item In a reduction to first differential order form by introducing a
momentum $\Pi$, according to the example \begin{equation} n^a \partial_a
\Phi = \Pi, \label{eq:advect} \end{equation}   there is a further
difficulty if $\Theta \ne 0$ at the boundary. The sign of $\Theta$
determines whether $n^a$ points outward or inward to ${\cal T}$, i.e
whether $\Phi$ is an ingoing or outgoing variable.   Thus the sign of
$\Theta$ determines whether such an advection equation requires a
boundary condition. This forces a Dirichlet condition on the normal
component of the shift in some 3+1 formulations.

\item There are also compatibility conditions between the initial data
and the boundary data at the edge ${\cal B}_0$. For the example of the
scalar wave problem (\ref{eq:swe1}) with a Dirichlet boundary condition
at $x=0$, the boundary data must satisfy
\begin{equation}
     \partial_t^2 \Phi |_{(t=0,x=0)}
     = (\partial_x^2+\partial_y^2+\partial_z^2)\Phi |_{(t=0,x=0)},
\end{equation} where the right hand side is determined by the initial
data. An infinite sequence of such conditions follow from taking time
derivatives of the wave equation. They must be satisfied if the solution
is required to be $C^\infty$. In simple problems, this sequence of
compatibility conditions can be satisfied by choosing initial data and
boundary data with support that vanishes in a neighborhood of the edge 
${\cal B}_0$. But in problems with elliptical constraints, such as occur
in general relativity, this simple approach is not possible. In numerical
relativity, these compatibility conditions are usually ignored, with the
consequence that some transient {\it junk} radiation emanating from the
edge is generated. In principle, this could be avoided by smoothly {\it gluing}
the initial data to an exterior region with Schwarzschild~\cite{corv} or
Kerr~\cite{corvsch} data. This gluing construction would avoid mathematical
difficulties but it is an implicit construction and in practice no
numerical algorithm for carrying it out has been proposed. In the
simulation of binary black holes, this edge effect combines with another
source of junk radiation which is hidden in the choice of initial data.
The tacit assumption is that the spurious radiation from these sources
is quickly flushed out of the simulation, with no significant effect after
a few crossing times. Since this issue is difficult to treat or quantify
in a useful way, I adopt the expedient assumption that all compatibility
conditions for a $C^\infty$ solution have been met.

\end{enumerate}

In order to resolve most of the above complications it appears that a
foliation ${\cal B}_t$ of the boundary ${\cal T}$ must be specified as
part of the boundary data. Such a foliation is a common ingredient of
all successful treatments to date. The foliation supplies the gauge
information which determines a unique outgoing null direction for a
Sommerfeld condition. In Sec.~\ref{sec:bare}, I specify ${\cal B}_t$
in terms of the choice of an evolution vector field on the boundary.

Most of the above complications stem from the fact the domain of
dependence determined by the boundary alone is empty. An initial value
problem for a hyperbolic system can be be consistently posed in the
absence of a boundary. But the opposite is not true for an IBVP. Without
an underlying Cauchy problem, an IBVP does not make sense. In an IBVP,
boundary data cannot determine a unique solution independently of the
initial Cauchy data and there is no domain in which the solution is
independent of the initial data. Thus a well-posed IBVP problem must be
based upon a well-posed Cauchy problem. 

\bigskip

\section{The bare manifold}
\label{sec:bare}

\bigskip

In constructing an evolution code for the gravitational field, the first
step is define a spatial grid and a time update scheme. This sets up the
underlying structure necessary to store the values of the various fields.
The analogous object at the continuum level corresponds to the bare
manifold on which the gravitational field is later painted. This is the
approach I will adopt here. It provides a useful way to order the
introduction of the basic geometric quantities which enter the IBVP. 

Setting up the spatial grid corresponds to the analytic specification of spatial 
coordinates $x^i$ on ${\cal S}_0$. The time update algorithm corresponds to
the introduction of an evolution field $t^a$ in ${\cal M}$ which is tangent to the
boundary ${\cal T}$. (In more complicated update schemes, which I won't
consider, the boundary might move through the grid.) The evolution field
must have the property that under its flow ${\cal S}_0$ is mapped into a
foliation ${\cal S}_t$ of ${\cal M}$, and its edge ${\cal B}_0$ is mapped
into a foliation ${\cal B}_t$ of ${\cal T}$.

If a time coordinate is initiated at $t=0$ on ${\cal S}_0$ then the flow
of $t^a$ induces  {\it adapted coordinates} $x^\mu=(t,x^i)$ on ${\cal M}$
by requiring
\begin{equation}
               {\cal L}_t t =1
\label{eq:fol}               
\end{equation}
\begin{equation}
               {\cal L}_t x^i=0,
\label{eq:xi}               
\end{equation}
where $ {\cal L}_t$ is the Lie derivative with respect to $t^a$. Note
that $t^a$ and the adapted coordinates $x^\mu$ are explicitly constructed
fields on ${\cal M}$ with no metric properties. They uniquely fix the
gauge freedom on ${\cal M}$ in precisely the same way that the numerical
grid and update scheme provide a unique evolution algorithm. If $x^A$ are the
coordinates on the edge ${\cal B}_0$, then the corresponding adapted
coordinates on the boundary are $(t,x^A)$, where $ {\cal L}_t x^A=0$. It
will be convenient throughout this review to let the manifold with boundary
be described by adapted coordinates
\begin{equation}
          x^\mu = (t\ge 0, x \ge 0, x^A).
          \label{eq:adapted}
\end{equation}

Under a diffeomorphism $\psi$ of ${\cal M}$ which maps ${\cal T}$ into
itself, $t^a \rightarrow \psi_* t^a$ where $\psi_* t^a$ can be
chosen to be any other possible evolution field. In particular, if
$\psi_* t^a =t^a$ in  ${\cal M}$ and $\psi x^i =x^i$ on ${\cal S}_0$ then the
diffeomorphism must be the identity. Thus, given a coordinate gauge on 
${\cal S}_0$, the choice of $t^a$ determines the remaining diffeomorphism
freedom. This allows a description of the evolution in a specific choice of
adapted coordinates without losing sight of the gauge freedom.

Note that any one-form normal to the boundary is proportional to
$\partial_a t$. However, at the bare manifold level the unit normal
cannot be specified since that involves metric information. The
projection tensor
\begin{equation}
              \pi^a_b=\delta^a_b -t^a\partial_b t
              \label{eq:tproj}
\end{equation}       
has the properties
\begin{equation}
          \pi^a_b v^b \partial_a t =0,
\end{equation}
i.e. it projects a vector field into the tangent space of ${\cal S}_t$, and 
\begin{equation}
  \pi^a_b w_a  t^b=0,
\end{equation}
i.e. it projects a 1-form into the space orthogonal to $\partial_t$.

These are the main structures that exist in the IBVP a priori to introducing
a geometry on ${\cal M}$. There is an alternative approach in which
geometrical concepts are introduced earlier. In the Cauchy problem, the
initial data can be specified  in geometrical form as tensor fields $\tilde
h_{ab}$ and $\tilde k_{ab}$ on a ``disembodied'' 3-manifold $\tilde {\cal
S}_0$ (cf. \cite{hawkel}). Only after the embedding of $\tilde {\cal S}_0$ in
${\cal M}$ is this data interpreted as the intrinsic metric $h_{ab}$ and
extrinsic curvature $k_{ab}$ of ${\cal S}_0$. The mean curvature
$k=h^{ab}k_{ab}$ can itself be interpreted as a variable determining the
location of ${\cal S}_0$. Similarly, in the IBVP the mean extrinsic
curvature of ${\cal T}$ can be interpreted as a wave equation determining
the geometric location of the boundary~\cite{fn}. However, these
interpretations assume knowledge of the spacetime geometry which is only
known after a solution is found. 

This order in which the basics objects are introduced is akin to the
question: Which came first - the geometry or the manifold (or some
combination)? Here I adopt the manifold approach, which is more akin to the
spirit of numerical relativity. I assume a priori, for the given choice
of evolution field $t^a$, that ${\cal M}$ is the domain of dependence of the
initial-boundary data, i.e. it is the manifold upon which the data
determines a unique evolution. Here ``a priori'' is used in the sense of a
spacetime geometry which exists only after the solution of the IBVP is
obtained.

\bigskip

\section{Initial data}
\label{sec:initial}

\bigskip

Since Einstein's equations are second differential order in the metric, any
evolution scheme must specify $g_{\mu\nu}$ and $\partial_t g_{\mu\nu}$ on
${\cal S}_0$.  The  classic result of the Cauchy problem is that a
geometrically unique solution of the Cauchy problem is determined by initial
data consisting of the intrinsic metric $h_{ab}$ of ${\cal S}_0$ and its
extrinsic curvature $k_{ab}$, subject to the constraints
(\ref{eq:hamc})--(\ref{eq:momc}).

The remaining initial data necessary to specify a unique spacetime metric
consist of gauge information, i.e. gauge data that affect the resulting solution only by
a diffeomorphism. One such quantity is the lapse $\alpha$, which relates the
unit future-directed normal to the time foliation according to
\begin{equation}
           n_a=-\alpha\partial_a t.
\label{eq:n}
\end{equation}
The embedding of ${\cal S}_0$ in ${\cal M}$ then gives rise to the
spacetime metric
\begin{equation}
   g_{ab} =- n_a n_b +h_{ab}
\label{eq:hng}
\end{equation}
and the interpretation of $k_{ab}$ as the extrinsic curvature
through the identification
\begin{equation}
     k_{ab} = h_a^c \nabla_c n_b,
\end{equation}   
where $\nabla_c$ is the covariant derivative associated with  $g_{ab}$.

The choice of evolution field $t^a$ supplies the remaining gauge data. It
is transverse but not in general normal to the Cauchy hypersurface so
that it determines a shift $\beta_a$ according to
\begin{equation}
     \beta_a = h_{ab} t^b.
\end{equation}
This relationship supplies the metric information 
$$ g_{ab} t^b =\alpha n_a +\beta_a
$$
relating $t^a$ to the unit normal $n_a$. 

In the adapted coordinates, the metric has components
\begin{equation}
       g_{tt}=-\alpha^2+h_{ij}\beta^i \beta^j
\end{equation}
\begin{equation}
       g_{ti}=\beta_i =h_{ij}\beta^j
\end{equation}
\begin{equation}
        g_{ij}=h_{ij}.
\end{equation}

The inverse metric is given by $g^{ab} = -n^a n^b +h^{ab}$, where
\begin{equation}
        h^{ab}n_b=0, \quad h^{ac}h_{bc}=\delta^a_b +n^a n_b.
\end{equation}
In the adapted coordinates,
\begin{equation}
       g^{tt}=-\alpha^{-2}
\end{equation}
\begin{equation}
       g^{ti}=\alpha^{-2}\beta^i
\end{equation}
\begin{equation}
        g^{ij}=h^{ij}\, , \quad h^{ik}h_{kj}=\delta^i_j.
\end{equation}

The implementation of the initial data into an evolution scheme depends
upon the details by which Einstein's equations are converted into a set
of PDEs governing $g_{\mu\nu}$ in the adapted
coordinates. All such schemes require specification of the initial values
of the lapse and shift, in addition to $h_{ij}$ and $k_{ij}$. Thus it can be
assumed that $g_{ab}$ is specified on ${\cal S}_0$. By Lie
transport along the streamlines of $t^a$, this then allows the
construction of a preferred  stationary background metric ${\gz}_{ab}$ on
${\cal M}$ picked out by the initial data.  Given the choice of evolution
field $t^a$ and the initial Cauchy data, this background metric is uniquely
and  geometrically determined by
\begin{equation}
 {\cal L}_t  {\gz}_{ab} =0 \, , 
       \quad {\gz}_{ab}|_{{\cal S}_0}=g_{ab}|_{{\cal S}_0} . 
 \label{eq:gz}
\end{equation}
In the adapted coordinates $ {\gz}_{\mu\nu}(t,x^i)=g_{\mu\nu}(0,x^i)$.
 
\bigskip

\section{Hyperbolic initial-boundary value problems}
\label{sec:hyperb}

\bigskip

There is an extensive mathematical literature on the IBVP for hyperbolic
systems. The major progress traces back to the formulation of maximally
dissipative boundary conditions for linear symmetric hyperbolic systems
due to Friedrichs~\cite{friedrichs} and Lax and Phillips~\cite{laxphil}.
There has been recent progress in obtaining results for quasilinear
systems where the boundary contains characteristics, as arises in some
formulations of Einstein's equations. Unfortunately, much of this
material is heavy on the mathematical side and not easy reading for
relativists coming from astrophysical or numerical backgrounds. In the
absence of the complications of shocks introduced by hydrodynamic
sources, relativists are content to deal with {\it smooth}, i.e.
$C^\infty$, solutions and forgo the Sobolev theory which enters a
complete discussion of the quasilinear IBVP. For relativists, the most
readable source on the theory of hyperbolic boundary problems is the
textbook by Kreiss and Lorenz~\cite{green-book}, which boasts:  ``In
parts, our approach to the subject is {\it low-tech}.... Functional
analytical prerequisites are kept to a minimum. What we need in terms of
Sobolev inequalities is developed in an appendix."
Taylor's~\cite{taylor1,taylor2} treatises on partial differential
equations contain a classic treatment of pseudo-differential theory but
are less readable for relativists. Fortunately, much of the critical
formalism pertinent to Einstein's equations appears as background
material in papers on the gravitational IBVP. The material I present here
is heavily based upon those sources,
namely~\cite{fn,stewart,reulsarrev,green-book,wpgs,wpe,isol}.

There are two distinct formulations of the IBVP, depending upon whether
you consider Einstein's equations as a natural second differential order
system of wave equations or whether you reduce it to a first order
system. While the second order approach is the most economical, it is not
applicable to all formulations of Einstein's equations, particularly
those whose gauge conditions do not have the semblance of wave equations.
The first order theory has been extensively developed because of its historic
importance to the symmetric hyperbolic formulation of hydrodynamics. The IBVP
for second order systems has received less attention and some new
techniques have originated in the consideration of the Einstein problem. 

There are also two distinct approaches to studying well-posedness - one
based upon energy estimates and the other based upon pseudo-differential
theory where Fourier-Laplace expansions are used to reduce the
differential operators to algebraic operators. The
pseudo-differential theory can be applied equally well to first or second
order systems. In the following, I give a brief account of the underlying
ideas in terms of some simple model problems. This will provide a
background for discussing the difficulties that arise when considering
constraint preservation in the gravitational IBVP.

The subclasses of hyperbolic systems consist of weak hyperbolicty, strong
hyperbolicity, symmetric hyperbolicity and strict hyperbolicity. These
subclasses are determined by the principal part of the system when
written as first differential order PDEs. Weakly hyperbolic systems do
not have a well-posed Cauchy problem, which turned out to be responsible
for the instabilities encountered in early attempts at numerical
relativity using naive ADM formulations with a prescribed lapse and shift.
Strong hyperbolicity is sufficient to guarantee a well-posed Cauchy
problem but not a well-posed IBVP. It is possible to base a well-posed
IBVP on either symmetric hyperbolic or strictly hyperbolic systems.
However, strictly hyperbolic systems arise very rarely and, to my
knowledge, not at all in numerical relativity. So I will limit my
discussion to the symmetric hyperbolic case.

I begin with the IBVP for a second order scalar wave equation, where the
underlying techniques are transparent rather than hidden in the
machinery of symmetric hyperbolic theory.
The generalization to systems of quasilinear wave equations,
described in Sec.~\ref{sec:swev}, can also be treated by the same techniques
as for symmetric hyperbolic systems.  However, for application to the
harmonic Einstein problem, the scalar treatment suffices since the
principal part of the system consists of a common wave operator acting on
the individual components of the metric. The mathematical
analysis which is necessary for a treatment of the quasilinear
IBVP in full rigor is beyond my competence and presumably
outside the interest of someone from a more physical or
computational background. I simply state the main results and give
references when such mathematical theory must be evoked. 

\bigskip

\subsection{Second order wave equations}

\bigskip

The ideas underlying the well-posedness of the IBVP are well illustrated
by the case of the quasilinear wave equation. I give some examples
which are are relevant to the harmonic formulation of Einstein's
equations and which illustrate the techniques behind both the energy
approach and the pseudo-differential approach.

\bigskip

\subsubsection{The energy method for second order wave equations}
\label{sec:enwave}

\bigskip

First consider first the linear wave equation for a scalar field,
\begin{equation}
     \Phi_{tt}=\Phi_{xx}+\Phi_{yy}+\Phi_{zz}+F
\label{eq:byp1}
\end{equation}
on the half-space
\begin{equation}
   x\ge 0,\quad -\infty <y<\infty,
     \quad -\infty < z < \infty, \nonumber
\end{equation}
with boundary condition at $x=0$
\begin{equation}
    \Phi_t-\alpha\Phi_x-\beta_2 \Phi_y-\beta_3 \Phi_z= q \, ,\quad 
    \alpha >0,~\beta_2^2+\beta_3^2<1 ,
\label{eq:byp2}
\end{equation}
with boundary data $q$, initial data of compact support
\begin{equation}
     \Phi=f_1,\quad \Phi_t=f_2,\quad t=0  
\label{eq:byp3}
\end{equation}
and forcing term $F(t,x,y,z)$.
The subscripts $(t,x,y,z)$ denote partial derivatives, e.g
$$\Phi_t={\partial \Phi \over \partial t}=\partial_t \Phi .
$$
All
coefficients and data are assumed to be real and
$\alpha >0,~\beta_2,~\beta_3$ are constants. The notation
$$ 
(\Phi,\Psi), \quad \|\Phi\|^2=(\Phi,\Phi);\quad (\Phi,\Psi)_B,\quad
\|\Phi\|^2_B=(\Phi,\Phi)_B,
$$
is used to denote the $L_2$-scalar product and norm over the half-space and 
boundary space, respectively.

In order to
adapt the standard definition of energy estimates to second order systems, the notation
${\bf \Phi}= (\Phi,\Phi_t,\Phi_x,\Phi_y,\Phi_z)$ is used
to represent the solution  and
its derivatives; and similarly ${\bf f}= (f_1,f_2,f_{1x},f_{1y},f_{1z})$ for the initial data.

For the scalar IBVP (\ref{eq:byp1})--(\ref{eq:byp3}), strong well-posedness
requires the existence of a unique solution satisfying the a priori estimate
\begin{equation}
  \|{\bf \Phi}(t)\|^2+\int_0^t\|{\bf \Phi}(\tau)\|^2_B d\tau
       \le K_T \left (\|{\bf f}\|^2 +\int_0^t\|F(\tau)\|^2 d\tau
       +\int_0^t\|q(\tau)\|^2_B d\tau \right ),
\label{eq:swp}
\end{equation}
in any time interval $0<t<T$, where the constant $K_T$ is 
independent of $F$, ${\bf f}$ and $q$.

It is important to note that (\ref{eq:swp}) estimates the derivatives of $\Phi$,
both in the interior and on the boundary, in terms of the data and the
forcing. This is referred to as ``gaining a derivative''. This property is
crucial in extending the local IBVP to global situations, e.g. where the
boundary is a sphere or where there is an interior and exterior boundary as in
a strip problem. Otherwise, reflection from the boundary could lead to the
``loss of a derivative'', which would lead to unstable behavior under
multiple reflections.

The usual procedure is to derive an energy estimate by integration by parts,
using for example $(\Phi_{yy},\Phi_z)=-(\Phi_y,\Phi_{yz})=0$.  Consider first
the estimates of the derivatives of $\Phi$ in the homogeneous case  $F\equiv
q\equiv 0$. Using the standard energy for a scalar field, integration by parts
gives
$$
 \partial_t (\|\Phi_t\|^2+ \|\Phi_x\|^2+ \|\Phi_y\|^2 + \|\Phi_y\|^2)=
    -2(\Phi_t,\Phi_x)_B.
$$
If  $\beta_2=\beta_3=0$ and $\alpha >0$ in the boundary condition
(\ref{eq:byp2}) then  $(\Phi_t,\Phi_x)_B \ge 0$, i.e. the boundary condition is
{\it dissipative}, and there is an energy estimate. Otherwise there is no
obvious way to estimate the boundary flux. Instead, it is possible to use a
non-standard energy $E$  for the scalar wave equation (\ref{eq:byp1}) which
does provide the key estimate if $\beta_2^2+\beta_3^2>0$.

The first step is to show that
\begin{equation}
E:=\|\Phi_t\|^2+\|\Phi_x\|^2+\|\Phi_y\|^2+\|\Phi_z\|^2-
2(\Phi_t,\beta_2 \Phi_y+\beta_3 \Phi_z)
\label{eq:byp4}
\end{equation}
is a norm for the derivatives $(\Phi_t,\Phi_x,\Phi_y,\Phi_z)$.
Since $\beta_2^2+\beta_3^2< 1$, this follows, after a rotation,
from the inequality $\Phi_t^2 + \Psi^2 -2\beta \Phi_t \Psi \ge 0$
for $\beta^2<1$.

This leads to

\medskip

{\it Lemma 1:} 
The solution of  (\ref{eq:byp1})--(\ref{eq:byp3}) satisfies the energy estimate
\begin{equation}
\partial_t E  +\alpha \|\Phi_x\|^2_B\le E + \|F\|^2 +\frac{1}{\alpha} \|q\|^2_B. 
\label{eq:byp5}
\end{equation}

\medskip

\noindent
{\it Proof.} Integration by parts gives
\begin{equation}
    \partial_t \|\Phi_t\|^2=2(\Phi_t,\Phi_{tt})=
    -\partial_t (\|\Phi_x\|^2+\|\Phi_y\|^2+\|\Phi_z\|^2)+
   2(\Phi_t,F)-2(\Phi_t,\Phi_x)_B  
\label{eq:byp6}
\end{equation}
and
\begin{equation}
2\partial_t (\Phi_t,\beta_2 \Phi_y+\beta_3 \Phi_z)
  =2(\Phi_{tt}, \beta_2 \Phi_y+\beta_3 \Phi_z)
= -2(\Phi_{x},\beta_2 \Phi_y+\beta_3 \Phi_z)_B 
      +2 (F, \beta_2 \Phi_y+\beta_3 \Phi_z).
\label{eq:byp7}
\end{equation}
Since (\ref{eq:byp2}) implies
$$
2(\Phi_t,\Phi_x)_B=2\alpha\|\Phi_x\|^2_B+
2(\Phi_{x},\beta_2 \Phi_y+\beta_3 \Phi_z)_B+
2 (\Phi_x,q)_B,$$
subtraction of (\ref{eq:byp7}) from (\ref{eq:byp6}) leads to
\begin{eqnarray}
   \partial_t E &=& 2(\Phi_t-\beta_2 \Phi_y -\beta_3 \Phi_z,F)-2\alpha\|\Phi_x\|^2_B
      -2(\Phi_x,q)_B  \nonumber \\
         & \le&  \|\Phi_t-\beta_2 \Phi_y -\beta_3 \Phi_z\|^2 + \|F\|^2 
       -\alpha\|\Phi_x\|^2_B +\frac{1}{\alpha}\|q\|^2_B. \nonumber
\end{eqnarray}

The identity
$$
  \| \Phi_t -\beta_2 \Phi_y -\beta_3 \Phi_z\|^2 = E-\| \Phi_x\|^2 -\| \Phi_y\|^2 
   -\| \Phi_z\|^2  + \| \beta_2 \Phi_y+\beta_3 \Phi_z \|^2
$$
then implies (\ref{eq:byp5}) and proves the lemma.

\medskip

By integration, the lemma estimates 
$$
E(T)~\hbox{and}~\int_0^T \|\Phi_x\|^2_B dt \quad \hbox{in terms of}\quad
E(0),~\int_0^T \|F\|^2dt~ \hbox{and}~
\int_0^T \|q\|^2_B dt.
$$
Strong well-posedness (\ref{eq:swp}) also requires estimates
of the boundary norms $\|\Phi_t \|_B$, $\|\Phi_y \|_B$
and $\|\Phi_z\|_B$.  
First, a calculation similar to that above gives
\begin{eqnarray}
\partial_t (\Phi_x,\Phi_t)&=&(\Phi_{xt},\Phi_t)+(\Phi_x,\Phi_{tt})
     \nonumber \\
   &=&-\frac{1}{2}\|\Phi_t\|^2_B+(\Phi_x,\Phi_{xx})+(\Phi_x,\Phi_{yy})
   +(\Phi_x,\Phi_{zz})+(\Phi_x,F) \nonumber \\
    &=&-\frac{1}{2}\|\Phi_t\|^2_B-\frac{1}{2}\|\Phi_x\|^2_B
    +\frac{1}{2}\|\Phi_y\|^2_B+\frac{1}{2}\|\Phi_z\|^2_B +(\Phi_x,F) .
\label{eq:byp8}
\end{eqnarray}
Estimates of $\|\Phi_t \|_B$ in terms of $\|\Phi_y \|_B$, $\|\Phi_z\|_B$,
$\|\Phi_x\|_B$ and $\|q\|_B$ can be obtained from
the boundary conditions (\ref{eq:byp2}), which give,
for any $\delta$ with $0<\delta<1$,
\begin{eqnarray}
\|\Phi_t\|^2_B&=& \|\beta_2 \Phi_y+ \beta_3 \Phi_z
    +\alpha \Phi_x+q\|^2_B \nonumber \\
&\le&
 \|\beta_2 \Phi_y+ \beta_3 \Phi_z\|^2_B+
 2\|\beta_2 \Phi_y+ \beta_3 \Phi_z\|_B \,
 \|\alpha \Phi_x+q\|_B+ \|\alpha \Phi_x+q\|^2_B \\
&\le& (1+\delta)
 \|\beta_2 \Phi_y+ \beta_3 \Phi_z\|^2_B
 +(1+{1\over\delta})\|\alpha \Phi_x+q\|^2_B \nonumber \\
&\le& (1+\delta)\left(\beta_2^2+\beta_3^2\right)
(\|\Phi_y\|^2_B+\|\Phi_z\|^2_B)+
(1+{1\over \delta})\|\alpha\Phi_x+q\|^2_B. \nonumber
\end{eqnarray}
Next, since $\beta_2^2+\beta_3^2<1$, $\delta$ can be chosen
such that $(1+\delta)(\beta_2^2+\beta_3^2)
\le (1-\delta).$ Therefore, by (\ref{eq:byp8}),
$$
\delta(\|\Phi_y\|^2_B+\|\Phi_z\|^2_B)\le 
(1+{1\over\delta})\| \alpha \Phi_x+q\|^2_B +  \|\Phi_x\|^2_B
+2\partial_t (\Phi_x,\Phi_t) -2 (\Phi_x,F). 
$$

Since $(\Phi_x,\Phi_t)$ can be estimated by $E$, there follows  

\medskip

{\it Lemma 2:}
\begin{eqnarray}
 & \int_0^T & \left(\|\Phi_t\|^2_B+\|\Phi_x\|^2_B+\|\Phi_y\|^2_B
   +\|\Phi_z\|^2_B\right)dt
   \nonumber \\
 &\quad & \le const (E(0)+\int_0^T \|F\|^2dt+\int_0^T \|q\|^2_B dt). \nonumber
\end{eqnarray}

\medskip

An estimate for $\Phi$ itself can easily be obtained by the change of variable
$\Phi\to e^{\mu t}\Phi$, as described in Sec.~\ref{sec:quasiw}
or in Appendix 1 of~\cite{wpe}. 
Together with the results of Lemma 1 and Lemma 2, this establishes

\medskip

{\it Theorem 1:} The IBVP (\ref{eq:byp1})--(\ref{eq:byp3}) is
strongly well-posed in the sense
of (\ref{eq:swp}).

\medskip

The result can also be generalized to half-plane problems for wave
equations of the general constant coefficient form
\begin{equation}
\Phi_{tt}= 2b^i \Phi_{it}+h^{ij} \Phi_{ij},
  \quad x_1\ge 0,\quad -\infty < (y,z) <\infty ,~x^i=(x,y,z) 
\label{eq:byp9}
\end{equation}
where $h^{ij}$ is is a metric of $(+++)$ signature.
By coordinate transformation, (\ref{eq:byp9}) can be
transformed into (\ref{eq:byp1}) and the appropriate boundary conditions
formulated.  See~\cite{wpe} for details. 

This example illustrates from the PDE perspective the constructions
necessary to establish strong well-posedness. For the purpose of
establishing the strong well-posedness of the IBVP for the wave equation
on a general curved space background, it is also instructive to take
advantage of the geometric nature of the problem.

In terms of standard relativistic notation, consider the wave equation
\begin{equation}
   g^{ab}\nabla_a \nabla_b\Phi = F
\label{eq:byp21}
\end{equation}
for a massless scalar field propagating on a Lorentzian spacetime ${\cal M}$
foliated by compact, $3$-dimensional time-slices ${\cal S}_t$, with boundary
${\cal T}$  foliated by ${\cal B}_t$. Here $\nabla$ denotes the covariant
derivative associated with the spacetime metric $g^{ab}$. For notational
simplicity, let $\Phi_a = \nabla_a\Phi$.

The IBVP consists in finding solutions of (\ref{eq:byp21}) subject to the
initial Cauchy data
\begin{equation}
\left. \Phi \right|_{{\cal S}_0} = f_1, \qquad
\left. n^b\Phi_b \right|_{{\cal S}_0} = f_2 
\label{eq:byp21b}
\end{equation}
and the boundary condition
\begin{equation}
\left[ (T^b + a N^b) \Phi_b \right]_{\cal T} = q ,
\label{eq:byp21c}
\end{equation}
with data $q$ on ${\cal T}$. Here $n^b$ is the future-directed unit normal to
the time-slices ${\cal S}_t$ and and $N^b$ is the outward unit normal to ${\cal
T}$; $T^b$ is an arbitrary future-directed timelike vector field which is
tangent to ${\cal T}$; and $a > 0$. The motion of the boundary is described
geometrically by the hyperbolic angle $\tanh\Theta=N^b n_b$. Without loss of
generality, assume the normalization $g_{bc} T^b T^c = -1$. A Sommerfeld
boundary condition then corresponds to the choice $a=1$ for which $T^b +N^b$
points in an outgoing null direction.

In order to establish estimates, consider the energy momentum tensor
of the scalar field  
$$
        \Theta_b^a = \Phi_b \Phi^a -{1\over 2}\delta_b^a \Phi^c \Phi_c.
$$
The essential idea is the use of an energy associated with a timelike vector
$u^a=T^a+\delta N^a$, where $0<\delta<1$, so that $u^a$
points outward from the boundary. The corresponding energy $E(t)$ and the
energy flux $P(t)$ through ${\cal B}_t$ are
\begin{equation}
     E(t) =\int_{{\cal S}_t } u^b \Theta_b^a n_a, 
     \label{eq:ewave}
\end{equation} 
which is a covariant version of  the non-standard energy (\ref{eq:byp4}),    
and 
\begin{equation}
     P(t) =\int\limits_{{\cal B}_t} u^b \Theta_b^a N_a .
\end{equation}
It follows from the timelike property of $u^a$  that $E(t)$ is a norm for
$\Phi_a(t)$.

Energy conservation for the scalar field, i.e. integration by parts, gives
$$
     \partial_t E =  P
          -\int\limits_{{\cal S}_t} (\Theta_{ab}\nabla^a u^b +u^a \Phi_a F) ,
$$
so that
\begin{equation}
     \partial_t E \le   P + const
         (E+ \int\limits_{{\cal S}_t} F^2 ) .
\label{eq:byp22}
\end{equation}

The required estimates arise from an identity satisfied by the flux density
\begin{equation}
   u^b \Theta_b^a N_a =
   -{\delta\over 2} \left (
       (N^a \Phi_a )^2 + (T^a \Phi_a )^2  + Q^{ab}\Phi_a \Phi_b \right)
          +N^a \Phi_a T^b \Phi_b +\delta (N^a \Phi_a )^2
            + \delta(T^a \Phi_a )^2  
             \nonumber
\end{equation}
where $Q_{bc} = g_{bc} + T_b T_c - N_b N_c$ is the positive definite
2-metric in the tangent space of the boundary orthogonal to $T^a$. By
using the boundary condition to eliminate $T^a \Phi_a$ in the last group
of terms, there follows
\begin{eqnarray}
  u^b \Theta_b^a N_a &=&  -{\delta\over 2} \left (
       (N^a \Phi_a )^2 + (T^a \Phi_a )^2  
     + Q^{ab}\Phi_a \Phi_b \right) \nonumber \\
       &+&\left (-a+\delta(1+a^2)\right)(N^a \Phi_a )^2 
      +(1-2a\delta)N^a \Phi_a q +\delta q^2    \nonumber \\
   &=&  -{\delta\over 2} \left (
       (N^a \Phi_a )^2 + (T^a \Phi_a )^2  + Q^{ab}\Phi_a \Phi_b \right) \nonumber \\
  &-&\left (a-\delta(1+a^2)\right )
      \left (N^a \Phi_a -\frac{\left(1-2a\delta \right)q}  
        {2 \left (a-\delta(1+a^2)\right )}\right )^2  
   +\left (\delta+ \frac {(1-2a \delta)^2} {4\left (a-\delta(1+a^2) \right )}
    \right )q^2. \nonumber
\end{eqnarray} 
The choice
$$  0< \delta  < \frac{a}{1+a^2} 
$$
(which also guarantees that $\delta<1$ so that $u^a$ is timelike),
gives the inequality                  
\begin{equation}
 u^b \Theta_b^a  N_a \le  -{\delta\over 2} \left (
       (N^a \Phi_a )^2 + (T^a \Phi_a )^2  + Q^{ab}\Phi_a \Phi_b \right)   
     + const \, q^2 .
\label{eq:byp23}
\end{equation}

It now follows from (\ref{eq:byp22}) and (\ref{eq:byp23}) that
\begin{eqnarray}
    \partial_t E &+& \int_{{\cal B}_t}{\delta\over 2} \bigg(
     (N^a \Phi_a )^2 + (T^a \Phi_a )^2  + Q^{ab}\Phi_a \Phi_b \bigg)
\nonumber \\
    &\le &  const \bigg (E+ \int_{{\cal S}_t} F^2 +  
       \int_{{\cal B}_t} q^2     \bigg ). 
\label{eq:byp24}
\end{eqnarray}
This is the required estimate of the gradient $\Phi_a$ on the boundary
(as well as usual estimate of the energy $E$) to prove that the problem
is strongly well-posed. As in the previous example, an estimate of $\Phi$
itself follows from the change of variable $\Phi\to e^{\mu t}\Phi$, which
introduces a mass term in  (\ref{eq:byp21}). 

\bigskip

\subsubsection{The quasilinear case}
\label{sec:quasiw}

\bigskip

The estimate (\ref{eq:byp24}) is sufficient to establish the criterion
(\ref{eq:swp}) for strong well-posedness of the IBVP for the linear
wave equation with constant metric coefficients. In order to extend the
result to the quasilinear case on a curved space background, where the
metric depends upon $\Phi$ and $\Phi_a$, it is necessary to show that the
corresponding estimates hold for arbitrarily high derivatives of
$\Phi$. In the process, this also requires stability of the system
under the addition of lower differential order terms, which arise under the
differentiation of the wave equation. These requirements
are sometimes neglected or misunderstood in the relativity literature.

More generally, local existence theorems for variable coefficient or
quasilinear equations follow by iteration of solutions of the linearized
equations with frozen coefficients. The energy estimates for the frozen
coefficient problem establish the existence of a unique solution which
depends continuously on the data. The extension of this result to the
quasilinear case first requires that the problem with variable
coefficients be strongly well-posed. For this, it already necessary to
obtain estimates for arbitrarily high derivatives of the solution to the
linearized problem. 

For the purpose of illustrating the procedure, it suffices to consider
the IBVP for the 2(spatial)-dimensional wave equation with variable
coefficients 
\begin{equation} \Phi_{tt}=P\Phi+R\Phi+F,\quad x\ge 0,\quad 
-\infty<y<\infty,  \label{eq:byp15} 
\end{equation}
with smooth initial data
\begin{equation}
     \Phi=f_1 \, ,\quad \Phi_t=f_2 \,  , \quad t=0 ,
\label{eq:byp17}
\end{equation}
and boundary condition
\begin{equation}
\alpha(t,y) \Phi_t=\Phi_x-\mu  \Phi+r(t,y) \Phi+q(t,y)\, ,\quad
 \alpha(t,y) \ge const  >0\, , \,  \mu =const >0
\label{eq:byp16}
\end{equation}
with smooth, compatible boundary data $q$.
Here
$$ P\Phi=(a\Phi_x)_x+(b\Phi_y)_y-2\mu \Phi_t-\mu^2\Phi\, , 
    \quad a>a_0=const>0, \, b>b_0=const>0\, ,
$$
is an elliptic operator which has been modified by terms which  arise in
(\ref{eq:byp15}) by the transformation $\Phi \rightarrow e^{\mu t} \Phi$,
where $\mu$ introduces a mass term; and
$$
R\Phi=c_1\Phi_t+c_2\Phi_x+c_3\Phi_y+c_4\Phi
$$
are terms of lower (zeroth
and first) differential order. The coefficients $a$, $b$ and $c_i$ are smooth functions
of $(t,x,y)$.

Consider the norm for the Cauchy data
$$
    E= \|\Phi_t\|^2+(\Phi_x,a\Phi_x)+(\Phi_y,b\Phi_y)+\mu^2\|\Phi\|^2.
$$
Integration by parts leads to
\begin{eqnarray}
 \partial_t E& =& -4\mu\|\Phi_t\|^2+2(\Phi_t,F)+2(\Phi_t,R\Phi)-2(\Phi_t,a \Phi_x)_B 
       + (a_t\Phi_x,\Phi_x)+(b_t\Phi_y,\Phi_y) \nonumber \\
&\le & const (\|F\|^2+E)-2(\Phi_t,a\Phi_x)_B .
\label{eq:byp18}
\end{eqnarray}
The boundary condition gives
\begin{eqnarray}
  -(\Phi_t,a\Phi_x)_B&=& -(\Phi_t,\alpha a\Phi_t)_B
    -\mu(\Phi_t, a\Phi)_B+(\Phi_t,ar\Phi+aq)_B \nonumber \\
    &=&-(\Phi_t,\alpha a\Phi_t)_B-\mu(\Phi_t, a_0 \Phi)_B
    -\mu\left(\Phi_t,(a-a_0)\Phi\right)_B +(\Phi_t,ar \Phi+aq)_B \\
    &\le & -\frac{1}{2} \mu a_0\partial_t \|\Phi\|^2_B
     -\frac{1}{2}(\Phi_t,\alpha a\Phi_t)_B+ const (\|\Phi\|^2_B+\|q\|^2_B).
\end{eqnarray} 
Therefore, from (\ref{eq:byp18}),
\begin{eqnarray}
&\partial_t (E+\mu a_0\|\Phi\|^2_B )+(\Phi_t,\alpha a \Phi_t)_B  \\
&\le const \left(E+\|\Phi\|^2_B+\|F\|^2+\|q\|^2_B\right),
\label{eq:byp19} 
\end{eqnarray}

This establishes that the energy estimate is stable against lower order
perturbations. Now it is possible to estimate the derivatives. For the
derivatives  $Y=\Phi_y$ and $T=\Phi_t$ tangential to the boundary,
differentiation of the wave equation gives
\begin{eqnarray}
Y_{tt}&=&PY+RY+R_y \Phi+(a_y \Phi_x)_x+(b_y Y)_y+F_y 
\label{eq:Y} \\
T_{tt}&=&PT+RT+R_t \Phi+(a_t\Phi_x)_x+(b_t Y)_y+F_t. 
\label{eq:T} 
\end{eqnarray}
Here $R_y \Phi$ and $R_t \Phi$ are linear combinations of first
derivatives of $\Phi$ which have already been estimated and can be
considered part of the forcing term $F$. Also, the wave equation
(\ref{eq:byp15}) implies
$$ a \Phi_{xx}=T_t-bY_y+~ \hbox{terms that have already been estimated}, $$
so that $\Phi_{xx}$ is also lower order with respect to (\ref{eq:Y}) and
(\ref{eq:T}). Thus, except for lower order terms, $Y$ and $T$ solve the
same wave equations (\ref{eq:Y}) and (\ref{eq:T}) as $\Phi$, with the
same boundary conditions up to lower order terms. Therefore all second
derivatives $\Phi_{ay}$ and $\Phi_{at}$ can be estimated as well as
$\Phi_{xx}$. Repetition of this process gives estimates for any number of
derivatives.

In order now to show the existence of a solution to the variable
coefficient problem, one approach, which is particularly familiar to
numerical relativists,  is to approximate the PDE by a stable finite
difference approximation. This approach is detailed in~\cite{green-book}
where summation by parts (SBP) is applied to the finite difference
problem in the analogous way that integration by parts is used above in
the analytic problem. The SBP  approach shows that the corresponding
estimates hold independently of gridsize. Existence of a solution of the
analytic problem then follows from the limit of vanishing gridsize.

It also follows from the estimates for arbitrary derivatives of the
variable coefficient problem that Sobolev's theorems can be used to
establish similar, although local in time, estimates for quasilinear
systems. Then the same iterative methods used for first order symmetric
hyperbolic systems can be used to show that well-posedness extends
locally in time to the quasilinear case. In this way it was shown
in~\cite{wpe} that  the general quasilinear wave problem
(\ref{eq:byp21})--(\ref{eq:byp21c}), where the spacetime metric now
depends upon $\Phi$ and $\Phi_a$, is strongly well-posed.  

\bigskip

\subsubsection{Systems of wave equations}
\label{sec:swev}

\bigskip

The strong well-posedness of the IBVP for the quasilinear scalar wave
(\ref{eq:byp21})--(\ref{eq:byp21c}) can be generalized to a system of
coupled wave equations
\begin{equation}
  g^{ab}(\Phi) \nabla_a\nabla_b\Phi^A = F^A(\Phi,\nabla\Phi),  
    \quad A=1,2,...N
\label{Eq:WaveSystemEq}
\end{equation}
with smooth initial data
\begin{equation}
   \left. \Phi^A \right|_{{\cal S}_0} = f_1^A\; , \qquad
       \left. n^b\nabla_b\Phi^A \right|_{{\cal S}_0} = f_2^A\; ,
\label{Eq:WaveSystemID}
\end{equation}
and the boundary condition
\begin{equation}
\left. \left( T^b + \alpha N^b \right )\nabla_b\Phi^A \right|_{\cal T}
 = c^{a\, A}{}_B\left. \nabla_a\Phi^B \right|_{\cal T} 
 + d^A{}_B\left. \Phi^B \right|_{\cal T} + q^A,
\label{Eq:WaveSystemBC}
\end{equation} where $a = a(x,\Phi) > 0$, $q^A = q^A(x)$, $c^{a\, A}{}_B
= c^{a\, A}{}_B(x,\Phi)$ and $d^A{}_B = d^A{}_B(x,\Phi)$ are smooth
functions of their arguments. All data are compatible. As before, each
time-slice ${\cal S}_t $ is spacelike, with future-directed unit normal
$n^a(x,\Phi)$, the boundary ${\cal T}$ is timelike with outward unit
normal  $N^a(x,\Phi)$ and   $T^a = T^a(x,\Phi)$ is an arbitrary
future-directed timelike vector field  tangent to ${\cal T}$.

In~\cite{isol}, it was shown that the IBVP
(\ref{Eq:WaveSystemEq})--(\ref{Eq:WaveSystemBC}) is
strongly well posed given certain restrictions on $c^{a\, A}{}_B$, i.e.
there exists a solution locally in time which satisfies  (\ref{eq:swp}) in
terms of the corresponding $L_2$ norms for $\Phi^A$ and its gradient.
One important situation in which the restrictions on $c^{a\, A}{}_B$ are
satisfied is  when it can be transformed into the upper diagonal form
\begin{displaymath}
   c^{a\, A}{}_B = 0,\qquad B \leq A.
\label{eq:triang}
\end{displaymath}
This has important applications to the constrained systems of wave
equations obtained by formulating Maxwell's equations in the Lorentz
gauge or Einstein's equations in the harmonic gauge~\cite{isol}, as
discussed in Sec.~\ref{sec:harm}.

\bigskip

\subsubsection {Pseudo-differential theory}
\label{sec:fourlap}

\bigskip

The previous scalar wave problems required a non-standard energy to
obtain the necessary estimates. In more general problems, an
effective choice of energy might not be obvious or even exist.
Pseudo-differential theory provides an alternative treatment of such
cases. The approach is based upon a Fourier transform for the spatial
dependence and a Laplace transform in time. It can be applied equally
well to first or second order systems. Here I illustrate how it applies
to the second order wave equation. The more general theory for a system
of equations is usually presented in first order form and is reviewed in
Sec.~\ref{sec:pseudodiff}.

As an illustrative example, consider the 2(spatial)-dimensional version
of the IBVP for the scalar wave considered in Sec.~\ref{sec:enwave},
\begin{equation}
   \Phi_{tt}= \Phi_{xx}+ \Phi_{yy}+F,\quad x\ge 0,~ -\infty <y<\infty ,
\label{eq:wp1}
\end{equation}
with boundary condition at $x=0$
\begin{equation}
\Phi_t-\alpha \Phi_x-\beta \Phi_y =q, \quad \alpha>0, \,  \, |\beta|<1 ,
\label{eq:wp2}
\end{equation}
with compact boundary data $q$  and initial data
\begin{equation}
 \Phi(0,x,y)=f_1(x,y),\quad  \Phi_t(0,x,y)=f_2(x,y) .
\label{eq:wp3}
\end{equation}

The following simple observation reveals the underlying idea. The system
cannot be well-posed if the homogeneous version of
(\ref{eq:wp1})--(\ref{eq:wp3}) with $F=q=0$ admits arbitrarily fast
growing solutions. The homogeneous system has solutions of the form 
\begin{equation}
 \Phi(t,x,t)=e^{st+i\omega y} \varphi(x),\quad |\varphi|_\infty <\infty ,
\label{eq:wp4}
\end{equation}
where 
\begin{eqnarray}
\varphi_{xx} &-& (s^2+\omega^2)\varphi=0, \label{eq:wp5} \\
  s\varphi(0)&=&\alpha \varphi_x(0)+i\beta\omega \varphi(0) .
\label{eq:wp6}
\end{eqnarray}
Here $\varphi(x)$ is a smooth bounded function, so that its maximum
norm $|\varphi|_\infty$ is finite, $\omega$ is a real constant
and $s=\eta +i\xi$ a complex constant. This poses an eigenvalue problem.
If there are solutions with $\eta=\Re\, s >0$ then 
$$
 \Phi_\nu =e^{s\nu t+i\nu\omega y} \varphi(\nu x)
$$
is also a solution for any $\nu >0$, so that there are solutions which
grow  arbitrarily fast exponentially. Therefore, a necessary condition
for a well-posed problem is that solutions with $\Re\, s >0$ must be
ruled out by the boundary condition.

The general solution of the ordinary differential equation (\ref{eq:wp5})
is
\begin{equation}
  \varphi=\sigma_1 e^{\kappa_+ x}+ \sigma_2 e^{\kappa_- x},
\label{eq:wp7}
\end{equation}
where 
\begin{equation}
\kappa_{\pm}=\pm \sqrt{s^2+\omega^2}
\label{eq:eigenv}
\end{equation}
solve the characteristic equation
$$ \kappa^2-(s^2+\omega^2)=0. 
$$
Here $ \Re\,\kappa_+>0$ and $ \Re \, \kappa_-<0$  for $\Re \, s>0$. By
assumption $\varphi$ is bounded which requires that $\sigma_1=0,$ i.e.
\begin{equation}
 \varphi=\sigma_2 e^{\kappa_- x}. 
\label{eq:wp10}
\end{equation}
Introducing (\ref{eq:wp10}) into the boundary condition (\ref{eq:wp6}) gives
\begin{equation}
 (s-\alpha\kappa_- -i\beta\omega )\sigma_2=0 .
\label{eq:wp11}
\end{equation}
\smallskip\noindent
Since $\Re \, \kappa_-<0$ and by assumption $\alpha >0$, there  
are no solutions with
$\Re \, s>0$.
Thus this necessary condition for a well-posed
problem is satisfied.

In order to proceed further
it is technically convenient
to assume that the initial data (\ref{eq:wp3}) vanishes. This
may always be achieved by the transformation 
\begin{equation}
      \Phi \rightarrow \Phi-e^{-t} f_1 - t e^{-t}(f_2+f_1),
\label{eq:homcd}
\end{equation}
so that the initial data gets swept into the forcing term $F$.
Then (\ref{eq:wp1})--(\ref{eq:wp3}) can be solved by
Fourier transform with respect to $y$ and Laplace transform with respect to $t$,
i.e. in terms of
\begin{equation}
  \hat \Phi (s,x,\omega) =\frac{1}{\sqrt{2\pi}}\int_{-\infty}^{\infty} dy
      e^{-i\omega y} \int_0^\infty dt e^{-st} \Phi(t,x,y)  \, , \quad \Re \, s >0,
\end{equation}
where $\omega$ is real and $s$ is complex.
The inhomogeneous versions of (\ref{eq:wp5}) and (\ref{eq:wp6}) imply
that the coefficients satisfy 
\begin{eqnarray}
\hat \Phi_{xx}&-&(s^2+\omega^2)\hat \Phi =-\hat F  \nonumber \\
 (s&-&i\beta \omega)\hat \Phi(s,0,\omega)-\alpha\hat \Phi_x(s,0,\omega) =\hat q(s,\omega).
\label{eq:wp12}
\end{eqnarray}

Since it has already been shown that the homogeneous system
(\ref{eq:wp5})--(\ref{eq:wp6}) has no eigenvalues for $\Re \,  s>0$
and $\Re \, \kappa<0$, it follows that (\ref{eq:wp12}) has a unique solution
$\hat \Phi$. Inversion of the Fourier-Laplace transform then
gives a unique solution for $\Phi$.

The well-posedness of a variable coefficient or quasilinear wave
problem also requires estimates of the higher derivatives of $\Phi$. The
system of equations for the derivatives are obtained by differentiating the
wave equation and the boundary condition. In that process, any variable
coefficient terms in the boundary condition lead to inhomogeneous
boundary data for the derivatives. It is possible to transform
the boundary data $q$ to $0$ by a transformation analogous to
(\ref{eq:homcd}), which sweeps $q$ and its derivatives into the forcing
term $F$. If this inhomogeneous boundary data is continually subtracted out
of the boundary condition, inhomogeneous terms of higher differential order
appear in the forcing term. As a consequence, the resulting estimates would
bound lower derivatives of the solution in terms of higher derivatives of 
the data, a process referred to as ``losing'' derivatives.

Instead, a different approach is necessary to establish well-posedness.
It is simple to calculate  the solution for $\hat F = 0$.
Corresponding to (\ref{eq:wp10}) and (\ref{eq:wp11}), there follows
\begin{equation}
\hat \Phi(s,x,\omega)=e^{\kappa_- x}\hat \Phi(s,0,\omega)
\label{eq:wp13}
\end{equation}
where
$$
(s-\alpha\kappa_- -i\beta \omega)\hat \Phi (s,0,\omega)=\hat q(s,\omega).
$$

It is now possible to establish~\cite{wpgs}

\medskip

\noindent {\bf Boundary stability:} The solution (\ref{eq:wp13})
satisfies the estimates
\begin{eqnarray}
   |\hat \Phi_x(s,0,\omega,)|&\le& K |\hat q(s,\omega)|, \nonumber \\
       \sqrt{|s|^2+\omega^2}\cdot
     |\hat \Phi(s,0,\omega)|&\le& K |\hat q(s,\omega)|,
\label{eq:wbstab}
\end{eqnarray}
where the constant $K$ is independent of $s$ and $\omega$.
Similar estimates hold for all the derivatives.
\medskip

The estimates (\ref{eq:wbstab}) follow from purely algebraic consequences of the eigenvalue
relations (\ref{eq:eigenv}) and (\ref{eq:wp11}).
The essential steps are to show

\begin{enumerate}

\item There is a constant $\delta_1>0$ such that$|\Re \, \kappa_- | > \delta_1 \Re \, s$ 

\item For all $\omega$ and $s$ with $\Re s \ge 0$,
there is a constant $\delta_2>0$ such that
$|s-\alpha\kappa_- -i\beta\omega|\ge\delta_2\sqrt{|s|^2+|\omega|^2}$.

\end{enumerate}
See~\cite{wpgs} for the details.

\medskip

Boundary stability allows the application of the theory of pseudo-differential operators
to show that the IBVP is {\em strongly well-posed in the generalized sense}, 

\medskip

\begin{equation}
\int_0^t  \|{\bf \Phi}(\tau)\|^2d\tau+\int_0^t \|{\bf \Phi}(\tau)\|^2_B d\tau
       \le K_T \left ( \int_0^t\|F(\tau)\|^2 d\tau
       +\int_0^t\|q(\tau)\|^2_B d\tau \right ) , \quad 0 \le t \le T,
\label{eq:swpg}
\end{equation}
where ${\bf \Phi}=(\Phi, \Phi_a)$.

The theory is discussed in Sec.~\ref{sec:pseudodiff} in the standard
context of first order systems. But the first order theory is flexible enough to apply
to second order systems. 
In particular, in Sec.~\ref{sec:pseudodiff} it is applied to show that the IBVP
for the quasilinear version of the second order wave equation with boundary
condition (\ref{eq:wp1})--(\ref{eq:wp2})
is well-posed in the generalized sense.

Strong well-posedness in the generalized sense is similar to strong
well-posedness (\ref{eq:swp}) except now the initial data has been swept into the forcing
term and the estimate for ${\bf \Phi}$ in the interior involves
a time integral. In both cases, the gradients at the boundary are estimated
by the boundary data and the forcing, i.e. a derivative is gained at the boundary.
This ensures that the well-posedness of the local halfplane problem can be
extended globally to include boundaries that lead to multiple reflections.

\bigskip

\subsubsection{Generalized eigenvalues}

\bigskip

Strong well-posedness in the generalized sense not only rules out eigenvalues
of (\ref{eq:wp6}) with $\eta=\Re(s)>0$ but also {\it generalized eigenvalues}
for which $\eta=0$. This is implicit in the estimates for boundary stability
(\ref{eq:wbstab}) in which the constant $K$ is independent of $s$.
However, generalized eigenvalues can exist in well-behaved physical systems. 
A prime example is a surface wave which travels
tangential to the boundary with periodic time dependence. See~\cite{stewart}
for the treatment of such an example from Maxwell theory.

Generalized eigenvalues are ruled out by the boundary conditions
required for strong well-posedness in the generalized sense and historically
have been treated on an individual basis. However, a new approach
to this problem has recently been formulated by H-O. Kreiss~\cite{heinzpc}. 
This approach splits the problem into two subproblems:

\begin{enumerate}

\item One in which the forcing vanishes, $F=0$.

\item  One in which the boundary data is homogeneous, $q=0$. 

\end{enumerate}

A second order wave problem is  called
{\em well-posed in the generalized sense} if these subproblems satisfy
the corresponding  estimates
\begin{equation}
   \int_0^t \|\Phi(\tau)\|^2_B d\tau
       \le K_T \int_0^t \|q(\tau)\|^2_B d\tau \, , \quad 0\le t \le T,
\label{eq:wpg1}
\end{equation} 
\begin{equation}
\int_0^t  \| {\bf \Phi}(\tau)\|^2d\tau     \le K_T  \int_0^t \|F(\tau)\|^2 d\tau \, , 
   \quad 0\le t \le T.
\label{eq:wpg2}
\end{equation}

Here it is only required that $\Phi$, and not its gradient ${\bf \Phi}$,
be estimated by the boundary data $q$. Thus the solution no longer gains a
derivative at the boundary. However, no global problems arise from multiple
reflections because the estimate (\ref{eq:wpg2}) implies the gain of one
derivative in the interior.

As examples, consider the scalar wave problem (\ref{eq:wp1}) with the two
choices of boundary conditions at $x=0$,
\begin{eqnarray}
 ({\bf A})&\quad & \Phi_x -i\beta \Phi_y =q,  \nonumber \\
 ({\bf B})& \quad & \Phi_x -\beta \Phi_y =q, \nonumber
\end{eqnarray}
where $\beta$ is real, with $|\beta|<1$. In case ({\bf A}), $\Phi$ is
complex. Introducing these boundary conditions  into the homogeneous
system (\ref{eq:wp4})--(\ref{eq:wp6}) gives
\begin{eqnarray}
 ({\bf A})&\quad & \kappa =\omega \beta,  \nonumber \\
 ({\bf B})& \quad &\kappa=-i\omega \beta, \nonumber
\end{eqnarray}
with $s^2=\kappa^2-\omega^2$.

In neither case is there a solution with $\Re s>0$ but both cases possess
generalized eigenvalues,
\begin{eqnarray}
 ({\bf A})& \quad& s^2=-\omega^2(1- \beta^2), \quad \Re s =0, \nonumber \\
 ({\bf B})& \quad &s^2=-\omega^2(1+\beta^2), \quad \Re s =0. \nonumber
\end{eqnarray}
The corresponding eigenfunctions are the {\it surface waves}
\begin{equation} ({\bf A}) \quad e^{i\omega
(\pm\sqrt{1-\beta^2}t+y)-|\omega\beta|x}, 
\end{equation}
and the oscillatory waves
\begin{equation} ({\bf B})  \quad e^{i\omega
(\pm\sqrt{1+\beta^2}t+\beta x+y)},
\end{equation}
which give rise to {\it glancing waves} for $\beta=0$.
By investigating the inhomogeneous problem, it can be shown that
both choices of boundary condition give rise to an IBVP which satisfies
(\ref{eq:wpg1}) and (\ref{eq:wpg2}) and is well-posed in the generalized
sense~\cite{heinzpc}.

\bigskip

\subsection{First order symmetric hyperbolic systems}
\label{sec:firstord}

\bigskip

Most of the work on the IBVP for hyperbolic systems has been directed
toward fluid dynamics, where a first order formulation is natural. Here I
describe the essentials of the two main approaches, pseudo-differential
theory and the theory of symmetric hyperbolic systems.

\bigskip

\subsubsection{Pseudo-differential theory}
\label{sec:pseudodiff}

\bigskip

In order to summarize the pseudo-differential theory for a
first order symmetric hyperbolic system consider first  the
constant coefficient system
\begin{equation}
   u_t={\cal P}(\partial_x)u+F,\quad
    {\cal P}(\partial_x)={\cal P}^i \partial_{x_i}
       =A\, \partial_{x_1}+\sum_{j=2}^m B_j \partial_{x_j}
\label{eq:A1}
\end{equation}
on the half-space
$$ t\ge 0,\quad x_1\ge 0,\quad -\infty < x_j <\infty\, , ~j=2,\ldots, m ,
$$
with initial data
\begin{equation}
 u(0,x)= f(x). \label{eq:A4} 
\end{equation}
Here $u(t,x)=\left(u^{(1)}(t,x),\ldots,u^{(N)}(t,x)\right)$ is a vector
valued function of the real variables $(t,x)=(t,x_1,\ldots, x_m)$  and
$A,B_j$ are constant $N\times N$ matrices. In applications to spacetime,
$m=3$ but the number of spatial dimensions does not complicate the theory.
The notation $ \langle u, v \rangle $ and $|u|^2= \langle u, u \rangle$
denotes the inner product and norm in the $N$-dimensional linear space.
All data are smooth, compatible and have compact support. 

The {\it symbol} representing the principal part of the system, 
\begin{equation}
 {\cal P}(i\omega)=iA\omega_1+iB(\omega_-),\quad B(\omega_-)
     =\sum_{j=2}^m B_j \omega_j 
      \, \quad |\omega |=1,
\label{eq:A2} 
\end{equation}
is obtained by replacing $\partial_x$ by its Fourier representation 
$i\omega=(i\omega_1, i\omega_-),~\omega_-=(\omega_2,\ldots,\omega_m)$.
Symmetric hyperbolicity requires that $A$ and $B$ be self-adjoint
matrices so that the eigenvectors of ${\cal P}$ form a complete set with
purely imaginary eigenvalues for all real $\omega$. More precisely, there
exists a symmetric, positive definite {\it symmetrizer} $H$ such that
$HA$ and $HB_j$ are self-adjoint. See~\cite{agran} for more
general applications.

Here it is also assumed that the {\it boundary matrix}
$A$ is nonsingular so that it can be transformed into the form
\begin{equation}
 A=\pmatrix{-\Lambda^I & 0\cr 0& \Lambda^{II}},
\label{eq:A3} 
\end{equation}
where $\Lambda^I,\Lambda^{II}$ are real positive definite diagonal
matrices acting on the $P$ dimensional subspace $u^I$ and the $(N-P)$
dimensional subspace $u^{II}$, respectively. The theory also applies to
the singular case where the boundary is uniformly characteristic, i.e.
the kernel of $A$ has constant dimension~\cite{majda}.
See Sec.~\ref{sec:md} for a treatment of the singular case by
the energy method.

The IBVP requires $P$  boundary conditions at $x_1=0$, corresponding to
the $P$ ingoing modes in the plane wave decomposition carried out below
in conjunction with (\ref{eq:A7}) and (\ref{eq:A11}). They are prescribed
in the form
\begin{equation}
    u^I(t,0,x_{-})=Su^{II}(t,0,x_{-})+q(t,x_{-}),\quad x_{-}=(x_2,\ldots,x_m).
 \label{eq:A5} 
\end{equation}

The main ingredient of a definition of well-posedness is the estimate of
the  solution in terms of the data. See Sec.~7.3 of~\cite{green-book}. By
a transformation analogous to (\ref{eq:homcd}), the IBVP (\ref{eq:A1}),
(\ref{eq:A4}), (\ref{eq:A5})  reduces to a problem with homogeneous
initial data $f=0$. The required estimate is the first order version of
the estimate (\ref{eq:swpg}) for the second order wave equation. The
first order problem is called {\em strongly well-posed in the generalized
sense} if  there is a unique solution $u$  such that
\begin{equation}
\int_0^t \|u(\tau)\|^2 d\tau + 
\int_0^t \|u(\tau)\|^2_B d\tau  \le
K_T \bigg ( \int_0^t \|F(\tau)\|^2 d\tau +
\int_0^t \|q(\tau)\|^2_B d\tau \bigg ) \, , \quad 0\le t\le T,
\label{eq:A6}
\end{equation} where the constant $K_T$ does not depend on $F$ or $q$. 
Here $ \|u\|$ and $ \|u\|_B$ are the $L_2$ norms of $|u|$ over the
half-space and the boundary, respectively.

As for the scalar wave problem considered in Sec.~\ref{sec:fourlap}, the
IBVP is not well-posed if the homogeneous system $F=q=0$ admits wave
solutions
\begin{equation}
u(t,x_1,x_-)=e^{st+i\langle \omega,x\rangle_-} \varphi(x_1),\quad
\langle \omega,x\rangle_-=\sum_{j=2}^m \omega_j x_j, \quad \Re \, s >0,
\label{eq:A7}
\end{equation}
with $|\varphi|_\infty <\infty$. The existence of such homogeneous
solutions would imply the existence of solutions which grow  arbitrarily
fast exponentially.

In order to decide whether such homogeneous waves exist,  introduce
(\ref{eq:A7}) into (\ref{eq:A1}) and (\ref{eq:A5}) to obtain
\begin{eqnarray}
 s\varphi&=A\varphi_{x_1}+iB(\omega_-)\varphi,\quad x_1 \ge 0, \nonumber \\
\varphi^I(0)&=S\varphi^{II}(0),\quad |\varphi|_\infty <\infty. 
\label{eq:A9} 
\end{eqnarray}
This is an eigenvalue problem for a system of ordinary differential
equations which can be solved in the usual way. Let $\kappa$ denote the
solutions of the characteristic equation
\begin{eqnarray}
   {\rm Det}|A\kappa-\left(sI-iB(\omega_-)\right)|=0,
\label{eq:A10} 
\end{eqnarray}
obtained by setting $\varphi(x_1)=e^{\kappa x_1}\varphi (0)$.

\medskip

It can be shown that
\begin{enumerate}

\item There are exactly $r$ eigenvalues with $\Re \, \kappa <0$
and $n-r$ eigenvalues with $\Re \, \kappa>0.$

\item There is a constant $\delta>0$ such that, for all $s$
with real $\Re s >0$
and all $\omega_-$,
$$ |\Re \, \kappa |>\delta \Re s. $$
In particular, for $\Re \, s >0,$ there are no $\kappa$ with  $\Re \, \kappa =0.$

\end{enumerate}
See~\cite{kreiss2} for the proof. 

\medskip
 
If all eigenvalues $\kappa_j$ are distinct, the general solution of (\ref{eq:A9})
has the form
\begin{equation}
   \varphi=\sum_{\Re \, \kappa_j<0} \sigma_j e^{\kappa_jx_1} h_j+
\sum_{\Re \, \kappa_j>0} \sigma_j e^{\kappa_jx_1} h_j ,
\label{eq:A11}
\end{equation}
where $h_j$ are the corresponding eigenvectors.
(If the eigenvalues are degenerate, the usual modifications apply.)
For bounded solutions, all $\sigma_j$ in the second
term are zero. Introducing $\varphi$ into the boundary conditions
(\ref{eq:A9}) at $x_1=0$ gives a linear system of $r$ equations for
the $r$ unknowns $(\sigma_1,\ldots,\sigma_r)= {\underline{\sigma}}$
of the form
\begin{equation}
 C(s,\omega_-) {\underline{\sigma}} =0. 
\label{eq:A12} 
\end{equation}
Therefore the problem is not well-posed if for some $\omega_-$ 
there is an eigenvalue $s_0$ with $\Re \,s_0>0$,  i.e. 
${\rm Det} \,C(s_0,\omega_-)=0$. In that case, the linear system
(\ref{eq:A12}), and therefore also (\ref{eq:A9}), has a nontrivial solution.
Thus the determinant condition
\begin{equation}
   {\rm Det} \,C(s_0,\omega_-)\ne 0, \quad \Re \,s_0>0 ,
\label{eq:detcond}
\end{equation}
is necessary for a well-posed problem.

Thus in order to consider a well-posed problem assume that
${\rm Det} \,C\ne 0$ for $\Re \, s >0.$ Then
the inhomogeneous IBVP (\ref{eq:A1}), (\ref{eq:A4}), (\ref{eq:A5})
can be solved by Laplace transform in time and Fourier
transform in the tangential variables. Again set $u(0,x)=f(x)= 0$. Then 
\begin{eqnarray}
s\hat u&=&A\hat u_x+iB(\omega_-) \hat u+\hat F, 
\label{eq:A13a} \\
\hat u^I(0)&=&S\hat u^{II}(0)+\hat q.
\label{eq:A13}
\end{eqnarray}
Since, by assumption, (\ref{eq:A9}) has only the trivial solution for $\Re \, s>0$,
(\ref{eq:A13}) has a unique solution. Inverting the Fourier and Laplace
transforms gives the solution in physical space.

As in the scalar wave case, it is simple to solve (\ref{eq:A13a}) for $\hat F = 0$. 
From the inhomogeneous versions of (\ref{eq:A11}) and (\ref{eq:A12}),
$$ \hat u(s,x_1,\omega_-)=\sum_{\Re \, \kappa_j <0} \sigma_j e^{\kappa_j  x_1} h_j, 
$$
where the $\sigma_j$ are determined by
$$ C(s,\omega_-) {\underline{\sigma}}=\hat q. 
$$

\medskip

It is now possible to establish the pseudo-differential version of boundary
stability:

\medskip

\noindent For all
$\omega,s$ with $\Re s>0,$ there is a constant $K$ independent
of $\omega,s$ and $\hat q$ such that the solutions of
(\ref{eq:A13a})--(\ref{eq:A13}) with $\hat F = 0$ satisfy 
\begin{equation}
   |\hat u(s,0,\omega)|\le K|\hat q(s,\omega)|.  
\label{eq:A14} 
\end{equation}

\medskip

Boundary stability is equivalent to the requirement
that the eigenvalue problem (\ref{eq:A9}) has
no eigenvalues for $\Re \, s\ge 0$ or that ${\rm Det} \, C(\omega_-,s)\ne 0$
for $\Re \, s\ge 0$. In particular, it rules out generalized eigenvalues.
It is essential to establish the

\medskip

\noindent {\bf Main Theorem}: If the half-space problem is boundary stable
then it is  strongly well-posed in the generalized sense of (\ref{eq:A6}).

\medskip

\noindent See~\cite{kreiss2} for the proof, where boundary
stability is used to construct a 
{\it symmetrizer} in the Fourier-Laplace representation
which leads to the estimate
\begin{equation}
\eta \|\hat u(s,x_1,\omega)\|^2+|\hat u(s,0,\omega)|^2\le
const\left( {1\over\eta} \|\hat F\|^2+c|\hat q|^2\right).
\label{eq:A16} 
\end{equation}
Inversion of the Fourier-Laplace transform proves the theorem.

\medskip

The pseudo-differential theory
has far reaching consequences. In particular, the computational rules
for pseudo-differential operators imply:

\begin{enumerate}

\item The Laplace transform only requires that the estimates hold for
$\eta>\eta_0>0$, where $\eta_0$ is sufficiently large to allow for (controlled)
exponential growth due to lower order terms.
This is essential for extending strong well-posedness to
systems with variable coefficients.

\item  Boundary stability is also valid if the symbol depends smoothly on $(t,x)$
and is not destroyed by lower order terms. Therefore the problem
can be localized and well-posedness in general domains can be
reduced to the study of the Cauchy problem and half-space problems.
 
\item The {\it principle of frozen coefficients} holds. 
The properties of the pseudo-differential operators give rise to estimates of
derivatives in the same way as for standard partial differential equations.
Therefore strong well-posedness in the generalized  sense can
be extended to linear problems
with variable coefficients and, locally in time,
to quasilinear problems.

\end{enumerate}

\medskip

Since pseudo-differential operators are much more flexible than standard
differential operators,  they can be applied to second order systems as
well as first order systems. Consider, for example, the problem
(\ref{eq:wp1})--(\ref{eq:wp3}) discussed in Sec.~\ref{sec:fourlap}. After
transforming the initial data to zero, the Fourier-Laplace transform
becomes
\begin{eqnarray}
&\hat \Phi_{xx}&=(s^2+\omega^2)\hat \Phi-\hat F, \nonumber \\
& s\hat \Phi&-\alpha \hat \Phi_x- i \beta \omega \hat \Phi =\hat q.
\label{eq:A17}
\end{eqnarray}
Introduction of a new variable $\hat \Phi_x=\sqrt{|s|^2+\omega^2}\,\hat
\Psi$ gives the first order system
\begin{eqnarray}
\hat u_x &=&
\frac{1}{\sqrt{|s|^2+\omega^2}}\,
\pmatrix{ 0 & |s|^2+\omega^2\cr s^{2}+\omega^{2} & 0 \cr}
 \hat u -\tilde F,\quad 
 \hat u=\pmatrix{\hat \Phi\cr \hat \Psi\cr}, \cr
&\cr
&{}&\frac{s}{\sqrt{|s|^2+\omega^2}}\hat \Phi-\alpha \hat \Psi
     - \frac{i\beta\omega}{\sqrt{|s|^2+\omega^2}}\hat \Phi=\tilde q,
\label{eq:A18}
\end{eqnarray}
where
\begin{eqnarray}
\tilde F={1\over\sqrt{|s|^2+\omega^2}} \pmatrix{0\cr \hat F\cr},\quad
\tilde q={1\over\sqrt{|s|^2+\omega^2}} \hat q.
\label{eq:A19}
\end{eqnarray}

\medskip

The second order problem (\ref{eq:wp1})--(\ref{eq:wp3}) is strongly
well-posed in the generalized sense if the corresponding first order
problem (\ref{eq:A18}) with general data $\tilde F,\tilde q$ has this
property. The second order version of boundary stability  (\ref{eq:A17})
established in Sec.~\ref{sec:fourlap}, rewritten  in terms of the first
order variables, implies
\begin{eqnarray}
|\hat \Phi(s,0,\omega)|+|\hat \Psi(s,0,\omega)| \le K \, |\tilde q(s,\omega)|, 
\label{eq:A20}
\end{eqnarray}
i.e. the first order order version of boundary stability (\ref{eq:A14}).
Thus the Main Theorem applies and the second order problem is
strongly well-posed in the generalized sense in the first order version
(\ref{eq:A6}), which is equivalent to the second order version
(\ref{eq:swpg}).

\bigskip

\subsubsection{The energy method for first order symmetric hyperbolic
systems}
\label{sec:md}

\bigskip

Energy estimates for first order symmetric hyperbolic systems were first
applied to Einstein's equations in harmonic coordinates by Fischer and
Marsden~\cite{fischmarsd} to give an alternative derivation of the
results of Choquet-Bruhat for the Cauchy problem. The energy method
extends to the quasilinear IBVP with {\em maximally dissipative} boundary
conditions.

Again, begin by considering  the constant coefficient system
(\ref{eq:A1})
\begin{equation}
   u_t={\cal P}^i\partial_i u+F,\quad {\cal P}^i\partial_i
   =A\, \partial_{x_1}+\sum_{j=2}^m B_j \partial_{x_j}
\label{eq:B1}
\end{equation}
on the half-space
$$ t\ge 0,\quad x_1\ge 0,\quad -\infty < x_j <\infty\, , ~j=2,\ldots, m ,
$$
with initial data $ u(0,x)= f(x)$. As before, $A$ and $B_j$ are symmetric
$N\times N$ matrices so that the eigenvectors of $P^i$ form a complete
set with real eigenvalues.  In matrix notation, there is a symmetric
positive definite symmetrizer  $H_{MN}$ such that  $H_{MP}A^P_N$ and
$H_{MP} B_{j}{}^{P}_N$ are symmetric. Here, the boundary matrix $A$ is
allowed to be  singular, cf.~\cite{majda,rauch,secchi3}. With an
appropriate choice of symmetrizer, it can be put in the form
\begin{equation}
 A=\alpha \pmatrix{-I^P & 0 &0\cr 0& O^Q& 0\cr 0&0& I^R} ,
     \,  \alpha>0, \quad
     \quad u= \pmatrix{u^I  \cr u^O \cr  u^{II}} 
\label{eq:B3} 
\end{equation} where $I^P$and $I^R$ are identity matrices acting on the
$P$-dimensional subspace $u^I$ and the $R$-dimensional subspace $u^{II}$,
respectively,and $O^Q$ is a zero matrix acting on $Q$-dimensional kernel
$u^O$, with $N=P+Q+R$. No boundary condition can be imposed on the
components of  $u^{II}$, which are the outgoing modes, or the components
of $u^O$, The  components of $u^O$ satisfy
PDEs intrinsic to the boundary. They are referred to as
{\it zero velocity modes} since they have no velocity relative
to the boundary. As in  the discussion of the
non-characteristic case in Sec.~\ref{sec:pseudodiff}, there are $P$
ingoing modes and the IBVP requires $P$  boundary conditions at $x_1=0$.
They are prescribed in the {\em maximal} form
\begin{equation}
   u^I(t,0,x_{-})=S u^{II}(t,0,x_{-})+q(t,x_{-}),\quad
         x_{-}=(x_2,\ldots,x_m) ,
 \label{eq:B5} 
\end{equation}
where the $P\times R$ matrix $S$ satisfies the {\em dissipative}
condition that, for homogeneous data $q=0$, the local energy flux out of
the boundary be positive,
\begin{equation}
       {\cal F}: =  \langle u, A u \rangle \ge 0.
\end{equation}
In the simplified form (\ref{eq:B3}), this leads to the requirement 
\begin{equation}
      - |Su^{II}(t,0,x_{-})|^2  + |u^{II}(t,0,x_{-})|^2  \ge 0,
 \label{eq:B6} 
\end{equation}
where $|u|^2 =  \langle u, u \rangle$ in terms of the linear space inner
product.

The rationale for these {\em maximally dissipative} boundary conditions
results from the energy estimate for the case of homogeneous boundary
data $q=0$. Beginning with
\begin{equation}
            \partial_t  \langle u,u \rangle
          =  2\langle u,A\, \partial_{x_1}u
     +\sum_{j=2}^m B_j \partial_{x_j}u +F \rangle ,
 \label{eq:B7} 
\end{equation}
integration over the half-space gives
\begin{eqnarray}
  \partial_t E:=  \partial_t  \| u \|^2
          &= & 2(u,A\, \partial_{x_1}u) +2(u,F) \nonumber \\
          &= & - (u,Au)_B +2(u,F), \nonumber \\
            &\le &2(u,F) \le  \| u \|^2+ \| F \|^2 =E + \| F \|^2 ,
 \label{eq:B8} 
\end{eqnarray}
where the $(u,v)$ denotes the integral of $\langle u, v\rangle $, etc.
Thus the maximally dissipative boundary conditions provide the required
energy estimate. Inhomogeneous boundary data $q$ can be transformed to
zero by the change of variable $u\rightarrow u-q$ to obtain
an analogous estimate. 

More generally, if the boundary is uniformly characteristic so that the
kernel of the boundary matrix $A$ has constant dimension $Q$, then the
quasilinear IBVP problem with maximally dissipative boundary conditions
is strongly well-posed: a solution exists locally in time which satisfies
(\ref{eq:A6}). See~\cite{shuxing,ohkubo,gues,secchi2,secchi3} for
details.

The theory can also be recast in the covariant space-time form
\begin{equation}
     A^\mu \partial_\mu u =F
     \label{eq:fosh}
\end{equation}
where $A^\mu=(A^t,A^i)$ are symmetric matrices and $A^t$ is positive
definite. As an illustration of this and the various choices of boundary
conditions,  consider the IBVP for the scalar wave equation
\begin{equation}
         g^{\mu\nu}\partial_\mu \partial_\nu \Phi = 0, 
         \quad x\ge 0, \quad t \ge 0, \quad g^{xx}>0.
\end{equation}
This can be rewritten in the first order symmetric hyperbolic form
(\ref{eq:fosh}) for a 5-component field $u$ by introducing the auxiliary
variables 
\begin{equation}
 u =\left( 
\begin{array}{c}
u_0  \\
u_t  \\          
u_i   
\end{array}
\right)
= \left( 
\begin{array}{c}
\Phi  \\
\partial_t \Phi \\          
\partial_i \Phi  
\end{array}
\right) .
\end{equation}  
The matrices $A^\mu$ are then given by
\begin{equation}
A^t = \left( 
\begin{array}{ccc}
1 & 0 & 0 \\
0 & -g^{tt}  & 0  \\          
0 & 0 &  g^{jk}  
\end{array}
\right) ,
\quad
A^i = \left( 
\begin{array}{ccc}
0 & 0 & 0 \\
0 & -2g^{ti}  & -g^{ji}  \\          
0 & -g^{ij} &  0   
\end{array}
\right),
\end{equation}
with
\begin{equation}
 F=  \left( 
\begin{array}{c}
 u_t \\
0  \\          
0   
\end{array}
\right).
\end{equation} 

In this first order form, the Cauchy data consist of $u|_{t=0}=f$ subject
to the constraints
\begin{equation}
  {\cal C}_i:=  u_i - \partial_i u_0 =0.
\label{eq:sconstr}
\end{equation}
The evolution system implies that the constraints satisfy
\begin{equation}
  \partial_t   {\cal C}_i=0,
\end{equation}
so that they propagate up the timelike boundary at $x=0$ and present no
complication. 

The boundary matrix $(-A^x)$ (oriented in the outward normal direction)
has a 3-dimensional kernel, whose transposed basis
consists of the zero-velocity modes
\begin{equation}
 (1,0,0,0,0), \quad
 (0,0,-g^{xy},g^{xx}, 0), \quad \mbox{and} \quad
 (0,0,-g^{xz},0, g^{xx} ) .
\end{equation}
In addition, there is one positive eigenvalue and one negative eigenvalue
\begin{equation}
 \lambda_{\pm}= \pm \lambda +g^{xt},
\end{equation}
where 
\begin{equation}
 \lambda= \sqrt{({g^{xt}})^2+\delta_{ij}g^{xi}g^{xj}} .
\end{equation}
Thus precisely one boundary condition is required.

In terms of the normalized eigenvectors
\begin{equation}
 e_{\pm} = \frac{1}{\sqrt{ \pm 2 \; \lambda \; \lambda_\pm }} \left( 
\begin{array}{c}
0 \\
 \pm \lambda +g^{xt}  \\          
+g^{xi}  ,
\end{array}
\right).
\end{equation}  
 $u=u_+ e_ + +u_- e_- +u_O$, where
$u_O$ lies in the kernel. The boundary condition takes the form $u_+
-Su_- =q$, subject to the dissipative condition
\begin{equation}
  {\cal F}=-\langle u,A^x u \rangle \ge 0.
\end{equation}
For the case of homogeneous data $q=0$, this requires that 
\begin{eqnarray}
    S^2 \le - {\lambda_-} \; / \; {\lambda_+} \; .
\end{eqnarray}

The limiting cases $S = \pm \sqrt{ - \; {\lambda_-} \; / \; {\lambda_+}}$
lead to reflecting boundary conditions.  In the case of a standard
Minkowski metric $g^{\mu\nu}=\eta^{\mu\nu}$, these correspond  to the
homogeneous Dirichlet condition and Neumann conditions
\begin{equation}
       \partial_t \Phi|_B = 0 , \quad \partial_x \Phi |_B=0,
\end{equation}
and the choice $S=0$ corresponds to the Sommerfeld condition
\begin{equation}
      ( \partial_t -\partial_x) \Phi |_B = 0.
\end{equation}

More generally, the geometric interpretation of these boundary conditions
is obscured because the energy $E$  (\ref{eq:B8}) standardly
used in the first order theory is constructed with the linear space inner
product, as opposed to the geometrically defined energy  (\ref{eq:ewave})
natural to the second order theory. It is possible to reformulate the
covariant theory of the wave equation in first order symmetric form with
boundary conditions based upon the covariant energy
(\ref{eq:ewave}).  But without such a guide to begin with, a
first order symmetric hyperbolic formulation can lose  touch with
the underlying geometry.

This scalar wave example also illustrates how the number of evolution
variables and constraints escalate upon reduction to first order form. As
a result, in the case of Einstein's equations, the advantages of
utilizing symmetric hyperbolic theory is counterbalanced by the increased
algebraic complexity which is further complicated by the wide freedom in
carry out a first order reduction. See Sec.~\ref{sec:firstord}.

\bigskip

\subsection{The characteristic Initial-boundary value problem}
\label{sec:civp}

\bigskip

There is another IBVP which gained prominence after Bondi's~\cite{bondi}
success in using null hypersurfaces as coordinates to describe
gravitational waves. In the null-timelike IBVP,  data is given on an
initial characteristic hypersurface and on a timelike worldtube to
produce a solution in the exterior of the worldtube.  The underlying
physical picture is that the worldtube data represent the outgoing
gravitational radiation emanating from interior matter sources, while
ingoing radiation incident on the system is represented by the initial
null data. 

The characteristic IBVP received little attention before its
importance in general relativity was recognized. Rendall~\cite{rend}
established well-posedness of the double null problem for the quasilinear
wave equation, where data is given on a pair of intersecting
characteristic hypersurfaces. He did not treat the characteristic problem
head-on but reduced it to a standard Cauchy problem with data on a
spacelike hypersurface  passing through the intersection of the
characteristic hypersurfaces so that well-posedness followed from
the result for the Cauchy problem. He extended this approach to establish
the well-posedness of the double-null formulation of the full Einstein
gravitational problem. Rendall's approach cannot be applied
to the null-timelike problem, even though the double null problem is a
limiting case. 

The well-posedness of the null-timelike problem for the gravitational
case remains an outstanding problem. Only recently has it been shown
that the quasilinear problem for scalar waves is well-posed~\cite{nullt}.
The difficulty unique to this problem can be illustrated in
terms of the 1(spatial)-dimensional wave equation
\begin{equation}
             ( \partial_{\tilde t}^2 -\partial_{\tilde x}^2)\Phi=0,
             \label{eq:tilde1d}
\end{equation}
where $(\tilde t,\tilde x)$ are standard space-time coordinates. The
conserved energy
\begin{equation}
       \tilde  E(\tilde t)= \frac{1}{2} \int d\tilde x  \bigg( (
        \partial_{\tilde t}\Phi)^2  +(\partial_{\tilde x}\Phi)^2 \bigg )
       \label{eq:tildeE}
\end{equation}
leads to the well-posedness of the Cauchy problem. In characteristic
coordinates $(t=\tilde t -\tilde x, \, x=\tilde t +\tilde x)$, the wave
equation transforms into
\begin{equation}
   \partial_t \partial_x \Phi =0.
   \label{eq:1dphi}
\end{equation}
The conserved energy evaluated on the characteristics $t={\rm const}$,
\begin{equation}
       \tilde   E(t) = \int dx (\partial_ x \Phi)^2,
\end{equation}
no longer controls the derivative $\partial_t \Phi$.

The usual technique for treating  the IBVP is to split the problem into a
Cauchy problem and local half-space problems and show that these
individual problems are well posed. This works for hyperbolic systems
based upon a spacelike foliation, in which case signals propagate with
finite velocity. For (\ref{eq:tilde1d}), the solutions to the Cauchy
problem with compact initial data on $\tilde t=0$ are square integrable
and well-posedness can be established using the $L_2$ energy norm 
(\ref{eq:tildeE}).

However, in characteristic coordinates the 1-dimensional wave equation
(\ref{eq:1dphi}) admits signals traveling in the $+x$-direction with
infinite coordinate velocity. In particular, initial data of compact
support $\Phi(0,x)=f(x)$ on the characteristic $t=0$ admits the solution
$\Phi = g(t) +f(x)$, provided that $g(0)=0$.  Here $g(t)$ represents the
profile of a wave which travels from past null infinity ($x\rightarrow
-\infty$) to future null infinity  ($x\rightarrow +\infty$). Thus,
without a boundary condition at past null infinity, there is no unique
solution and the Cauchy problem is ill posed. Even with the boundary
condition $\Phi(t,-\infty)=0$,  a source of compact support $S(t,x)$
added to (\ref{eq:1dphi}), i.e.
\begin{equation}
   \partial_t \partial_x \Phi =S,
   \label{eq:1dphis}
\end{equation}
produces waves propagating to $x=+\infty$ so that although the solution
is  unique it is still not square integrable.

On the other hand, consider the modified problem obtained by setting
$\Phi=e^{ax}\Psi$,
\begin{equation} 
 \partial_t (\partial_x+a) \Psi=F \, ,  \quad \Psi(0,x)
    = e^{-ax}f(x), \quad  \Psi(t,-\infty)=0 \,  ,\quad a>0 ,
       \label{eq:1dpsi}
\end{equation} 
where $F=e^{-ax}  S$. 
The solutions to (\ref{eq:1dpsi}) vanish at $x=+\infty$ and are square
integrable. As a result, the problem (\ref{eq:1dpsi}) is well
posed with respect to an $L_2$ norm. For the simple case where $F=0$,
multiplication of (\ref{eq:1dpsi}) by $(2a \Psi+\partial_x
\Psi+\frac{1}{2}\partial_t \Psi)$ and integration by parts gives
\begin{equation}
       \frac{1}{2}\partial_t  \int dx \bigg( 
             (\partial_ x \Psi)^2+2a^2 \Psi^2 \bigg)
            =\frac{a}{2} \int dx \bigg(2(\partial_ t \Psi)\partial_ x \Psi 
           -  (\partial_ t \Psi)^2 \bigg)   
         \le \frac{a}{2} \int dx  (\partial_ x \Psi)^2 .
\end{equation} 
The resulting inequality
\begin{equation}
       \partial_t  E \le {\rm const} E 
\end{equation} 
for the energy
\begin{equation}
      E=\frac{1}{2}  \int dx \bigg( (\partial_ x \Psi)^2+2a^2 \Psi^2 \bigg)
      \label{eq:1denergy}
\end{equation} 
provides the estimates for $\partial_x \Psi$ and $\Psi$ which are
necessary for well-posedness. Estimates for $\partial_t \Psi$, and other
higher derivatives, follow from applying this approach to the derivatives
of  (\ref{eq:1dpsi}). The approach can be extended to include the source
term $F$ and other generic lower differential order terms. This allows
well-posedness to be extended to the case of variable coefficients and,
locally in time, to the quasilinear case.

Although well-posedness of the problem was established in the 
modified form (\ref{eq:1dpsi}), the energy estimates can be translated back to the
original problem (\ref{eq:1dphis}).  The modification in going from  
(\ref{eq:1dphis}) to  (\ref{eq:1dpsi}) leads to an effective modification
of the standard energy for the problem.  Rewritten in terms of the
original variable $\Phi=e^{ax}\Psi$, (\ref{eq:1denergy}) corresponds to
the energy 
\begin{equation}
     E=\frac{1}{2}  \int dx e^{-2ax}  \bigg( (\partial_ x \Phi)^2
     +a^2 \Phi^2 \bigg ).
     \label{eq:enorm}
\end{equation}
Thus while the Cauchy problem for (\ref{eq:1dphis}) is ill posed with
respect to the standard $L_2$ norm it is well posed with respect to the
exponentially weighted norm (\ref{eq:enorm}). 

This technique can be applied to a wide range of characteristic
problems. In particular, it has been applied to the quasilinear wave
equation for a scalar field $\Phi$ in an asymptotically flat curved space
background with source $S$,
\begin{equation}
        g^{ab}\nabla_a \nabla_b \Phi = S (\Phi,\partial_c \Phi, x^c),
        \label{eq:qw}
\end{equation}
where the metric $g^{ab}$ and its associated covariant  derivative
$\nabla_a$ are explicitly prescribed functions of $(\Phi,x^a)$. In
Bondi-Sachs coordinates~\cite{bondi,sachs} based upon outgoing null
hypersurface $u={\rm const}$, the metric has the form 
\begin{equation}
   g_{\mu\nu}dx^\mu dx^\nu = -(e^{2\beta}W-r^{-2}h_{AB}W^A W^B)du^2 
        -2e^{2\beta}dudr
      -2h_{AB}W^B dudx^A   +r^2h_{AB}dx^A dx^B,
      \label{eq:nullmet}
\end{equation}
where $x^A$ are angular coordinates, such that $(u,x^A)={\rm const}$
along the outgoing null rays, and $r$ is an areal radial coordinate. Here
the metric coefficients $(W,\beta,W^A,h_{AB})$ depend smoothly upon
$(\Phi,u,r,x^A)$ and fall off in the radial direction consistent with
asymptotic flatness. The null-timelike problem consists of determining
$\Phi$ in the exterior region given data on an initial null hypersurface
and on an inner timelike worldtube,
\begin{equation}
     \Phi(0,r,x^A)=f(r,x^A)\,  ,\quad \Phi(u,R,x^A) =q(u,x^A), 
        \quad R\le  r<\infty  , \quad u\ge 0.
         \label{eq:ndata}
\end{equation}

\bigskip

It is  shown in~\cite{nullt}:

\medskip

\noindent {\it The null-timelike IBVP
(\ref{eq:qw})--(\ref{eq:ndata}) is strongly well-posed subject to a
positivity condition that the principal part of the wave operator reduces
to an elliptic operator in the stationary case. }

\bigskip

The proof is based upon energy estimates obtained in compactified
characteristic coordinates extending to ${\cal I}^+$. 

\bigskip

\section{Historical developments}
\label{sec:history}

\bigskip

Early computational work in general relativity focused on the Cauchy
problem. The IBVP only recently received serious attention 
and is still a work in progress.
Although there are two formulations where strong well-posedness has been
established (see Sec's~\ref{sec:fn} and~\ref{sec:harm}), several
important questions remain. Along the way,  there have been partial
successes based upon ideas of potential value in guiding future work.

\bigskip

\subsection{The Frittelli-Reula formulation}
\label{eq:fritreul}

\bigskip

The first extensive treatment of the IBVP for Einstein equations was
carried out by Stewart~\cite{stewart}, motivated at the time by the
tremendous growth in computing power which made numerical relativity a
realistic approach for applications to relativistic astrophysics. His
primary goal was to investigate how to formulate an IBVP for Einstein's
equations  which would allow unconstrained numerical evolution. Stewart
focused upon a formulation of Einstein's equations due to Frittelli and
Reula~\cite{fritrel1,fritrel2}, although his approach is sufficiently
general to have application to other formulations.

The Frittelli-Reula system was chosen because it is symmetric hyperbolic
for certain choices of the free parameters. It is based upon the ADM
formalism, with metric (\ref{eq:admmet}), in which all second derivatives
are eliminated by the introduction of auxiliary variables. There is a
2-parameter freedom in the choice of first order variables. The
densitized lapse $h^p \alpha$, where $p$ is an additional adjustable
parameter, and the shift are treated as explicitly prescribed variables.
In addition to the Hamiltonian and momentum constraints
(\ref{eq:hamomc}), the integrability conditions arising from the
auxiliary variables lead to 18 additional constraints.
Another adjustable parameter controls the
freedom of mixing the constraints with the evolution system.

This net result is that the vacuum Einstein equations reduce to a first
order evolution system of the form (\ref{eq:A1}) consisting of 30
equations governing the metric variables and 22 equations governing the
constraints. This is further complicated by the lack of geometric or
tensorial properties of the evolution variables. Frittelli and Reula
analyzed the principal part of the system and showed that it is symmetric
hyperbolic for certain values of the adjustable parameters. 

The well-posedness of the IBVP for such symmetric hyperbolic reductions
of Einstein's equations depends upon whether proper constraint preserving
boundary conditions can be imposed. Stewart analyzed the eigenvalues of
the boundary matrix for the linearized system using the Fourier-Laplace
method described in Sec.~\ref{sec:fourlap}. For the evolution system
governing the metric variables, he identified 6 ingoing modes, 6 outgoing
modes and 18 zero-velocity modes in the kernel which propagate tangential
to the boundary. Thus this system requires exactly 6 boundary conditions.
From the boundary matrix for the system governing constraint evolution,
he identified 3 ingoing modes, 3 outgoing
modes and 16 zero velocity modes, so that 3 boundary conditions are
required for constraint enforcement.

The Fourier-Laplace analysis of the constraint system showed that
the determinant condition  (\ref{eq:detcond}) was satisfied, i.e.  the
homogeneous problem had only the trivial solution, and that the linearized
system had a well-posed IBVP. The three boundary conditions
could be satisfied by requiring that  the Cauchy momentum constraints
(\ref{eq:momc}) vanish on the boundary.

The analysis of the evolution system governing the metric showed that the
constraints could be enforced by a particular choice of boundary data for
the metric variables. In this way, the evolution could be freed from the
constraint system. The details are hidden in the symbolic algebra scripts
necessary to analyze the complexity of the Fourier-Laplace modes.
No discussion was given of the estimates for the derivatives which would be
necessary for the well-posedness of the nonlinear problem. There has
apparently been no further results for the Frittelli-Reula formulation
and no attempt at a numerical evolution code.

Although the results were inconclusive for the nonlinear case,  Stewart's
treatment provided the first example of  how to apply pseudo-differential
techniques to the IBVP for Einstein's equations
and served as the basis for much of the following work. The recognition that
constraint preservation for this system could be achieved by enforcing
the Cauchy momentum constraints on the boundary suggests a possible wider
application but whether there is any universal procedure for enforcing
the constraints in the  $3+1$ formulation remains an open issue. See
Sec.~\ref{sec:constr} for a discussion. 

\bigskip

\subsection{The BSSN and NOR formulations}

\bigskip

Codes based upon the Baumgarte-Shapiro-Shibata-Nakamura (BSSN)
formulation~\cite{bssn1,bssn2} have been used by the majority of
groups~\cite{mCcLpMyZ06,jBjCdCmKjvM06b,jGetal,pDfHdPeSeSrTjTjV,bssnsom}
carrying out simulations of  binary black hole and neutron star systems.
The successful development of the BSSN formulation proceeded through an
interplay between educated guesses and feedback from code performance.
Only in hindsight has its success spurred mathematical analysis, which
showed that certain versions were strongly hyperbolic and thus had a
well-posed Cauchy problem~\cite{sartig,nor,gundl1}.  Although significant
progress has been made in establishing some of the necessary conditions
for well-posedness and constraint preservation  of the
IBVP~\cite{gundl2,horst,nunsar},  there is still no satisfactory
mathematical theory on which to base a numerical treatment of the
boundary. 

Similar to the Frittelli-Reula system, the BBSN system reduces the
Einstein equations to first order form by the introduction of auxiliary
variables.  In addition, there is a conformal-traceless decomposition of
the 3-geometry. The Nagy-Ortiz-Reula (NOR) system~\cite{nor} is similar
but without the conformal decomposition.  Again, a number of free
parameters enter into the first order reduction and into the way that the
constraints are mixed with the evolution system. For a  particular choice
of parameters the linearization off Minkowski space is symmetric
hyperbolic~\cite{nunsar} and leads to a well-posed IBVP for the
linearized problem. However, the corresponding nonlinear system is no
longer symmetric hyperbolic and there is no well-posed boundary
treatment.

Gundlach and Martin-Garcia~\cite{gundl2} studied simplified second order
versions of of the BSSN and NOR systems which were symmetric hyperbolic in
a sense they defined in~\cite{gundl3}. They were able to confirm and
generalize many of the results found in~\cite{sartig} for the first order
reduction. Of particular interest, they found that all the characteristic
modes propagated causally, in contrast to the superluminal modes present
in the first order system. The chief shortcoming of their treatment is
the incompatibility of the constraints with the dissipative boundary
conditions necessary for well-posedness.

The strong well-posedness of the IBVP for $3+1$ formulations remains an
outstanding problem. The strategy in current numerical practice for BSSN
evolution systems is to apply naive, homogeneous Sommerfeld boundary
conditions, where needed, to each evolution variable (cf.~\cite{bssnsom})
and place the boundary out far enough so that its harmful effects are
limited.

\bigskip

\subsection{Other $3+1$ studies}
\label{sec:other}

\bigskip

Many of the other early investigations on the well-posedness of the IBVP
for $3+1$ formulations centered about linearized systems~\cite{szilschbc,
calpulreulbc, calsarbc, nagysar}, spherically symmetric and 1D
spacetimes~\cite{iriond,bardbuch,calehnt, fritgombcss} or other model
problems~\cite{novak,lindbc,reulsarmod, gundl3} which simplified the
treatment. In particular, well-posedness of the linearized problem is a
necessary condition for extension to the nonlinear case.

One promising approach was based upon a generalization of the
Frittelli-Reula and Einstein-Christoffel (EC)~\cite{andy} systems, which
Kidder, Scheel and Teukolsky (KST) showed was symmetric hyperbolic for
certain values of the free parameters~\cite{kst}. 
In~\cite{calpulreulbc}, energy estimates for  maximally dissipative
boundary conditions were used to formulate a well posed IBVP for the
linearization of this system off Minkowski space. However, constraint
preservation limited the allowed boundary conditions to the reflecting
Dirichlet or Neumann type. 

The geometric analogy between the Hamiltonian and momentum constraints
(\ref{eq:hamomc})  of the Cauchy problem and the {\it boundary constraints}
\begin{equation}
               G^{ab}N_b =0,
               \label{eq:bconstr}
\end{equation}
where $N_b$ is the normal to the boundary, led Frittelli and Gomez to
propose that (\ref{eq:bconstr}) be enforced as boundary
conditions~\cite{fritgombc, fritgombc2,fritgombc4}. They showed for the
EC system, with vanishing shift and certain choices of the free
parameters, that enforcing three linearly dependent combinations of the
boundary constraints would lead to preservation of the Hamiltonian and
momentum constraints. Furthermore, they showed that these linear combinations
could be used to formulate boundary conditions for three of the ingoing
metric variables. They did not study the boundary stability of the
resulting IBVP. 

In~\cite{calsarbc}, the determinant condition (\ref{eq:detcond}) of the
Fourier-Laplace method was applied to the linearized (EC) system to
identify ill-posed modes arising from various choices of boundary
conditions and free parameters. Several noteworthy results were found.
For parameter choices in which the EC system was strongly but not
symmetric hyperbolic, they found that maximally dissipative boundary
conditions gave rise to ill posed modes. This is in accord with the
general theory which requires both maximally dissipative boundary
conditions and a symmetric system to guarantee a well posed IBVP. In
addition, for a range of parameters giving rise to a symmetric system,
ill-posed modes were found for boundary conditions based upon the
boundary constraints (\ref{eq:bconstr}). As in the case of the Cauchy
momentum constraints proposed by Stewart, this casts doubt on whether
such boundary constraints are universally applicable. The approach
in~\cite{calsarbc} was effective for ferreting out what doesn't work but
did not go beyond the results of~\cite{calpulreulbc} for establishing a
well-posed IBVP.

In a later study~\cite{sarbtigl}, the EC system was further generalized
to include a dynamical lapse of the Bona-Masso type~\cite{bonmass} and
fixed shift. The IBVP was analyzed in the high frequency limit, again
using the determinant condition of the Fourier-Laplace method to
determine ill-posed modes. It was found that constraint preserving
boundary conditions that were based upon the Newman-Penrose~\cite{np}
Weyl curvature component $\Psi_0$ satisfied the determinant condition
provided the evolution system was strongly hyperbolic and the constraint
propagation system was symmetric hyperbolic. Boundary conditions based
upon the Newman-Penrose $\Psi_0$ component were first introduced in the
Friedrich-Nagy system~\cite{fn}. Other $\Psi_0$ boundary conditions for
the EC system were tested
in~\cite{lindbc,caltechbc,rinne,nagysar,oRlLmS07,sarbuch,sarbuch2,
ruizhbc}. They have been improved to be highly effective absorbing
boundary conditions for gravitational waves (see Sec.~\ref{sec:absorbc})
and have performed well in numerical tests. See Sec.~\ref{sec:num}.

\bigskip

\subsection{The harmonic and Z4 formulations}

\bigskip

The IBVP for general relativity takes on one of its simplest forms in the
harmonic formulation, in which the Einstein equations reduce to 10
quasilinear wave equations. Nevertheless, progress on this problem was
not straightforward. Difficulties arose in handling the harmonic
constraints (\ref{eq:harmcond}), in which derivatives of the metric
tangential to the boundary prohibited use of standard dissipative
boundary conditions for the wave equation. An early well-posed
treatment was based upon the observation that the harmonic Cauchy problem
is well-posed~\cite{harl,mBbSjW06}. Consequently, if locally smooth
reflection symmetry were imposed across the boundary then the
well-posedness of the Cauchy problem would imply well-posedness of
resulting IBVP on either side of the boundary. The reflection symmetry
forces the troublesome tangential derivatives to vanish but it also
forces homogeneous boundary conditions of the Dirichlet or Neumann type.
Although these boundary conditions satisfy the dissipative criterion for
a well-posed IBVP, they were too restrictive for use in practical
numerical applications and did not allow
large boundary data. It took a different approach to formulate a
strongly well-posed harmonic IBVP with Sommerfeld type boundary
conditions. See Sec.~\ref{sec:harm}.

The $Z4$ formalism~\cite{z4} aims at a covariant version of hyperbolic
reduction by expressing the vacuum Einstein equations in the form
\begin{equation}
    G^{\mu\nu} -\nabla^{(\mu}Z^{\nu)} 
       +\frac{1}{2}g^{\mu\nu}\nabla_\rho Z^\rho =0 .
       \label{eq:z4reduced}
\end{equation}
When the vector field $Z^\mu$ is identified with the (generalized)
harmonic conditions ${\cal C}^\mu$, this reduces to the harmonic
formulation. However, the freedom is retained to introduce other gauge
conditions which force $Z^\mu =0$. When only 6 components of
(\ref{eq:z4reduced}) are required to vanish, it has been
shown~\cite{z4sb,constrdamp} that the $Z4$ formalism encompasses the
standard $3+1$ formulations, including the ADM,  NOR, BSSN and KST
systems.

It is possible that the close analogue between the $Z4$ and harmonic
formulations might be used to shed light on the IBVP for $3+1$ systems.
Constraint preserving boundary conditions for such $Z4$ systems have been
proposed~\cite{bonabc, bonabc2,hilditch}. However, as for other $3+1$
formulations, the boundary stability necessary for a strongly
well-posed IBVP has not been established. Nevertheless, the results of
numerical tests are promising. See Sec.~\ref{sec:num}.

\bigskip

\section{Non-reflecting outer boundary conditions}
\label{sec:absorbc}

\bigskip

The correct physical description of an isolated system involves
asymptotic conditions at infinity which ensure that the radiation fields
have the proper $1/r$ falloff and that the total energy and radiative
energy loss are finite. This can be achieved by locating the outer
boundary at ${\cal I}^+$. Instead, current simulations of binary black
holes are carried out with an outer boundary at a large but finite
distance in the wave zone, i.e. many wavelengths from the source. This is
in accord with the standard practice in  computational physics to impose
an artificial boundary condition (ABC) which attempts to approximate the
proper behavior of the exterior region.

At the analytic level, many ABCs are possible, even Dirichlet or Neumann
conditions, provided the proper boundary data is known to allow outgoing
radiation to pass through. However, the determination of the correct
boundary data is a global problem, which requires extending the solution
to ${\cal I}^+$ either by matching to an exterior (linearized or
nonlinear) solution obtained by some other means. The matching approach
has been reviewed elsewhere~\cite{winrev}. As shown by Gustafsson and
Kreiss, the construction of a non-reflecting boundary condition for an
isolated system in general requires knowledge of the solution in a
neighborhood of infinity~\cite{guskreis}.

Even if the outgoing radiation data for the analytic problem were known,
at the numerical level a Dirichlet or Neumann condition would reflect 
waves generated by the numerical error and trap them in the grid. The
alternative approach is an ABC which is non-reflecting for homogeneous
data. Artificial boundary conditions for an isolated radiating system for
which homogeneous data is approximately valid are commonly called
absorbing boundary conditions (see e.g.~\cite{Eng77,Hig86,Tre86,
Bla88,Jia90,Ren}), or non-reflecting boundary conditions (see e.g.
\cite{Hed79,giv,kell,agh}) or radiation boundary conditions (see e.g.
\cite{Bay80,hag1}). Such boundary conditions are advantageous for
computational use. However, local ABCs are in general not perfect.
Typically they cause some partial reflection of an outgoing wave
back into the system~\cite{Lind1,Orsz,Hig86,Ren}. It is only required
that there be no spurious reflection in the limit that the boundary
approaches an infinite sphere.

A traditional ABC for the wave equation is the Sommerfeld condition. For
a scalar field $\Phi$ satisfying the Minkowski space wave equation
(\ref{eq:byp1}) with compact source,  the exterior retarded field has the
form
\begin{equation}
    \Phi=\frac{f(t-r,\theta,\phi)}{r} +\frac{g(t-r,\theta,\phi)}{r^2}
    +\frac{h(t,r,\theta,\phi)}{r^3} ,
    \label{eq:outgoing}
\end{equation}
where $f$, $g$ and $h$ are smooth bounded functions. The simplest case is
the monopole radiation field
\begin{equation}
    \Phi=\frac{f(t-r)}{r}
\end{equation}
which satisfies $(\partial_t+\partial_r) (r\Phi)=0$. This motivates the
use of the Sommerfeld condition
\begin{equation}
  \frac{1}{r}(\partial_t+\partial_r)(r\Phi)|_R= q(t,R,\theta,\phi)
  \label{eq:somm1}
\end{equation}
on a finite boundary $r=R$. However a homogeneous Sommerfeld condition,  
i.e. $q=0$, is exact only for the spherically symmetric monopole  field.
The Sommerfeld boundary data $q$ for the general case (\ref{eq:outgoing})
falls off as $1/R^3$, so that a homogeneous Sommerfeld condition
introduces an error which is small only for large $R$. As an example,
\begin{equation}
    q= \frac{f(t-R)\cos\theta}{R^3}
    \label{eq:qdip}
\end{equation}
for the dipole solution
\begin{equation}
    \Phi_{\mathrm{Dipole}}=\partial_z{\frac{f(t-r)}{r}}
                 =-\left( \frac{f'(t-r)}{r} + \frac{f(t-r)}{r^2} \right)
        \cos\theta .
\end{equation}
A homogeneous Sommerfeld condition at $r=R$ leads to a solution $\tilde
\Phi_{Dipole}$ containing a reflected ingoing wave. For large $R$, it is
given by
\begin{equation}
  \tilde \Phi_{\mathrm{Dipole}} \sim \Phi_{\mathrm{Dipole}}
  + \kappa \frac{F(t+r-2R)\cos\theta}{r},
\end{equation}
where $\partial_t f(t)=F(t)$ and the reflection coefficient has
asymptotic behavior $\kappa=O(1/R^2)$. More precisely, the Fourier mode
\begin{equation}
  \tilde \Phi_{\mathrm{Dipole}}(\omega) 
       = \partial_z\bigg (\frac{e^{i\omega (t-r)}}{r}
         + \kappa_\omega \frac{e^{i\omega (t+r-2R)}}{r} \bigg )
\end{equation}
satisfies the homogeneous boundary condition
$(\partial_t+\partial_r)(r\tilde\Phi_{Dipole}(\omega))|_R=0$ with
reflection coefficient
\begin{equation}
          \kappa(\omega) =\frac{1}{2\omega^2 R^2 +2i\omega R-1}
                 \sim \frac{1}{2\omega^2 R^2} .
\label{eq:skappa}
\end{equation}
Note, from  (\ref{eq:qdip}) and  (\ref{eq:skappa}),
\begin{equation}
           \kappa \sim q R.
           \label{eq:kappaq}
\end{equation}           
Use of this relationship simplifies the determination of the asymptotic
falloff of the reflection coefficient by avoiding an explicit calculation
of the reflected wave. Also note, that if (\ref{eq:somm1}) is replaced by
the second order Sommerfeld condition
\begin{equation}
  \frac{1}{r^3}(\partial_t+\partial_r) r^2(\partial_t+\partial_r)
      (r\Phi)|_R= q_2
  \label{eq:somm2}
\end{equation}
then dipole radiation has homogeneous data $q_2=0$. In this way,
the falloff rate of the reflection coefficient is reduced from
(\ref{eq:skappa}) to $\kappa_2 \sim 1/R^3$.

The exponent $n$ of the $O(1/R^n)$ falloff of the reflection coefficient
is an important measure of the accuracy of an ABC. Such reflection
coefficients can be calculated for linearized gravitational waves on a
Minkowski background, analogous to the above scalar wave example,
using either the Regge-Wheeler-Zerilli~\cite{regge,zerilli,oSmT01p}
perturbative method, as carried out in~\cite{sarbuch,sarbuch2,ruizhbc},
or by the Bergmann-Sachs~\cite{bergsachs} gravitational Hertz
potential method, as carried out in~\cite{isol}. See Sec.~\ref{sec:isol}. 
The main difference in the gravitational case arises
from the gauge modes, which exist 
along with the radiative degrees of freedom. In first order formulations,
this is further complicated by the modes introduced by the auxiliary
variables. The second order harmonic formulation, in which all modes
propagate on the light cone,  is simplest to analyze. See
Sec.~\ref{sec:harm} for a discussion of reflection
coefficients in the harmonic case.

Local ABCs have been extensively applied to linear problems with varying
success~\cite{Lind1,Eng77,Bay80,Tre86,Hig86,Bla88,Jia90}. Some  are local
approximations to exact integral representations of the solution in the
exterior of the computational domain~\cite{Eng77}, while others are based
on approximating the dispersion relation of the so-called one-way wave
equations~\cite{Lind1,Tre86}. Higdon~\cite{Hig86} showed that this last
approach is essentially equivalent to specifying a finite number of
angles of incidence for which the ABCs yield perfect transmission. Local
ABCs have also been derived for the linear wave equation by considering
the asymptotic behavior of outgoing wave solutions~\cite{Bay80}, thus
generalizing the Sommerfeld condition. Although this type of ABC is
relatively simple to implement and has a low computational cost, the final
accuracy can be limited if the assumptions about the behavior of the
waves are oversimplified. See~\cite{giv,Ren,tsy,ryab} for general
discussions.

The disadvantages of local ABCs have led to consideration of nonlocal
versions based on integral representations of the infinite domain
problem~\cite{Tin86,giv,tsy,agh}. Even when the Green
function is known, such approaches were initially dismissed as
impractical~\cite{Eng77}; however, the rapid development of computer
power and numerical techniques has made it possible to implement exact
nonlocal ABCs for the linear wave equation and Maxwell's equations in
3D~\cite{deM,kell2}. If properly implemented, this method can yield
numerical solutions to a linear problem which converge to the exact
infinite domain problem in the continuum limit, while keeping the
artificial boundary at a fixed distance. However, due to nonlocality, the
computational cost per time step usually grows at a higher power with
grid size, $\mathcal{O} (N^4)$ per time step in three spatial dimensions,
than for a local approach~\cite{giv,deM,tsy}.

The extension of ABCs to nonlinear problems is much more difficult. The
problem is normally treated by linearizing the region between the outer
boundary and infinity, using either local or nonlocal linear
ABCs~\cite{tsy,ryab}. The neglect of the nonlinear terms in this region
introduces an unavoidable error at the analytic level. However, even larger
errors are typically introduced in prescribing the outer boundary data.
The correct boundary data must correspond to the continuity of fields and
their normal derivatives when extended across the boundary into the
linearized exterior. This is a subtle global requirement for any
consistent boundary algorithm, since discontinuities in the field or its
derivatives would otherwise act as a spurious sheet source on the
boundary that would contaminate both the interior and the exterior
solutions. However, the fields and their normal derivatives constitute an
over determined set of data for the boundary problem. So it is necessary
to solve a global linearized problem, not just an exterior one, in order
to find the proper data. An expedient numerical method
to eliminate back reflection is the use of {\it sponge layers},
cf.~\cite{sponge}, in which damping terms are introduced into the
evolution equations near the outer boundary.

The designation ``exact ABC'' is given to an ABC
for a nonlinear system whose only error is due to linearization of the
exterior. An exact ABC requires the use of global techniques, such as the
boundary potential method, to eliminate back reflection at the
boundary~\cite{tsy,agh}.  Furthermore, nonlinear waves intrinsically
backscatter, which makes it incorrect to try to entirely eliminate
incoming radiation from the outer region.  For the nonlinear wave
equation, test results presented in~\cite{holvorc1,holvorc2} showed that
Cauchy-characteristic matching outperformed all ABC's in the existent
literature.

It is an extra challenge to apply ABCs to strongly nonlinear hydrodynamic
problems~\cite{giv}. Thompson~\cite{Tho87} generalized a previous
nonlinear ABC of Hedstrom~\cite{Hed79} to treat 1D and 2D problems in gas
dynamics. These boundary conditions performed poorly in some situations
because of difficulties in adequately modeling the field outside the
computational domain~\cite{Tho87,giv}. Hagstrom and
Hariharan~\cite{Hag88} have overcome these difficulties in 1D gas
dynamics by their use of Riemann invariants. They proposed, at the
heuristic level,  a generalization of their local ABC to 3D.

In order to reduce the analytic error, an artificial boundary for a
nonlinear problem must be placed sufficiently far from the strong-field
region. This can increase the computational cost in multi-dimensional
simulations~\cite{Eng77}. There is no ABC which converges
(as the discretization is refined) to the infinite domain exact solution
of a strongly nonlinear wave problem in multi-dimensions, while keeping
the artificial boundary fixed.  When the system is nonlinear and not
amenable to an exact solution, a finite outer boundary condition must
necessarily introduce spurious effects. Attempts to use
compactified Cauchy hypersurfaces which extend the domain to spatial
infinity have failed because the phase of short wavelength radiation
varies rapidly in spatial directions~\cite{Orsz}. In fact, in his pioneering
simulation of binary black holes, Pretorius~\cite{pret1,pret2} used this
effect as a numerical expedience by applying artificial dissipation to
diminish short wavelength error arising from the use of a compactified
outer boundary at spatial infinity. For a recent review of ABCs in the
computational mathematics literature, see~\cite{hagrev}.

The situation for the gravitational IBVP is not as severe as for
hydrodynamics, especially for formulations in which the gauge modes and
radiation modes propagate with the same speed. However, due to
nonlinearities, there is always some error of an analytic nature
introduced by a finite boundary which is independent of
discretization. In general, a systematic reduction of this error can only
be achieved by moving the computational boundary to larger and larger
radii.  There has been recent progress in designing absorbing boundary
conditions for the gravitational field. Buchman and
Sarbach~\cite{sarbuch} have developed higher order local boundary
conditions based upon derivatives of $\Psi_0$ which are non-reflecting up
to any given multipole mode for linearized gravitational waves,
analogous to the switch from a first order
Sommerfeld condition (\ref{eq:somm1}) to the second order condition
(\ref{eq:somm2}). These boundary conditions {\it freeze} the value of
$\Psi_0$ to its initial value, i.e.
\begin{equation}
        \partial_t \Psi_0 =0,
        \label{eq:freeze}
\end{equation}
in order to avoid a $O$th order violation of the compatibility condition
between the initial and boundary data. They have extended this approach
to include quadrupole gravitational waves on a Schwarzschild background 
and also to account for $O(M/R)$ back scatter using a nonlocal
version~\cite{sarbuch2}. For a review of this approach see~\cite{oS07}.
It has been applied to the harmonic formulation in~\cite{ruizhbc} and to
the Z4 formulation in~\cite{hilditch}. These are possibly the best
possible local boundary conditions for the gravitational radiation modes.

Lau~\cite{lau1,lau2,lau3} has formulated an exact ABC
for linearized gravitational waves on a Schwarzschild background. Based
upon the flat space work of~\cite{agh}, he reduces the calculation of the
Green function incorporating the boundary condition for the perturbed
metric to the integration of a radial ODE by using a combined spherical
harmonic and Laplace transform of the Regge-Wheeler-Zerilli equations. He
discusses the trade-off in computational cost for this nonlocal ABC
versus the larger computational domain required by a local condition.
This is similar to the trade-off between characteristic matching and the
application of local boundary conditions.

\bigskip

\section{The Friedrich-Nagy system}
\label{sec:fn}

\bigskip

Friedrich and Nagy~\cite{fn} have presented a theorem establishing the
first strongly well-posed IBVP for Einstein's equations with the
generality to handle an outgoing radiation boundary condition. The
approach uses the energy method for first order symmetric hyperbolic
systems described in Sec.~\ref{sec:md}. Their formulation is based upon
the Einstein-Bianchi system of equations with evolution variables
consisting of an orthonormal tetrad ${\bf e}_{\tilde a}$,  $\tilde a =
(0,1,2,3)$, the associated connection coefficients $\Gamma^{\tilde
a}_{\tilde b \tilde c}$ and Weyl curvature tetrad components $C_{\tilde a
\tilde b \tilde c \tilde d}$. Although it differs from the metric based
formulations used in numerical relativity, the success of their treatment
suggests that many of the underlying ideas should be universally
applicable. In their words, ``There are certainly many possibilities to
discuss the initial boundary value problem and there will be as many ways
of stating boundary conditions. However, all of these should be just
modifications of the boundary conditions given in our theorem''. Perhaps
this is overstated since their formulation is 3rd differential order in
the metric, as opposed to the 2nd order $3+1$ or harmonic  formulations.
Yet, all successful formulations must have the common property of
prescribing data which produces a unique solution to Einstein's
equations.

The tetrad vector ${\bf e}_0$ is chosen to be timelike and tangent to the
boundary. It is used to construct adapted coordinates
$x^\mu=(t,x^i)=(t,x,y,z)$ satisfying
\begin{equation}
       {\cal L}_{{\bf e}_0} t =1\, , \quad   {\cal L}_{{\bf e}_0} x^i =0,
       \label{eq:fncoord}
\end{equation}
so that $e^\mu_0$ plays the role of the evolution vector field $t^\mu$
introduced in Sec.~\ref{sec:bare}. However, the evolution field
now has the metric property of being a timelike unit vector so, as a
reminder, I denote it by $T^\mu=e^\mu_0$. In accord with the notation in
Sec.~\ref{sec:bare}, let the initial hypersurface ${\cal S}_0$ be given
by $t=0$ and the boundary ${\cal T}$ be given by $x=0$, with adapted
coordinates $(t,x^A)$. The tetrad vectors are adapted to the geometry as
follows. Let $N^\mu ={e_1}^\mu \propto -\nabla^\mu x$ be the unit outer
normal to ${\cal T}$.

Extend $N^\mu$ throughout the spacetime manifold ${\cal M}$ by requiring
that it be the unit normal to the hypersurfaces ${\cal T}_c$ given by
$x=c=const>0$. On ${\cal S}_0$, the remaining tetrad vectors $e_A^\mu$,
$A=(2,3)$,  are chosen to be an orthonormal dyad for the  $(t=0, x=c)$
subspaces. They are then propagated throughout ${\cal M}$ by Fermi-Walker
transport along the integral curves of $T^\mu$, which lie in ${\cal
T}_c$. The connection components intrinsic to ${\cal T}_c$ are considered
to be freely specifiable gauge source functions. In addition, the mean
extrinsic curvature $K$ of ${\cal T}_c$ is also considered to be a gauge
source function. Moreover, it is shown that the equation $K=q(x^\mu)$ can
be cast as a quasilinear wave equation which determines ${\cal T}_c$
given initial data corresponding to $x=c$ and $\partial_t x=0$ at $t=0$.

By design, this choice of adapted coordinates and tetrad gauge greatly
simplify the IBVP. The evolution system consists of the gauge conditions
governing the tetrad, the equations relating the connection to the
tetrad, the vacuum equations relating the Weyl curvature to the
connection (which imply the vanishing of the Einstein tensor) and the
vacuum Bianchi identities, i.e. the tetrad version of the equations
\begin{equation}
           \nabla_\mu {C^\mu}_{\nu\rho\sigma}=0.
\end{equation}
The system is over determined and subject to constraints arising from
integrability conditions. Remarkably, it can be reduced to a system with
the properties:

\begin{itemize}

\item The evolution system is symmetric hyperbolic.

\item The boundary matrix (\ref{eq:B3}) admits two ingoing variables
corresponding to combinations of the $ \Psi_0 $ and $ \Psi_4$ 
Newman-Penrose components of the Weyl tensor.

\item Maximally dissipative boundary conditions take the form
\begin{equation}
      \Psi_0 +  \alpha \Psi_4 + \beta \bar \Psi_4 = q,
      \label{eq:fnpsi}
\end{equation}
where $q$ is the (complex) boundary data and $\alpha$ and $\beta$ are
coefficients subject to a dissipative condition. (Conventions here are
chosen to be consistent with Newman-Penrose conventions and differ
from~\cite{fn}.)

\item The subsidiary system governing the constraints is  symmetric
hyperbolic and intrinsic to the ${\cal T}_c$ hypersurfaces, i.e. all
derivatives are tangential to ${\cal T}_c$. This gives rise to constraint
preservation without requiring any further restrictions on the boundary
conditions.

\end{itemize}

The resulting IBVP is strongly well-posed. Given
initial data on ${\cal S}$, including the hyperbolic angle $\Theta$ in
(\ref{eq:hangle}) at the edge ${\cal B}_0$, the boundary data $q$,  a
choice of gauge source functions and a dissipative choice of boundary
condition, there exists a unique solution locally in time. Furthermore,
the solution depends continuously on the data.

Several important points should be noted:

\begin{itemize}

\item The specification of the mean extrinsic curvature $K$ of the
boundary geometrically determines the location of the boundary.   

\item The choice of unit timelike vector $T^a$ tangent to the boundary
represents gauge freedom in the construction of the solution. It induces
a corresponding foliation of the spacetime and the boundary
according to ${\cal L}_T t=1$.

\item The geodesic curvature of the integral curves of $T^a$ constitute
gauge source functions required on the boundary. This gauge freedom feeds
into the adapted coordinates $(t,x^A)$ of the boundary. As a result, the
functional specification of $K(t,x^A)$ becomes gauge dependent. This
complication could be avoided by choosing $T^a$ to be geodesic but at the
expense of possible coordinate  focusing singularities which would affect
the long term existence of the solution.

\item The outgoing null vector $K^a$ and ingoing null vector $L^a$ used
in defining $\Psi_0$ and $\Psi_4$ are determined by the choice of $T^a$
(gauge) and the boundary normal $N^a$ by
\begin{equation}
           K^a=T^a+N^a\, , \quad   L^a=T^a-N^a.
           \label{eq:fnk}
\end{equation}
Friedrich and Nagy are careful to point out that gauge freedom prevents
any meaningful interpretation of  $\Psi_0$ and $\Psi_4$ as either purely
outgoing or ingoing radiation. 

\end{itemize}

\bigskip

\section{The harmonic IBVP}
\label{sec:harm}

\bigskip

The first successful treatment of the harmonic IBVP for Einstein's
equations was carried out using  the pseudo-differential theory described
in Sec.~\ref{sec:fourlap}, which established strong well-posedness in a
generalized sense~\cite{wpgs}. The theory was developed for the second
order formulation of the generalized harmonic formulation
(\ref{eq:harmcond})--(\ref{eq:reduced}). The boundary conditions were
given in Sommerfeld form
in terms of the outgoing null vector $K^a=T^a+N^a$ normal to the
foliation of the boundary.  Here, as in the Friedrich-Nagy approach
(\ref{eq:fnk}), $T^a$ is a future directed timelike unit vector tangent
to the boundary but now it is also chosen to be normal to its foliation
${\cal B}_t$. Recall  that $n^a$,  the normal to the Cauchy foliation
${\cal S}_t$, is not necessarily tangent to the boundary. The motion of
the boundary, characterized by the hyperbolic angle (\ref{eq:hangle}),
 distinguishes $T^a$ from $n^a$.

In the adapted coordinates $x^\mu=(t\ge 0,x\ge 0,x^A)$ described in
(\ref{eq:adapted}), six Sommerfeld boundary conditions for the densitized
metric $\gamma^{\mu\nu} =\sqrt{-g}g^{\mu\nu}$ were given by
\begin{eqnarray}
  K^\mu \partial_\mu  \gamma^{AB} &=&q^{AB}(t,x^A) \label{eq:sombc1} \\
  K^\mu \partial_\mu( \gamma^{tA}+\gamma^{xA} )
          &=&q^{A}(t,x^A) \label{eq:sombc2} \\
  K^\mu \partial_\mu ( \gamma^{tt}+2\gamma^{tx}+\gamma^{xx} )
       &=&q(t,x^A), \label{eq:sombc3}
\end{eqnarray} 
where the $q$'s are freely prescribed Sommerfeld data. The harmonic
constraints (\ref{eq:harmcond}) were used to supply four additional
boundary conditions, which could be expressed in the Sommerfeld form 
\begin{eqnarray}
   \sqrt{-g}{\cal C}^A& = &
      \frac{1}{2}(\partial_t -\partial_x)( \gamma^{tA}-  \gamma^{xA})   
                \nonumber \\
     &+&\frac{1}{2}(\partial_t +\partial_x)( \gamma^{tA} +\gamma^{xA})
       +\partial_B \gamma^{AB}  - \sqrt{-g}\hat \Gamma^A =0
          \label{eq:scbc1} \\                      \nonumber \\
  \sqrt{-g}( {\cal C}^t+{\cal C}^x) &=&
      \frac{1}{2}(\partial_t -\partial_x)(\gamma^{tt}- \gamma^{xx})  
              \nonumber \\
    &+& \frac{1}{2}(\partial_t +\partial_x)
  (\gamma^{tt}+2\gamma^{xt}+\gamma^{xx}) 
    +\partial_A (\gamma^{tA}+\gamma^{xA}) 
           - \sqrt{-g}(\hat \Gamma^t+\hat \Gamma^x)  =0
	     \label{eq:scbc2}  \\
   \sqrt{-g}{\cal C}^t& =&
         \frac{1}{2}(\partial_t -\partial_x)(\gamma^{tt}- \gamma^{tx})  \nonumber \\
   &+& \frac{1}{4}(\partial_t +\partial_x))\bigg((\gamma^{tt}+2
   \gamma^{tx}+\gamma^{xx})
       +(\gamma^{tt}-\gamma^{xx})\bigg)
   +\partial_A \gamma^{tA}  -\sqrt{-g}\hat \Gamma^t =0 .
       \label{eq:scbc3}
\end{eqnarray}
Constraint preservation then follows from the homogeneous wave equation
(\ref{eq:bianchi}). 

The key feature of (\ref{eq:sombc1})--(\ref{eq:scbc3}) is that they form
a sequential hierarchy of Sommerfeld boundary conditions for the metric
variables such that the source terms are given in
terms of derivatives of previous variables in the sequence. For instance,
the terms on the second line of  (\ref{eq:scbc1}) are derivatives of 
$\gamma^{AB}$ and $(\gamma^{tA}+\gamma^{xA})$, whose boundary 
conditions are prescribed previously in (\ref{eq:sombc1}) and (\ref{eq:sombc2}).
This pattern persists for the remaining boundary conditions in the sequence,
i.e.  (\ref{eq:scbc2}) and (\ref{eq:scbc3}). This structure gives rise to
a corresponding sequence of estimates for the variables in the hierarchy,
which is the key for establishing boundary stability and the strong
well-posedness of the IBVP. There is considerable freedom in the boundary
conditions provided this hierarchical structure is preserved.

The well-posedness of the harmonic IBVP was subsequently also established
using estimates for the non-standard energy (\ref{eq:ewave}) associated
with a timelike vector pointing outward from the boundary~\cite{wpe}. 
The hierarchical structure of the boundary conditions corresponds to the
upper triangular property (\ref{eq:triang}), which is sufficient
condition for a well-posed IBVP for a coupled system of quasilinear wave
equations. 

A more general and geometrical version of these results in
terms of a background metric $\gz_{ab}$ was presented in~\cite{isol}. 
The connection $\nablaz_a$ and curvature tensor $\Rz{}^d{}_{cab}$
associated with the background metric  $\gz_{ab}$ have the same tensorial
properties as the corresponding quantities $\nabla_a$ and $R^d{}_{cab}$
associated with $g_{ab}$. In particular, the difference $\nabla_a
-\nablaz_a$ defines a tensor field $C^d_{ab}$ according to
\begin{equation}
  (\nabla_a -\nablaz_a) v^d = C^d_{ab} v^b
  \label{eq:cconn}
\end{equation}
for any vector field $v^b$. In terms of the (nonlinear) perturbation
\begin{equation}
    f_{ab}=g_{ab}-\gz_{ab}
\label{eq:f}
\end{equation}
of the metric from the background,
\begin{equation}
  C^d_{ab}
   = \frac{1}{2} g^{dc}\left( \nablaz_{a} f_{bc} 
     + \nablaz_{a} f_{bc} - \nablaz_c f_{ab} \right).
     \label{eq:chris}
\end{equation}     
Since $\gz_{ab}$ is explicitly known, a solution for $ f_{ab}$ is
equivalent to a solution for $g_{ab}$.   Einstein's equations are given by
 \begin{equation}
     E^{ab}:= G^{ab} -\nabla^{(a}{\cal C}^{b)} 
       +\frac{1}{2}g^{ab}\nabla_d{\cal C}^d =0
       \label{eq:ceq}
\end{equation}
subject to the harmonic constraints
\begin{equation}
                  {\cal C}^d:= g^{ab} C^d_{ab} =0.
                  \label{eq:chc}
\end{equation}

In the adapted coordinates,  the harmonic constraints take the form
\begin{equation}
  { \cal C}^\rho: =
          g^{\mu\nu}( \Gamma^\rho_{\mu\nu}- \Gammaz^\rho_{\mu\nu} ) =0,
    \label{eq:charm}
\end{equation}
so that the background Christoffel symbols $\Gammaz^\rho_{\mu\nu}$ appear
as harmonic gauge source functions. When the harmonic constraints are
satisfied, the reduced Einstein equations 
form the desired quasilinear wave system for $f_{\mu\nu}$,
\begin{equation}
    g^{\rho\sigma}\nablaz_\rho\nablaz_\sigma f_{\mu\nu} 
 = 2 g_{\lambda\tau} g^{\rho\sigma} C^\lambda{}_{\mu \rho}C^\tau{}_{\nu\sigma}
   +4 C^\rho{}_{\sigma(\mu} g_{\nu)\lambda} 
    C^\lambda{}_{\rho \tau}g^{\sigma\tau}
    - 2 g^{\rho\sigma}\Rz^\lambda{}_{\rho\sigma(\mu} g_{\nu)\lambda} .
 \label{eq:beinst}
\end{equation}

The analogue of the Sommerfeld conditions
(\ref{eq:sombc1})--(\ref{eq:scbc3}) are prescribed in terms of the
boundary decomposition of
the metric 
\begin{equation}
      g_{ab} =N_a N_b+H_{ab}\, , \quad H_{ab}= -T_a T_b +Q_{ab},
      \label{eq:bdecom}
\end{equation}
This leads to an orthonormal tetrad $(T^a,N^a,Q^a,\bar Q^a)$ on ${\cal
T}$, where $Q^a$ is a complex null vector tangent to ${\cal B}_t$ with
normalization
\begin{equation}
  Q_{ab}= Q_{(a}\bar Q _{b)}\, , 
  \quad Q^a \bar Q_a=2 \, , \quad  Q^a Q_a=0.
\label{eq:qnorm}
\end{equation}
(The tetrad is unique up to the spin freedom $Q^a \rightarrow e^{i\theta}
Q^a$ which does not enter in any essential way.)  In terms of the
outgoing and ingoing null vector fields $K^a=T^a+N^a$ and $L^a=T^a-N^a$,
respectively, which are normal to ${\cal B}_t$, the metric has the null
tetrad decomposition
\begin{equation}
  g_{ab} = - K_{(a}L _{b)}+Q_{(a}\bar Q _{b)}.
  \label{eq:ntetrad}
\end{equation}

Six Sommerfeld boundary conditions which determine the components of the
outgoing null derivatives  $K^a \nablaz_a f_{bc}$ are then given by
\begin{eqnarray} 
     \frac{1}{2} K^b K^c K^a 
      \nablaz_a f_{bc}  &=& q^a K_a ,
      \label{eq:ak}   \\
    (Q^b K^c K^a
       -\frac {1}{2} K^b K^c Q^a) \nablaz_a f_{bc} &=& q^a Q_a    ,
   \label{eq:aq}\\ 
     (L^b K^c K^a  
   -\frac {1}{2} K^b K^c L^a)
    \nablaz_a f_{bc}&=& q^a L_a ,
  \label{eq:al} \\
          (\frac {1}{2}Q^b Q^c K^a
       - Q^b K^c Q^a) \nablaz_a f_{bc} &=& 2\sigma ,
   \label{eq:sigqq}       
\end{eqnarray}
in terms of boundary data $q^a$ and $\sigma$. The harmonic constraints
provide four additional boundary conditions which, in terms of the null
tetrad, can be expressed in the Sommerfeld form
\begin{eqnarray} 
 -2{\cal C}^a K_a =\left( Q^b\bar{Q}^c  K^a+  K^b K^c L^a
            - K^b \bar{Q}^c Q^a  -  K^b Q^c \bar{Q}^a \right) 
	    \nablaz_a f_{bc}  &=& 0 , 
\label{eq:hk}  \\
 -2{\cal C}^a Q_a  =\left(  L^b Q^c K^a + K^b Q^c  L^a
            - K^b L^c  Q^a+Q^b Q^c  \bar{Q}^a \right)
	    \nablaz_a f_{b c} &=& 0,
\label{eq:hq} \\
 - 2{\cal C}^a L_a  = \left( L^b L^c  K^a+  Q^b \bar{Q}^c L^a
            - \bar{Q}^b L^c Q^a-  Q^b L^c \bar{Q}^a \right)
	    \nablaz_a f_{bc}&=& 0.
 \label{eq:hl}
\end{eqnarray}
As before, constraint preservation follows from (\ref{eq:bianchi}).

Together, (\ref{eq:ak})-- (\ref{eq:hl}) provide Sommerfeld boundary
conditions for the  components of $K^a \nablaz_a f_{bc}$ in the
sequential order $(KK),(QK),(LK),(QQ),(Q\bar Q),(LQ),(LL)$ in terms of
the boundary data and the derivatives of preceding components in the
sequence. This hierarchy of  Sommerfeld boundary conditions satisfies the
requirements given in Sec.~\ref{sec:swev} (Theorem 1 of~\cite{isol}) for
a strongly  well-posed IBVP for the quasilinear hyperbolic system
(\ref{eq:beinst}). See Sec.~\ref{sec:geom} for a geometrical interpretation
of the boundary data and Sec.~\ref{sec:num} for numerical tests.

\bigskip

\subsection{Application to an isolated system}.
\label{sec:isol}

\bigskip

The main application of the gravitational IBVP is to the spherical outer
boundary used in the simulation of an isolated system emitting radiation.
As discussed in Sec.~\ref{sec:absorbc}, in the absence of an exterior
solution, the simplest approach is the use of a Sommerfeld boundary
conditions with homogeneous data. In doing so it is important to take
advantage of the freedom in the form of the  boundary conditions in order
to reduce back reflection.

Sommerfeld boundary conditions consistent with a well posed harmonic IBVP
have wide freedom regarding the addition of (i) partial derivative terms
consistent with the hierarchical structure and (ii) lower order algebraic
terms. Various choices were considered in~\cite{isol}. They were tested
by computing the resulting reflection coefficients for spherical waves in
a Minkowski space background. For this purpose the densitized metric is
approximated to linearized accuracy by  \begin{equation}
\sqrt{-g}g^{\mu\nu}=\eta^{\mu\nu}+ \gamma^{\mu\nu}, \end{equation} where
$\eta^{\mu\nu}$ is the Minkowski metric. The calculation of the
reflection coefficients proceeds as for the scalar wave example in
Sec.~\ref{sec:absorbc}, as modified to deal with the gauge modes. 

Linearized waves in the harmonic gauge can be constructed from the
gravitational analogue of  the Hertz potential~\cite{bergsachs}, which
has the symmetries
\begin{displaymath}
  H^{\mu\alpha\nu\beta}=  H^{[\mu\alpha]\nu\beta}=H^{\mu\alpha[\nu\beta]}
      =H^{\nu\beta\mu\alpha}
\end{displaymath}
and satisfies the flat space wave equation $\partial^\sigma
\partial_\sigma H^{\mu\alpha\nu\beta}= 0$. Then the perturbation
\begin{displaymath}
        \gamma^{\mu\nu} =\partial_\alpha \partial_\beta H^{\mu\alpha\nu\beta}. 
\end{displaymath}
satisfies the linearized Einstein equations $ \partial^\sigma
\partial_\sigma  \gamma^{\mu\nu}=0$ in the harmonic gauge $ \partial_\mu
\gamma^{\mu\nu}=0$. Outgoing waves can be generated from the potential
\begin{equation}
    H^{\mu\alpha\nu\beta}= K^{\mu\alpha\nu\beta}\frac{f(t-r)}{r}\, ,
 \quad  \gamma^{\mu\nu} =
    K^{\mu\alpha\nu\beta}  \partial_\alpha \partial_\beta \frac{f(t-r)}{r},
    \label{eq:Khertz}
\end{equation}
where $ K^{\mu\alpha\nu\beta}$ is a constant tensor.  (All higher
multipoles can be constructed by taking spatial derivatives.)
$K^{\mu\alpha\nu\beta}$ has 21 independent components but the choice
$K^{\mu\alpha\nu\beta}=\epsilon^{\mu\alpha\nu\beta}$ leads to
$\gamma^{\mu\nu}=0$ so there are only 20 independent waves. These can
reduced to 10 pure gauge waves for which the linearized Riemann tensor
vanishes, which correspond to the trace terms in $K^{\mu\alpha\nu\beta}$;
e.g. $K^{\mu\alpha\nu\beta}=\eta^{\alpha\nu}\eta^{\beta\mu}
-\eta^{\mu\nu}\eta^{\alpha\beta}$ leads to a monopole gauge wave. The
trace-free part gives rise to 10 independent quadrupole gravitational
waves, corresponding to spherical harmonics with $(\ell=2,-2\le m\le 2)$
in the two independent polarization states. 

For a boundary at $r=R$, the Sommerfeld derivative in
the outgoing null direction is
\begin{equation}
        K^\mu\partial_\mu = \partial_t  +\partial_r.
\end{equation}
In formulating boundary conditions which minimize back reflection, the
property $K^\mu \partial_\mu f(t-r) =0$ is used to introduce the
appropriate powers of $r$, analogous to the scalar example
(\ref{eq:somm1}). In~\cite{isol}, the optimal choice was found to be the
Sommerfeld hierarchy
\begin{eqnarray}
  \frac{1}{r^2}K_\alpha K_\beta K^\mu \partial_\mu (r^2\gamma^{\alpha\beta})
           &=& q_{KK}\; ,
 \label{eq:modasymKK}\\
  \frac{1}{r^2} K_\alpha Q_\beta K^\mu\partial_\mu(r^2 \gamma^{\alpha\beta} )    
       &=& q_{KQ}\; ,
 \label{eq:modasymKQ}  \\
   \frac{1}{r^2}Q_\alpha\bar Q_\beta
       K^\mu\partial_\mu(r^2\gamma^{\alpha\beta})
      -\frac{\gamma}{r}   &=& q_{Q\bar Q}\; ,
 \label{eq:modasymQbQ} \\
    Q_\alpha Q_\beta K^\mu\partial_\mu  \gamma^{\alpha\beta}
     - Q_{\alpha}K_\beta Q^\mu \partial_\mu \gamma^{\alpha\beta} &=& q_{QQ}\; .
\label{eq:modasymQQ}
\end{eqnarray}
It was found that the data $q_{..}=O(1/R^4)$ for the outgoing
gravitational quadrupole waves and  $q_{..} =O(1/R^3)$ for the outgoing
gauge waves. This implies, in accord with (\ref{eq:kappaq}), that
homogeneous Sommerfeld data gives rise to reflection coefficients
$\kappa=O(1/R^3)$ for the gravitational waves and $\kappa=O(1/R^2)$ for
the gauge waves. These results were confirmed using the
Regge-Wheeler-Zerilli perturbative formulation along with the metric
reconstruction method described in \cite{oSmT01p}. 

The analysis  of linearized  waves shows that $q_{QQ}$ controls the
amplitude of the gravitational radiation passing through the boundary. 
Higher order boundary conditions can be based upon replacing
(\ref{eq:modasymQQ}) by a condition on $\Psi_0$,  which in the linearized
theory controls the radiation in a gauge independent manner. In this
way, the $\Psi_0$ based boundary conditions discussed in
Sec.~\ref{sec:absorbc} can be used  to further increase the $1/R^n$
falloff rate of the reflection coefficients for gravitational waves. 

\bigskip

\section{Constraint preservation}
\label{sec:constr}

\bigskip

The IBVP for Einstein's equations is still not well understood due to a
great extent from complications arising from the constraints. The
Hamiltonian and momentum constraints on the Cauchy data take the
universal form (\ref{eq:hamomc}) in terms of the components of the
Einstein tensor normal to the initial hypersurface. However, there is no
common way to ensure constraint preservation for the various formulations
of Einstein's equations.  Even the constraints themselves take on
different forms.

The Friedrich-Nagy system (see Sec.~\ref{sec:fn}) is based upon the
Einstein-Bianchi equations which is third differential order in terms of
the metric. In that case, by a cleverly designed choice of adapted
coordinates and gauge, constraint propagation is governed by a system
tangential to the boundary. Thus there are no ingoing constraint modes
and constraint preservation is straightforward.

In the strongly well-posed harmonic system described in
Sec.~\ref{sec:harm}, the harmonic conditions ${\cal C}^a$ became
surrogates for the Hamiltonian and momentum
constraints. Because ${\cal C}^a$ satisfies a homogeneous wave
equation, there are four ingoing constraint modes. These could be
eliminated by dissipative boundary conditions with homogeneous data. In
Sec.~\ref{sec:harm}, homogeneous Dirichlet conditions on
${\cal C}^a$ were chosen. This allowed
the constraints to be enforced in terms of first differential Sommerfeld
conditions on the metric. Homogeneous Neumann or Sommerfeld conditions
on the constraints would also ensure constraint preservation but at the
expense of a more complicated coupling with the evolution system for the
metric.

The worldtube constraints which arise in the gravitational version of the
null-timelike  IBVP discussed in Sec.~\ref{sec:civp} present an entirely
different aspect. In that problem, boundary data on a worldtube ${\cal
T}$ and initial data on an outgoing null hypersurface ${\cal N}_0$
determine the exterior spacetime by integration along the outgoing null
geodesics. The worldtube constraints impose conditions on the integration
constants. The Bondi-Sachs formalism~\cite{bondi,sachs} introduces
coordinates  $x^\alpha=(u,r,x^A)$ based upon a family of outgoing null
hypersurfaces ${\cal N}_u$, where $u$ labels the null hypersurfaces,
$x^A$  are angular labels for the null rays and $r$ is a surface area
coordinate.  The evolution system is composed of radial propagation
equations along the outgoing null rays consisting of the hypersurface
equations $G_\mu^u=G_\mu^\nu \nabla_\nu u=0$, which only contain
derivatives tangent to ${\cal N}_u$, and the evolution equations
$G_{AB} - \frac{1}{2} g_{AB} g^{CD}G_{CD}=0$.

The components of Einstein's equations independent of the hypersurface
and evolution equations are worldtube constraints (called supplementary
conditions by Bondi and Sachs),
\begin{eqnarray}
        g^{AB}G_{AB}&=&0 \label{eq:triv} \\
        G_A^r&=&0 \label{eq:consu}\\
        G_u^r&=&0. \label{eq:consA} 
\end{eqnarray}
When the hypersurface and evolution equations are satisfied, the
contracted  Bianchi identity 
\begin{equation}
         \nabla_\nu G_\mu^\nu =0
         \label{eq:bianch}
\end{equation}
implies that these equations need only be satisfied on the worldtube
${\cal T}$. The identity for $\mu=r$ reduces to  $g^{AB}G_{AB}=0$ so that
(\ref{eq:triv}) becomes trivially satisfied. (Here it is necessary that
the worldtube have nonvanishing expansion so that the areal radius $r$ is
a non-singular coordinate.) The identity for $\nu=A$ then reduces to the
radial ODE
\begin{equation}
         \partial_r (r^2 G_A^r) =0,
\end{equation}
so that $G_A^r$ vanishes if it vanishes on ${\cal T}$. When $G_A^r=0$, 
the identity for $\nu=u$ then reduces to
\begin{equation}
        \partial_r  (r^2 G_u^r)=0,
\end{equation}
so that $G_u^r$ also vanishes if it vanishes on ${\cal T}$.

Thus the worldtube constraints reduce to (\ref{eq:consu}) and (
\ref{eq:consA} ), which are equivalent to the condition that the Einstein
tensor satisfy
\begin{equation}
    \xi^{\mu}G_{\mu}^{\nu}N_\nu=0,
    \label{eq:Gcon}
\end{equation}
where $\xi^{\mu}$ is any vector field tangent to the worldtube, whose normal
is $N_\nu$.   These are the boundary analogue of the momentum
constraints for the Cauchy problem. In Stewart's treatment of the $3+1$
IBVP (see Sec.~\ref{eq:fritreul}), it was the Cauchy momentum constraints
which were enforced on the boundary. In the characteristic IBVP, it is
the worldtube constraints (\ref{eq:Gcon}) which must be enforced. They
form three components of the boundary constraints (\ref{eq:bconstr})
proposed by Frittelli and Gomez for the $3+1$ problem. (See
Sec.~\ref{sec:other}.)

This worldtube constraints (\ref{eq:Gcon}) can be interpreted as flux
conservation laws for the $\xi$-momentum contained in the
worldtube~\cite{tam},
\begin{equation}
           P_\xi(u_2)-P_\xi(u_1) = \int_{u_1}^{u_2} dS_\nu \{
     \nabla^\nu \nabla_\mu \xi^\mu - \nabla_\mu \nabla^{(\nu}\xi^{\mu)} \}
\end{equation}
where
\begin{equation}
           P_\xi =\oint dS_{\mu\nu} \nabla^{[\nu}\xi^{\mu]} 
\end{equation} and $dS_{\mu\nu}$ and $dS_\nu$ are the 2-surface and
3-volume elements on the worldtube. When $\xi^\mu$ is a Killing vector
for the intrinsic 3-metric of the world-tube, this gives rise to a strict
conservation law. For the limiting case at ${\cal I}^+$, these flux
conservation laws govern the energy-momentum, angular momentum and
supermomentum corresponding to the asymptotic symmetries~\cite{tam}. For
an asymptotic time translation, they give rise to the Bondi's
famous result~\cite{bondi} relating the mass loss to the square of the
news function.

In terms of the intrinsic metric of the worldtube
\begin{equation}
   H_{\mu\nu}=g_{\mu\nu}-N_\mu N_\nu,
\end{equation}
its intrinsic  covariant derivative ${\cal D}_\mu$ and its extrinsic
curvature
\begin{equation}
        K_{\mu\nu}=H_\mu^\rho \nabla_\rho N_\nu,
\end{equation}
the worldtube constraints (\ref{eq:Gcon}) can be rewritten as
\begin{equation}
  H_\nu^\mu 
    G_{\mu\rho} N^\rho ={\cal D}_\mu (K^\mu_\nu 
          - \delta^\mu_\nu K^\rho_\rho) =0.
    \label{eq:bmom}
\end{equation}    
These are the analogue of (\ref {eq:momc}) for  the Cauchy problem but they
now form a symmetric hyperbolic system because of the timelike nature
of the worldtube. In terms of a dyad (\ref{eq:qnorm}) adapted to the
foliation of the worldtube, this gives rise to the {\bf Worldtube
Theorem}~\cite{josh}:

\bigskip

\noindent {\em Given $H_{ab}$, $Q^a Q^b K_{ab}$ and $K$, the worldtube
constraints constitute a well-posed initial-value problem which
determines the remaining components of the extrinsic curvature $K_{ab}$}.

\bigskip

The theorem constrains the integration constants for the
nullcone-worldtube IBVP. Similarly, they constrain the boundary data
for a $3+1$ IBVP subject to the Frittelli-Gomez conditions.
Unfortunately, for neither of these IBVPs has it been possible to combine
the boundary constraints with the evolution system in a manner consistent
with a strongly well-posed IBVP.

The enforcement of the boundary constraints is an indirect way to enforce
the Hamiltonian and momentum constraints constraints $H=G^{\mu\nu} n_\mu
n_\nu$ and $P^\mu= h^\mu_\rho G^{\rho\nu} n_\nu $, where in the $3+1$
decomposition with respect to the Cauchy hypersurfaces
$$
   h_{\mu\nu}=g_{\mu\nu } + n_\mu n_\nu.
$$
 The more direct
approach commonly used to investigate constraint preservation in the
$3+1$ Cauchy problem is to cast the contracted  Bianchi identity 
(\ref{eq:bianch})  into a hyperbolic system. The results depend upon the
particular formulation.

As a first example, consider the ADM system (\ref{eq:adm}) in which only
the 6 Einstein equations
\begin{equation}
       h_\mu^\rho h_\sigma^\nu {R_\rho}^\sigma=0
      \label{eq:admev}
\end{equation}
are evolved. Application of the contracted Bianchi identity gives rise to
the symmetric hyperbolic constraint propagation system
\begin{eqnarray}
     n^\gamma \partial_\gamma H - \partial_j P^j 
          &=& B^\gamma G_{\nu \gamma }n^\nu \nonumber \\      
     n^\gamma \partial_\gamma P^i  - h^{ij} \partial_j H 
        &=& B^{\mu\gamma}  G_{\nu \gamma }n^\nu,
        \label{eq:adnc}
\end{eqnarray} 
where the coefficients $B^\gamma$ and $B^{\mu\gamma}$ arise from 
Christoffel symbols and do not enter the principal part.  When applied to
the IBVP, a complication arises from the component of the shift normal
to the boundary,
\begin{equation}
    \beta^N=\beta^\mu N_\mu = -\alpha \sinh \Theta 
\end{equation}
in terms  of the lapse $\alpha$ and the hyperbolic angle $\Theta$
(\ref{eq:hangle}) governing the velocity of the boundary. Here $\beta^N
<0$ ($\beta^N >0$) for a boundary which is moving inward (outward) with
respect to the Cauchy hypersurfaces.  An analysis of (\ref{eq:adnc})
shows that only one boundary condition is allowed provided $\beta^N \le
0$, i.e provided the boundary is moving inward. In that case, the theory of
symmetric hyperbolic systems guarantees that all the constraints
would be preserved if the single constraint
\begin{equation}
     H+P^i N_i = G_{\mu\nu} n^\mu K^\nu =0
	   \label{eq:admcon} 
\end{equation}
is satisfied at the boundary, where $K^\mu$ is the outgoing null vector to
the foliation of the boundary. (Additional boundary conditions are
necessary for constraint preservation if  $\beta^N> 0$.)  By virtue of
the evolution system (\ref{eq:admev}), the constraint (\ref{eq:admcon})
is equivalent to
\begin{equation}
           G_{\mu\nu} K^\mu K^\nu =0.
	   \label{eq:raych} 
\end{equation}
This is the Raychaudhuri equation~(cf.~\cite{wald})
\begin{equation}
   K^\mu \partial _\mu \theta +\frac{1}{2}\theta^2+\sigma \bar\sigma =0,
\label{eq:drho}	
\end{equation}
where $\theta$ is the expansion and $\sigma$ is the shear of the outgoing
null rays tangent to $K^\mu$. Thus, for the ADM system, constraint
preservation can be enforced by the Sommerfeld boundary condition
(\ref{eq:drho}) for $\theta$. Unfortunately, although the constraint
system has these attractive properties, the ADM evolution system is only
weakly hyperbolic and consequently leads to unstable evolution.

Next consider the BSSN evolution system, which enforces the 6 Einstein
equations
\begin{equation}
         h_\mu^\rho h_\sigma^\nu {R_\rho}^\sigma
      -\frac{2}{3}h_\mu^\nu H=0.
      \label{eq:bssnmev}
\end{equation}
The contracted Bianchi identity now implies the constraint system
\begin{eqnarray}
     n^\gamma \partial_\gamma H 
     -\partial_j P^j  
          &=&  B^\gamma G_{\nu \gamma }n^\nu \\      
     n^\gamma \partial_\gamma P^i 
      + \frac{1}{3} h^{ij} \partial_j H 
        &=&  B^{\mu\gamma}  G_{\nu \gamma }n^\nu.
	\label{eq:bssnmc}
\end{eqnarray}
This is not symmetric hyperbolic and would not lead to stable constraint
preservation even for the Cauchy problem. This is remedied in the course
of introducing auxiliary variables which reduce the BSSN system to first
order form. Auxiliary constraints are mixed into the evolution system 
(\ref{eq:bssnmev}) and they combine with the constraint system
(\ref{eq:bssnmc}) to form a larger symmetric hyperbolic constraint
system. There is a large freedom in the constraint-mixing parameters  and
gauge conditions. For a particular choice made by  N{\' u}{\~ n}ez and
Sarbach~\cite{nunsar}, the linearization off Minkowski space yields a
symmetric hyperbolic evolution system.  The boundary conditions for this
system are complicated by the normal component of the shift. As discussed
in conjunction with (\ref{eq:advect}), the number of boundary conditions
required by the advection  equations introduced in the first order
reduction depends upon whether $\beta^N$ is positive or negative. This
forces use of a Dirichlet condition, e.g.  $\beta^N=0$, rather than a
Sommerfeld condition on the shift. Constraint preservation holds only in a
certain parameter range, $(b_1 \le1,b_2 \le 1)$ for the boundary conditions
given in equation (97) of~\cite{nunsar}. The particular choice $b_1=0$,
leads to the boundary condition~\cite{olivpc}
\begin{equation}
         H-3  P^i N_i =G_{\mu\nu} n^\mu (n^\nu - 3N^\nu) = {\cal Z},
	   \label{eq:bssncon} 
\end{equation}
where ${\cal Z}$ represents contributions from the auxiliary constraints,
or, by using the evolution system (\ref{eq:bssnmev}),
\begin{equation}
           G_{\mu\nu} L^\mu L^\nu ={\cal Z},
	   \label{eq:inraych} 
\end{equation}
where $L^\mu$ is the ingoing null vector to the boundary. It is a bizarre
feature of the $3+1$ problem that the constraint preserving boundary
condition switches from the outgoing Raychaudhuri form (\ref{eq:raych})
to the ingoing Raychaudhuri form (\ref{eq:inraych}) in going from the ADM
to the BSSN system. The Raychaudhuri equation for the outgoing null
direction cannot be imposed in the allowed range of $(b_1,b_2)$. 

The widely varying nature of constraint enforcement among different
formulations does not provide any apparent insight. However, one problem
common to many first order formulations arises from the advective
derivative $n^\mu \partial_\mu$, which determines whether the auxiliary
variables are ingoing or outgoing at the boundary, depending on the sign
of $\beta^N$.  This problem could be avoided by instead using the
derivative $t^\mu \partial_\mu$ determined by the evolution field, which
can always be chosen tangential to the boundary. That suggests that the
projection operator $ \pi^\mu_\nu$ associated with $t^\mu$, given in
(\ref{eq:tproj}), might be useful in separating out the evolution system
from the constraints, rather than the projection operator $h^\mu_\nu$
used in (\ref{eq:admev}) and (\ref{eq:bssnmev}). This gives rise to many
ways to obtain a symmetric hyperbolic constraint system whose boundary
treatment is independent of $\beta^N$. As a simple example, in adapted
coordinates the evolution system $G^{ij} =\lambda \delta^{ij} G^{tt}$,
with $\lambda>0$, leads via (\ref{eq:bianch}) to the symmetrizable
constraint system
\begin{eqnarray}
    \partial_t G^{tt} +\partial_j G^{tj} &=& \text {lower order terms }
  \nonumber \\
     \partial_t G^{ti} +\lambda \partial_i G^{tt} &=&
    \text {lower order terms} .
\end{eqnarray}
Independently of $\beta^N$, this system requires only 1 boundary
condition to preserve all the constraints.

\bigskip
      
\section{Geometric uniqueness of the IBVP}
\label{sec:geom} 

\bigskip

The solution of the Cauchy problem has the important property of  {\it
geometric uniqueness},  i.e. Cauchy data $(h_{ab},k_{ab})$  on ${\cal
S}_0$ determine a metric $g_{ab}$ which is unique up to
diffeomorphism.  Under a diffeomorphism $\psi$, the data  $(\psi^*
h_{ab},\psi^* k_{ab})$ determines an equivalent metric. As well as being
a pretty result, this has the practical application of allowing numerical
simulations with the same initial data but carried out with different
formulations and different gauge conditions to  produce geometrically
equivalent spacetimes. Friedrich~\cite{hjuerg} has emphasized that this
property remains an unresolved issue for the IBVP.

There are different ways in which this property might be formulated for
the IBVP. The most demanding way would be to require that the data at a
point of the boundary be locally determined by the boundary geometry in
the neighborhood of that point. Such data might include the trace $K$ of
the extrinsic curvature of the boundary, which forms part of the data for
the Friedrich-Nagy system. However, it is clear that at least two more
pieces of data are necessary to prescribe the gravitational radiation
degrees of freedom. In the Friedrich-Nagy system, these two pieces of
data are supplied by the combination (\ref{eq:fnpsi}) of the Weyl tensor
components $\Psi_0$ and $\Psi_4$. However, the associated outgoing and
ingoing  null vectors  are not determined by the local geometry but
depend upon the choice of timelike evolution field $T^a$ tangent to the
boundary, according to (\ref{eq:fnk}). This could be avoided by requiring
these null vectors to satisfy the local geometric condition that they be
principle null directions of the Weyl tensor (cf.~\cite{wald}); but in a
general spacetime this would lead to four choices which would then have
to be incorporated somehow into the evolution system. An alternative,
suggested in~\cite{hjuerg}, is to base the data on the
eigenvector problem  determined by the trace free part of the extrinsic
curvature of the boundary,
\begin{equation} 
    (  K_{ab} -\frac{1}{3} H_{ab} K ) V^b = \lambda H_{ab} V^b.
 \end{equation} 
As in the preceding presentation, $H_{ab}$ is the intrinsic metric of the
boundary. For a spherical worldtube $r=R$ in
Minkowski space,
\begin{equation}
             K_{ab} -\frac{1}{3} H_{ab} K = \frac{1} {3R}( H_{ab} 
      +3 \tilde T_a \tilde T_b)
 \end{equation} where $\tilde T_a$ is a timelike eigenvector. This raises
the possibility of whether this eigenvector problem can be used to pick
out a locally preferred timelike direction $\tilde T_a$ in the curved space case.
Similar algebraic properties of the extrinsic curvature hold under roundness
conditions which are typically satisfied by the artificial outer boundary
of an isolated system. However, whether such an approach can be
incorporated into the evolution system and whether the two radiation
degrees of freedom can be encoded in the extrinsic curvature are not
obvious. 

Neither of the two strongly well-posed formulations of the IBVP described
in Sec's~\ref{sec:fn} and~\ref{sec:harm} are based upon purely local
geometric data. In both of them, a foliation of the boundary consistent
with a choice of evolution field plays an essential nonlocal role. This
suggests that a version of geometric uniqueness based upon purely local
data might not be possible.  The prescription of an evolution field
$t^a$ as part of the boundary data provides the necessary structure to pose
a version of geometric uniqueness~\cite{juerg,disem}. As explained in
Sec's~\ref{sec:bare} and~\ref{sec:initial},  the flow of the evolution
field carries the initial edge ${\cal B}_0$ into a foliation ${\cal B}_t$
of the boundary; and it carries the initial Cauchy data into a stationary
background metric $\gz_{ab}$ according to (\ref{eq:gz}). Thus the
evolution field provides the two essential structures to geometrize the
boundary data: the foliation ${\cal B}_t$ determines the outgoing null
direction $K^a$ and the preferred background metric allows the Sommerfeld
derivative to be expressed covariantly as $K^a \nablaz_a$ in terms of the
background connection. Under a diffeomorphism, the evolution field
transforms according to $t^a \rightarrow \psi_* t^a$ with the consequence
that  $\gz_{ab} \rightarrow \psi^* \gz_{ab} $. 

The boundary data $q^a$ and $\sigma$ for the covariant version of the
covariant Sommerfeld conditions (\ref{eq:ak})--(\ref{eq:hl}) then have
the geometric interpretation that
\begin{equation}
     q^a = K^b (\nabla_b - \nablaz_b) K^a
     \label{eq:qdata}
\end{equation}
is the acceleration of the outgoing null vector $K^a$ relative to the
background acceleration, and
\begin{equation}
      \sigma =  
     \frac{1}{2}Q^a Q^b(\nabla_a - \nablaz_a) K_b, \\
           \label{eq:shdata}
\end{equation} is the shear of $K^a$ relative to the background. The use
of the shear in posing geometrical boundary conditions for the harmonic
formulation was also suggested in~\cite{ruizhbc}.  The  rotation freedom
in the dyad dependence (\ref{eq:shdata}) can be removed by introducing
the rank-2 shear tensor
\begin{equation}
               \sigma^{ab} =\frac{1}{2}(Q^{ac}Q^{bd}
    - \frac{1}{2} Q^{ab}Q^{cd})(\nabla_c - \nablaz_c) K_d\, , \quad 
       \sigma = Q_a Q_b \sigma^{ab},
\end{equation}                        
with $\sigma^{ab}\nabla_b t=0$.

By construction, all quantities involved in the boundary conditions map
as tensor fields under a diffeomorphism $\psi$. As a result of the
covariant form of the generalized harmonic equations  (\ref{eq:ceq}) and
(\ref{eq:chc}), the solution $f_{ab}=g_{ab}-\gz_{ab}$ also maps
as a tensor field.  The metric $g_{ab}$ satisfies
the generalized harmonic condition (\ref{eq:charm})
with respect to the background $\gz_{ab}$ and the mapped
metric  $\psi^* g_{ab}$ satisfies the generalized harmonic condition with
respect to $\psi^*  \gz_{ab}$

This can be taken one step further~\cite{disem}. As characterized
in~\cite{hawkel}, the Cauchy data $h_{ab}$ and $k_{ab}$ can be
interpreted as fields $\tilde h_{ab}$ and $\tilde k_{ab}$ on a {\it disembodied}
3-manifold  $\tilde {\cal S}_0$  via its embedding  ${\cal S}_0$ in the
4-dimensional spacetime manifold ${\cal M}$.  A similar approach applies
to the boundary data. Let
\begin{equation}
        q^a = q_N N^a + q_{\cal T}^a,
\end{equation}
so that $q_{\cal T}^a$ is tangent to the boundary ${\cal T}$. Then the
fields $q_N$, $q_{\cal T}^a$ and $  \sigma^{ab}$ are intrinsic to the
boundary. Along with the Cauchy data and the hyperbolic angle $\Theta$ at the
edge ${\cal B}_0$, they can be induced by the embedding of
a disembodied version of data. This leads via the well-posedness of the IBVP with 
Sommerfeld data to a harmonic version of a

\medskip

\noindent Geometric Uniqueness Theorem:

\bigskip

\noindent  Consider the 3-manifolds $\tilde {\cal T}$ and $\tilde {\cal S}_0$
meeting in an edge $\tilde {\cal B}_0$. On  $\tilde {\cal S}_0$ prescribe
the symmetric tensor fields $\tilde h_{ab}$  and $\tilde k_{ab}$, subject
to the Hamiltonian and momentum constraints and the condition that
$\tilde h_{ab}$ be a Riemannian metric . On $\tilde{\cal B}_0$ prescribe
the scalar field $\tilde \Theta$. On $\tilde{\cal T}$ prescribe a smooth
foliation $\tilde{\cal B}_t$, parametrized by a scalar function $\tilde
t$, the scalar field $\tilde q_N$, the vector field $\tilde q_{\cal T}^a$
and the rank-2 tensor field $\tilde \sigma^{ab}$. Then, after embedding
$\tilde{\cal S}_0 \cup \tilde {\cal T}$ as the boundary ${\cal S}_0 \cup
{\cal T}$ of a 4-manifold ${\cal M}$ as depicted in
Fig.~\ref{fig:bound}, this provides the Sommerfeld boundary data for a
vacuum spacetime in a neighborhood of ${\cal B}_0$ which is unique up to
diffeomorphism.

\bigskip

Because the boundary data contain gauge information, this version of
geometric uniqueness is weaker than for the Cauchy problem. It is an open
question whether the boundary data can be prescribed purely in terms of
local geometric objects~\cite{hjuerg}. Note that the data contains no
information about the 3-metric of the boundary, not even that it is a
timelike 3-manifold. The geometrical interpretation of the data involves
the metric of the embedded spacetime, whose existence is in the content
of the theorem. The diffeomorphism freedom lies in the freedom in the
embedding and in the choice of evolution field $t^a$, which determines
the gauge and background geometry. See~\cite{reulsarrev} for a discussion
of these issues in the context of linearized gravitational theory.

The spatial locality of the solution can be extended, say to a boundary
with spherical topology, by patching solutions together. However, the
locality in time presents a more complicated problem regarding the
maximal development of the solution.  For instance, the time foliation
might develop a gauge pathology which prematurely stops the evolution.
That makes it unclear how the maximal development for the Cauchy
problem, as constructed by Choquet-Bruhat ~\cite{gerochbr}, might be
generalized to the IBVP. A restart of the evolution at an intermediate
time in order to extend the solution would introduce a new gauge, new
initial data, a new evolution field and thus a new background metric. A
maximal development based upon the original background would have to be
based upon a maximal choice of evolution field. 

The geometric nature of the Sommerfeld conditions
(\ref{eq:ak})--(\ref{eq:sigqq}) allows them to be formally applied to any metric
formulation. In a $3+1$ formulation, the metric has the decomposition
\begin{equation}
      g_{\mu\nu} = -n_\mu n_\nu +\hat N_\mu \hat N_\nu +Q_{\mu\nu},
      \label{eq:cdecom}
\end{equation}
where $\hat N_\mu$ is the unit normal to the boundary which lies in the
Cauchy hypersurfaces and, as before,  $Q_{\mu\nu}= Q_{(\mu}\bar Q
_{\nu)}$ is the 2-metric intrinsic to its foliation ${\cal B}_t$. The
Sommerfeld boundary conditions (\ref{eq:ak})-- (\ref{eq:aq}) and
(\ref{eq:sigqq}) supply boundary data for the $\hat N^\mu \hat N^\nu
k_{\mu\nu}$, $Q^\mu \hat N^\nu k_{\mu\nu}$ and $Q^\mu Q^\nu k_{\mu\nu}$
components of the extrinsic curvature of the Cauchy foliation. However,
(\ref{eq:al}) supplies the boundary data for the normal component of the
shift, which for many $3+1$ formulations would require a Dirichlet
condition that fixes its sign. The remaining
Sommerfeld boundary conditions (\ref{eq:hk})--(\ref{eq:hl}), which
enforce the harmonic constraints, would also require modification
depending upon the particular $3+1$ gauge conditions. See~\cite{nunsar}
for a discussion relevant to the BSSN formulation. Numerical tests would
be necessary to study whether application of 
(\ref{eq:ak})--(\ref{eq:aq}) and (\ref{eq:sigqq}) would improve the
performance over the present boundary treatment of $3+1$ systems.

\bigskip

\section{Numerical tests}
\label{sec:num}

\bigskip

Post and Votta~\cite{postvot} have emphasized that ``Verification and
validation establish the credibility of code predictions. Therefore it's
very important to have a written record of verification and validation
results.'' The {\em validation} of a code implies that its predictions
are in accord with observed phenomena. For the present status of
numerical relativity, in the absence of any empirical observations,
the burden falls completely on {\em verification}.  Post and Votta list five
verification techniques:

\begin{enumerate}

\item ``Comparing code results with an exact answer''.

\item ``Establishing that the convergence rate of the truncation error
with changing grid spacing is consistent with expectations''.

\item ``Comparing calculated with expected results for a problem
especially manufactured to test the code''.

\item ``Monitoring conserved quantities and parameters, preservation of
symmetry properties and other easily predictable outcomes''.

\item ``Benchmarking -- that is, comparing results from those with
existing codes that can calculate similar problems''.

\end{enumerate}

The importance of the first four techniques has now been recognized by
most numerical relativity groups and their implementation in practice has
improved the integrity of the field. Individual groups cannot easily
carry out the fifth technique independently. This was the motivation
behind formation of the Apples with Apples (AwA)
Alliance~\cite{awa}. 

The early attempts at developing numerical codes were primarily judged by
their ability to simulate black holes, understandably because of the
astrophysical importance of quantifying that system. When the difficulty with
numerical stability became apparent, there was increased focus on a better
mathematical and computational understanding of the analytic and
numerical algorithms. Only a few groups had based their codes upon
symmetric or strongly hyperbolic formulations of Einstein's
equations and fewer had even begun to worry about how to apply
boundary conditions.  The cross fertilization between computational
mathematics and numerical relativity was entering a productive stage.  At
the same time, standardized tests were developed by the AwA Alliance in 
order to isolate problems, calibrate accuracy and compare code
results, \url{http://www.ApplesWithApples.org}.

Such testbeds  have been historically used in computational
hydrodynamics. There are two fundamentally different types. One compares
simulations of a physically important process, such as the binary black
hole problem.  The second type involve idealized situations which isolate
problems, such as the  ``shock tube'' test  in computational fluid
dynamics. This is the type of testbed considered by the AwA Alliance.

The first tests were designed to study evolution algorithms in the
absence of boundaries~\cite{mex1,mex2}. Five tests were based upon a
toroidal 3-manifold (equivalent to periodic boundary conditions):

\begin{itemize} 

\item The {\it robust stability} test evolves random initial data in the
linearized regime. This is a pass/fail test designed as a screen to
eliminate unstable codes.

\item The {\it linearized wave} test propagates a periodic plane wave
either parallel or diagonal to an axis of the 3-torus. The test checks the
accuracy in tracking both the amplitude and phase of the wave.

\item The {\it gauge wave} test is a pure gauge version of the linearized
wave test, but with amplitude in the non-linear regime.

\item The {\it shifted gauge wave} test is based upon a gauge wave with
non-vanishing shift. Both the gauge wave and shifted gauge wave tests are
challenging because of exponentially growing modes in the analytic
problem~\cite{babev}.

\item The {\it Gowdy wave} test simulates an expanding or contracting
toroidal spacetime, which contains a plane polarized gravitational wave in a
genuinely curved, strong field context. 

\end{itemize}

The wave tests provide exact solutions which allow convergence
measurements. Instabilities are monitored by the growth of the
Hamiltonian constraint. Test results were carried out for codes based
upon numerous formulations: harmonic,  Friedrich-Nagy, NOR, and several
versions of BSSN and ADM. See~\cite{mex2} for the test results.

Tests of the Cauchy evolution algorithm cull out algorithms whose
boundary stability is doomed from the outset. A subsequent plan for
boundary tests was formulated by opening  up one axis of the 3-torus to
form a manifold with boundary. This has the advantage that the boundaries
are smooth 2-tori, thus avoiding the complication of sharp boundary
points. This could later be extended to opening up all three axes to test
performance with a cubic boundary. For the robust stability test the
boundary data consist of random numbers. For the wave tests, the boundary
data is supplied by the exact (or linearized) solution. See~\cite{awa}
for the detailed specifications of the five AwA boundary tests.

These boundary tests were first formulated and applied in the early
development of  boundary algorithms for harmonic codes. The robust
stability test~\cite{robust} was used to verify the stability of a code
based upon a well posed IBVP for linearized harmonic
gravity~\cite{szilschbc}. Subsequently, the gauge wave tests were carried
out with a harmonic code whose underlying IBVP was well-posed for
homogeneous Dirichlet and Neumann conditions~\cite{harl,mBbSjW06}. This
revealed problems in the very nonlinear regime, where the approximation
of small boundary data was violated. The shifted gauge wave test posed an
additional difficulty, beyond the unstable analytic modes that already
challenged the Cauchy evolution test~\cite{bab,babev}. The periodic time
variation of the shift produced an effective oscillation of the
boundaries which blue shifted and trapped the error resulting from the
reflecting boundary conditions. This led to unstable behavior in the
nonlinear regime.

After formulation of the strongly well-posed harmonic IBVP described in
Sec.~\ref{sec:harm},  the Sommerfeld boundary conditions were implemented
and tested in a harmonic code~\cite{harmsomm,harmexcis,seil,seilthesis}.
Numerical stability and accurate phase and amplitude tracking were
confirmed by the robust stability and linearized wave tests.  An
important attribute of strong well-posedness is the estimate of boundary
values provided by the energy conservation obeyed by the principal part
of the equations. This boundary stability extends to the semi-discrete
system obtained by replacing spatial derivatives by finite differences
obeying {\it summation by parts} (the discrete counterpart of integration
by parts), so that energy conservation caries over to the semi-discrete
problem~\cite{kreissch}. Stability then extends to the fully discretized
evolution algorithm obtained with an appropriate time integrator, such as
Runge-Kutta~\cite{Kreiss-Wu}. It was found that these discrete
conservation laws were both effective and essential in controlling the
exponential analytic modes latent in the gauge wave test.
Although the analytic proof of well-posedness given
in~\cite{wpe} was based upon a scalar wave energy differing by a small
boost from the standard energy, it is interesting that these successful
code tests were based upon the standard energy. This confirms the
robustness of the underlying approach.

The oscillating boundaries in the shifted gauge wave test excite a
different type of long wavelength instability which could not be
suppressed by purely numerical techniques.  Knowledge of the Sommerfeld
boundary data allows the wave to enter and leave the boundaries, but the
numerical error, although small and convergent, excites an exponential
mode of the analytic problem. However, because this instability violates
the harmonic constraints it was possible to suppress it by a harmonic
constraint adjustment of the form
(\ref{eq:creduced2})~\cite{harmsomm,babev}. This example emphasizes the
importance of understanding instabilities in the analytic problem in
order to control them in a numerical simulation.

Other wave solutions which have been used for numerical tests are
the Teukolsky waves~\cite{teuk}, which are linearized spherical waves
appropriate for testing a spherical boundary, and the nonlinear Brill
waves~\cite{brill},  which are useful for testing wave propagation during
collapse to a black hole. Teukolsky wave tests of the harmonic Sommerfeld
conditions confirm that the constraint violation error due to homogeneous
outer boundary data drops to numerical truncation error as the wave
propagates off the grid~\cite{seil}. Furthermore, when constraint damping
is applied to the interior of the grid, the error drops to machine round-off.
For Brill wave tests with the same code, homogeneous Sommerfeld
conditions lead to considerable back reflection off the boundary, as
expected from the discussion in Sec.~\ref{sec:absorbc}.  In~\cite{improved},
Teukolsky waves were also used to test an implementation of the improved
higher order harmonic boundary conditions proposed by Buchman and
Sarbach~\cite{sarbuch2}.

Rinne, Lindblom and Scheel~\cite{oRlLmS07} have developed a numerical
test for comparing back reflection of waves from a spherical boundary.
First, using perturbative techniques, they construct a reference solution
for a linearized gravitational wave propagating on an exterior
Schwarzschild background. Then the full numerical code is run with a
finite spherical outer boundary. The error with respect to the reference
solution is used to compare different choices of boundary conditions.
Using the first order harmonic code described in~\cite{lLmSlKrOoR06},
they used this test to compare several boundary conditions. The
comparisons of homogeneous Sommerfeld boundary conditions were
 in accord with the theoretical expectations discussed in
Sec's~\ref{sec:absorbc} and~\ref{sec:harm}. The higher order
Sommerfeld conditions and the
$\Psi_0$ freezing condition both produced less back refection than the
first order conditions (\ref{eq:modasymKK}) -(\ref{eq:modasymQQ}). The
test was also used to reveal the spurious effects arising from
a sponge boundary condition and also from the combination of 
numerical dissipation and spatial compactification used by
Pretorius~\cite{pret1,pret2}. In an independent follow up of this
test~\cite{ruizhbc}, the advantages of
higher order Sommerfeld conditions were confirmed.

Although $3+1$ codes have been predominant in binary black hole
simulations, boundary tests have not been very extensive as  compared to
harmonic codes. This is especially pertinent for the BSSN system.
Boundary tests based on a gravitational wave
perturbation of the Schwarzschild exterior have been carried out for the
KST system~\cite{caltechbc}. The results revealed instabilities although
they showed how improvements to the boundary conditions could reduce
constraint violation. The robust stability and Brill wave tests have been
applied to boundary conditions for a symmetric hyperbolic version of the
Einstein-Christoffel system~\cite{sarbtigl}. Although the Cauchy problem
for this system is well-posed, both tests revealed instabilities due to
the boundary algorithm. Again the tests were useful guides for
understanding the problems with the boundary treatment.

C. Bona and C. Bona-Casas~\cite{gowdy} have made the first application of
the Gowdy wave boundary test. They show how it can be applied to a
symmetric hyperbolic version of the first order Z4 formalism. Here, as in
many other studies, the Cauchy problem is well-posed but the strong
well-posedness of the IBVP depends on the boundary condition. Improper
boundary conditions can lead to instability and/or constraint  violation.
They first demonstrate that the test is effective at investigating
methods for preserving the energy constraint for the Z4 system in a
strong field environment. In a subsequent work~\cite{z4cpbc}, they apply 
both the Gowdy wave and robust stability boundary  tests in an expanded
study of constraint violation in the Z4 framework. The robust stability
case is applied both with opening up one axis of the 3-torus and with a
fully cubic boundary. This allows testing SBP algorithms at the corners
and edges. The results show numerical stability of the proposed boundary
algorithms in the linearized regime. The Gowdy wave test extends this
study to constraint preservation in the nonlinear regime. The results
provide further evidence of numerical stability and show that
constraint violation can be kept at the level of discretization error.

\section{Open questions}
\label{sec:quest}

\bigskip

The IBVP has analytical, computational, geometrical  and physical
aspects. The analytic goal is a strongly well-posed IBVP, which is the
prime necessity for the computational goal of an accurate evolution
algorithm. It is also the raison d'etre for the geometric goal of a
gauge invariant formulation of the boundary data. The prime physical
goal, at present, is the accurate simulation of binary black holes. The
binary black hole problem has taken a course of its own, which has been
remarkably successful in view of the gaps in our current understanding
of the other aspects of the IBVP.

Some important open questions which would help close those gaps
are:

\begin{itemize}

\item {\bf Question 1.} Is there a strongly well-posed IBVP based upon a
$3+1$ formulation?

\bigskip

Some insight into this question would be provided by the answer to

\item {\bf Question 2.} Can the necessary boundary data be represented by 
gauge invariant, local geometric objects?

\bigskip

In the Friedrich-Nagy treatment, there are three pieces of boundary
data which are not pure gauge: the trace $K$ of the extrinsic boundary
curvature, which determines the location of the boundary, and the
Weyl curvature components encoding the two radiation degrees
of freedom. In the harmonic system, the Sommerfeld data which
encode the radiation consist of the shear, or the Weyl curvature
components for a higher order condition. This leads to

\item {\bf Question 3.} In the harmonic formulation, can the
trace $K$ be used as gauge invariant boundary data? 

For the Cauchy problem, there exists a maximal
development of the solution~\cite{gerochbr}.

\item {\bf Question 4.} What is the proper formulation of the maximal
development of the IBVP?

\bigskip

\end{itemize}

The quote of Turing at the beginning of this review expresses the
relative degree of difficulty between the Cauchy problem and the IBVP.


\begin{acknowledgments}


This work was supported by NSF grant PHY-0854623 to the
University of Pittsburgh. I have benefited
from numerous discussions with H. Friedrich,
H-O. Kreiss, O. Reula, B. Schmidt and O. Sarbach.

\end{acknowledgments}

\end{document}